\documentclass[twocolumn,twoside]{article}

\usepackage{a4}
\usepackage{bm}
\usepackage{amssymb}
\usepackage{amsmath}
\usepackage[numbers,square,sort&compress]{natbib}
\usepackage{graphicx}
\usepackage{caption}
\newcommand{\beq}{\begin{equation}}
\newcommand{\eeq}{\end{equation}}
\newcommand{\bea}{\begin{eqnarray}}
\newcommand{\eea}{\end{eqnarray}}
\newcommand{\req}[1]{Eq.~(\ref{#1})}
\newcommand{\kB}{k_\mathrm{B}}
\newcommand{\aB}{a_\mathrm{B}}
\newcommand{\aion}{a_\mathrm{i}}
\newcommand{\alphaf}{\alpha_\mathrm{f}}
\newcommand{\am}{a_\mathrm{m}}
\newcommand{\ChiH}{\chi^\mathrm{H}}
\newcommand{\ChiA}{\chi^\mathrm{A}}
\newcommand{\chiH}{\chi^\mathrm{H}}
\newcommand{\chiA}{\chi^\mathrm{A}}
\newcommand{\dd}{\mathrm{d}}
\newcommand{\EF}{\epsilon_\mathrm{F}}
\newcommand{\emissivity}{\varepsilon_\omega}
\newcommand{\etal}{et al.}
\newcommand{\gcc}{\mbox{g~cm$^{-3}$}}
\newcommand{\Gamc}{\Gamma_{\mathrm{c}}}
\newcommand{\Gami}{\Gamma_{\mathrm{Coul}}}
\newcommand{\Gamr}{\Gamma_{\mathrm{r}}}
\newcommand{\Gamri}{\Gamma_{\mathrm{ri}}}
\newcommand{\gammar}{\gamma_\mathrm{r}}
\newcommand{\gfact}{g_\mathrm{i}}
\newcommand{\Kc}{K_\mathrm{c}}
\newcommand{\khat}{\hat{\mathbf{k}}}
\newcommand{\Kp}{K_\perp}
\newcommand{\lambdi}{\lambda_\mathrm{i}}
\newcommand{\low}{^{\phantom{}}}
\newcommand{\mel}{m_\mathrm{e}}
\newcommand{\mion}{m_\mathrm{i}}
\newcommand{\Mspin}{(2\sion+1)}
\newcommand{\NH}{N_\mathrm{H}}
\newcommand{\nion}{n_\mathrm{i}}
\newcommand{\Nion}{N_\mathrm{i}}
\newcommand{\omc}{\omega_\mathrm{c}}
\newcommand{\omci}{\omega_\mathrm{ci}}
\newcommand{\ompe}{\omega_{\mathrm{pe}}}
\newcommand{\opac}{\varkappa}
\newcommand{\pF}{p_\mathrm{F}}
\newcommand{\rhos}{\rho_\mathrm{s}}
\newcommand{\Rph}{R_\mathrm{ph}}
\newcommand{\sigmaT}{\sigma_\mathrm{T}}
\newcommand{\sion}{s_\mathrm{i}}
\newcommand{\sSB}{\sigma_\mathrm{SB}}

\newcommand{\Tbb}{T_\mathrm{bb}}
\newcommand{\Tc}{T_\mathrm{crit}}
\newcommand{\Teff}{T_\mathrm{eff}}

\newcommand{\Ts}{T_\mathrm{s}}
\newcommand{\xg}{x_\mathrm{g}}
\newcommand{\ycol}{y_\mathrm{col}}
\newcommand{\zete}{\zeta_\mathrm{e}}
\newcommand{\zeti}{\zeta_\mathrm{i}}
\newcommand{\Znuc}{Z_\mathrm{n}}

\begin{document}
\renewcommand{\figurename}{{\small\textbf{Fig.}}}
\renewcommand{\tablename}{{\small\textbf{Table}}}
\renewcommand{\contentsname}{Contents}

\title{\textbf{Atmospheres and radiating surfaces of neutron
stars}}

\author{Alexander Y. Potekhin$^{1,2,3}$}
\date{\normalsize$^1${Ioffe Physical-Technical Institute,
Politekhnicheskaya 26, 194021 Saint Petersburg,
Russian Federation\\
 E-mail: palex@astro.ioffe.ru}\\
$^2${Centre de Recherche Astrophysique de Lyon (CNRS, UMR 5574);
 Ecole Normale Sup\'{e}rieure de Lyon;
  Universit\'e de Lyon, Universit\'e
Lyon 1;
 Observatoire de Lyon, 9 avenue Charles Andr\'e,
69230 Saint-Genis-Laval, France}\\
$^2${Central Astronomical Observatory of RAS at Pulkovo,
Pulkovskoe Shosse 65, 196140 Saint Petersburg, Russia}
}

\maketitle

\begin{abstract}
The early 21st century witnesses a dramatic rise in the
study of thermal radiation of neutron stars. Modern space
telescopes have provided a wealth of valuable information
which, when properly interpreted, can elucidate the physics
of superdense matter in the interior of these stars. This
interpretation is necessarily based on the theory of
formation of neutron star thermal spectra, which, in turn,
is based on plasma physics and on the understanding of
radiative processes in stellar photospheres. In this paper,
the current status of the theory is reviewed with particular
emphasis on neutron stars with strong magnetic fields. In
addition to the conventional deep (semi-infinite)
atmospheres, radiative condensed surfaces of neutron stars
and "thin" (finite) atmospheres are considered.\\


\end{abstract}

\vspace*{1ex}
\tableofcontents
\thispagestyle{empty}

\section{Introduction}
\label{sect:intro}

Neutron stars are the most compact of all stars ever
observed: with a typical mass $M\sim (1$\,--\,$2)\,
M_\odot$, where $M_\odot=2\times10^{33}$~g is the solar
mass, their radius is $R\approx10$\,--\,13 km. The mean
density of such star is $\sim10^{15}$ \gcc, i.e., a few
times the typical density of a heavy atomic nucleus
$\rho_0=2.8\times10^{14}$~\gcc. The density at the
neutron-star center can exceed $\rho_0$ by an order of
magnitude. Such matter cannot be obtained in a laboratory,
and its properties still remain to be clarified. Even its
composition is not completely known, because neutron stars,
despite their name, consist not only of neutrons.  There are
a variety of theoretical models to describe neutron-star
matter (see~\cite{NSB1} and references therein), and a
choice in favor of one of them requires an analysis and
interpretation of relevant observational data. Therefore,
observational manifestations of the neutron stars can be
used for verification of theoretical models of matter in
extreme conditions \cite{Fortov}. Conversely, the progress
in studying the extreme conditions of matter provides
prerequisites for construction of  neutron-star models and
adequate interpretation of their observations. A more
general review of these problems is given in \cite{P10ufn}.
In this paper, I will consider more closely one of
them, namely the formation of thermal electromagnetic
radiation of neutron stars.

Neutron stars are divided into accreting and isolated ones.
The former ones accrete matter from outside,  while an
accretion onto the latter ones is negligible. There are also
transiently accreting neutron stars (X-ray transients), whose
active periods (with accretion) alternate with quiescent
periods, during which the accretion almost stops. The bulk
of radiation from the accreting neutron stars is due to the
matter being accreted, which forms a circumstellar disk,
accretion flows, and a hot boundary layer at the surface.
At contrast, a significant part of radiation from isolated
neutron stars, as well as from the transients in quiescence,
appear to originate at the surface or in the atmosphere. To
interpret this radiation, it is important to know the
properties of the envelopes that contribute to the spectrum
formation. On the other hand, comparison of theoretical
predictions with observations may be used to deduce these
properties and to verify theoretical models of the dense
magnetized plasmas that constitute the envelopes.

We will consider the outermost envelopes of the 
neutron stars -- their atmospheres. A stellar atmosphere is
the plasma layer in which the electromagnetic spectrum is
formed and from which the radiation escapes into space
without significant losses. The spectrum contains a valuable
information on the chemical composition and temperature of
the surface, intensity and geometry of the magnetic field,
as well as on the stellar mass and radius.

In most cases, the density in the atmosphere grows with
increasing depth gradually, without a jump, but stars
with a very low temperature or a superstrong magnetic
field can have a solid or liquid surface. Formation of the
spectrum with presence of such a surface will also be
considered in this paper.

\section{Basic characteristics of neutron stars}

\subsection{Masses and radii} 

The relation between mass $M$ and radius $R$ of a star is
given by a solution of the hydrostatic equilibrium equation
for a given equation of state (EOS), that is the dependence
of pressure $P$ on density $\rho$ and temperature $T$, along
with the thermal balance equation. The pressure  in neutron
star interiors is mainly produced by highly degenerate
fermions with Fermi energy $\EF\gg \kB T$ ($\kB$ is the
Boltzmann constant), therefore one can neglect the 
$T$-dependence in calculations of $R(M)$. For the central
regions of typical neutron stars, where $\rho\gtrsim\rho_0$,
the EOS and even composition of matter is not well known
because of the lack of the precise relativistic many-body
theory of strongly interacting particles. Instead of the
exact theory, there are many approximate models, which give
a range of theoretical EOSs and, accordingly, $R(M)$
relations (see, e.g., Chapt.~6 of \cite{NSB1}).

For a star to be hydrostatically stable, the density at the
stellar center has to increase with increasing mass. This
condition is satisfied in a certain interval
$M_\mathrm{min}<M<M_\mathrm{max}$. The minimum neutron-star
mass is rather well established,
$M_\mathrm{min}\approx0.1\,M_\odot$ \cite{HaenselZD02}. The
maximum mass until recently was allowed to lie in a wide
range $M_\mathrm{max}\sim(1.5$\,--\,$2.5)\,M_\odot$ by
competing theories (see, e.g., Table~6.1 in Ref.~\cite{NSB1}),
but the discoveries of neutron stars with masses
$M=1.97\pm0.04\,M_\odot$ \cite{Demorest} and
$2.01\pm0.04\,M_\odot$ \cite{Antoniadis} showed that
$M_\mathrm{max}>2\,M_\odot$.

Simulations of formation of neutron stars
\cite{ZhangWH08,PejchaTK12} show that $M$, as a rule,
exceeds $M_\odot$, the most typical values being in the range
(1.2\,--\,$1.6)\,M_\odot$. Observations generally agree with
these conclusions. Masses of several pulsars in double
compact-star systems are known with a high accuracy
($\lesssim1$\%) due to the measurements of the General
Relativity (GR) effects on their orbital parameters. All of
them lie in the interval from $1.3\,M_\odot$ to
$2.0\,M_\odot$ \cite{KramerStairs,Demorest,Antoniadis}.
Masses of other neutron stars that have been
measured with an accuracy better than 10\% cover the
range $M_\odot\lesssim M \lesssim 2\,M_\odot$
\cite{NSB1,Lattimer}.

Were radius $R$ and mass $M$ known precisely for at least a
single neutron star, it would probably ensure selecting one
of the nuclear-matter EOSs as the most realistic one.
However, the current accuracy of measurements of neutron
star radii leaves much to be desired. 

\subsection{Magnetic fields}
\label{sect:B}

Most of the known neutron stars possess strong magnetic
fields, unattainable in the terrestrial laboratories. 
Gnedin and Sunyaev \cite{GnedinSunyaev74} pointed out that 
spectra of such stars can contain the resonant
electron-cyclotron line. Its detection allows one
to obtain magnetic field $B$ by measurement of
the cyclotron frequency $\omc=eB/(\mel c)$, where $\mel$ and
$(-e)$ are the electron mass and charge, and $c$ is the
speed of light in vacuum (here and hereafter we use the
Gaussian system of units). The discovery of the cyclotron
line in the spectrum of the X-ray pulsar in the binary
system Hercules X-1 \cite{Truemper-ea78} gave a striking
confirmation of this idea. About 20 accreting X-ray pulsars
are currently known to reveal the electron cyclotron line
and sometimes several its harmonics at energies of tens keV,
corresponding to $B\approx(1$\,--\,$4)\times10^{12}$~G
(e.g., \cite{Coburn-ea02,RodesRoca,Pottschmidt,Boldin}). 

An alternative interpretation of the observed lines was
suggested in~\cite{BaushevBisno}. It assumes an anisotropic
distribution of electron velocities in a collisionless shock
wave  with large Lorentz factors (the ratios of the total
electron energy to $\mel c^2=511$~keV), $\gammar\sim40$. The
radiation frequency of such electrons strongly increases
because of the relativistic Doppler effect, which enables an
explanation of the observed position of the line by a much
weaker field than in the conventional interpretation. It was
noted in Ref.~\cite{ArayaHarding00} that the small width of
the lines (from one to several keV \cite{Coburn-ea02}) is
difficult to accommodate in this model. It also leaves
unexplained, why the  position of the line is usually almost
constant. For example, the measured cyclotron energy of the
accreting X-ray pulsar A~0535+26 remains virtually constant
while its luminosity changes by two orders of magnitude
\cite{Terada-ea06}. 

On the other hand, most X-ray pulsars do exhibit a
dependence, albeit weak, of the observed line frequency on
luminosity \cite{Becker-ea12}. In order to explain this
dependence, a model was suggested in \cite{Poutanen-ea13},
assuming that the cyclotron lines are formed by reflection
from the stellar surface, irradiated by the accretion column.
When luminosity increases, the bulk of reflection occurs at
lower magnetic latitudes, where the field is weaker than at
the pole, therefore the cyclotron frequency becomes smaller.
This model, however, does not explain the cases where the
observed frequency increases with luminosity and, as noted
in \cite{Nishimura14}, it does not reproduce X-ray pulses at
large luminosities.

A quantitative description of all observed dependences of
the cyclotron frequency on luminosity is developed in
Ref.~\cite{Becker-ea12}, based on a physical model of
cyclotron-line formation in the accretion column. The height
of the region above the surface, where the lines are formed,
$h\sim(10^{-3}$\,--\,$10^{-1})R$, correlates with
luminosity, the correlation being positive or negative
depending on the luminosity value. Then the line is centered
at the frequency $\omc/(1+h/R)^3$, where $\omc$ is the
cyclotron frequency at the base of the accretion column. In
\cite{Nishimura14}, variations of a polar cap diameter and
a beam pattern were additionally taken into account, which
has allowed the author to explain variations in the width
and depth of the observed lines in addition to their
frequencies.

When cyclotron features are not identified in the spectrum,
one has to resort to indirect estimates of the magnetic
field. For the isolated pulsars, the most widely used
estimate is based on the expression
\beq
B \approx 3.2\times10^{19} \,C\,
 \sqrt{\mathcal{P}\dot{\mathcal{P}}}\textrm{~~G},\label{PPdot} 
\eeq
where $\mathcal{P}$ is the period in seconds,
$\dot{\mathcal{P}}$ is the period
time derivative, and
$C$ is a coefficient, which depends on stellar
parameters. For the rotating magnetic dipole in vacuo
\cite{Deutsch55}
$C=R_6^{-3}\,(\sin\alpha)^{-1}\,\sqrt{I_{45}}$, where
$R_6\equiv R/(10^6\mbox{~cm})$, $I_{45}$ is the moment of
inertia in units of $10^{45}$ g~cm$^2$, and $\alpha$ is the
angle between the magnetic and rotational axes. In this case
\req{PPdot} gives the magnetic field strength at the pole. 
If $M\approx(1$\,--\,2)\,$M_\odot$, then
$R_6\approx 1.0$\,--\,1.3 and $I_{45}\approx1$\,--\,3 (see
\cite{NSB1}). For estimates, one usually sets $C=1$ in
\req{PPdot} (e.g.,  \cite{ManchesterTaylor}).

A real pulsar strongly differs from a rotating magnetic
dipole, because its magnetosphere is filled with plasma,
carrying electric charges and currents (see reviews
\cite{Beskin99,Michel04,Spitkovsky11,BeskinIF13} and recent
papers \cite{Petri12,TchekhovskoySL13,TimokhinArons13}).
According to the model by Beskin
\etal~\cite{BeskinGI83,BeskinGI93}, the magnetodipole
radiation is absent beyond the magnetosphere, while the
slowdown of rotation is provided by the current energy
losses. However, the the relation between $B$ and $\mathcal{P}\dot{\mathcal{P}}$
remains similar. Results of numerical simulations of plasma
behavior in the pulsar magnetosphere can be approximately
described by \req{PPdot} with $C\approx0.8
R_6^{-3}\,(1+\sin^2\alpha)^{-1/2}\,\sqrt{I_{45}}$
\cite{Spitkovsky06}. As shown in \cite{BeskinIF13}, this
result does not contradict to the model
\cite{BeskinGI83,BeskinGI93}.

Magnetic fields of the ordinary radio pulsars are
distributed near  $B\sim10^{12}$~G \cite{ATNF}, the
``recycled'' millisecond pulsars have
$B\sim(10^{8}$\,--\,$10^{10})$~G
\cite{ATNF,Bisno06,Lorimer}, and the fields of magnetars
much exceed $10^{13}$~G
\cite{DuncanThompson92,PopovProkhorov}. According to the
most popular point of view, anomalous X-ray pulsars (AXPs)
and soft gamma repeaters (SGRs)
\cite{DuncanThompson92,PopovProkhorov,Mereghetti08,Mereghetti13,Rea13}
are magnetars. For these objects, the estimate (\ref{PPdot})
most often (although not always) gives $B\sim10^{14}$~G, but
in order to explain their energy balance, magnetic fields
reaching up to $B\sim10^{16}$\,--\,$10^{17}$~G in the core
at the birth of the star are considered (see
\cite{DallOsso_ss} and references therein). Numerical
calculations \cite{Ardeljan-ea05} show that
magnetorotational instability in the envelope of a
supernova, that is a progenitor of a neutron star, can give
rise to nonstationary magnetic fields over $10^{15}$~G.
It is assumed that in addition to the poloidal
magnetic field at the surface, the magnetars may have
much stronger toroidal magnetic field
embedded in deeper layers
\cite{GeppertKP06,PerezAzopinMP06}. 
Indeed, for a characteristic poloidal component
$B_\mathrm{pol}$ of a neutron-star magnetic field to be
stable, a toroidal component $B_\mathrm{tor}$ must be
present, such that, by order of magnitude, $B_\mathrm{pol}
\lesssim B_\mathrm{tor} \lesssim
10^{16}\mbox{~G}\sqrt{B_\mathrm{pol}/(10^{13}\mbox{~G})}$
\cite{Akgun-ea13}.
Meanwhile, there is increasing evidence
for the absence of a clear distinction between AXPs and SGRs
\cite{GavriilKW02}, as well as between these objects and
other neutron stars \cite{Kaspi10,Mereghetti13,Rea13}. 
There has even appeared the paradoxical name ``a low-field
magnetar,'' applied to those AXPs and SGRs that have
$B\ll10^{14}$~G (e.g., \cite{Rea-ea14,Rea-ea13}, and
references therein).

For the majority of isolated neutron stars, the
magnetic-field estimate (\ref{PPdot}) agrees with 
other data (e.g., with observed properties
of the bow shock nebula in the vicinity of the star
\cite{KaspiRH06}). For AXPs and SGRs,
however, one cannot exclude alternative models, which do not
involve superstrong fields but assume weak accretion on a
young neutron star with  $B\sim10^{12}$~G from a
circumstellar disk, which could remain after the supernova
burst
\cite{Marsden-ea01,ErtanAXP,BisnoIkhsanov14,Truemper-ea13}.
There is also a ``drift model'', which suggests that the
observed AXP and SGR periods equal not to rotation periods
but to periods of drift waves, which affect the
magnetic-lines curvature and the direction of radiation in
the outer parts of magnetospheres of neutron stars with
$B\sim10^{12}$~G \cite{MalovMachabeli,Malov10}. Another
model suggests that the AXPs and SGRs are not neutron stars
at all, but rather massive ($M>M_\odot$) rapidly rotating
white dwarfs with $B\sim10^8$\,--\,$10^9$~G
(\cite{Boshkaev-ea13} and references therein).

The measured neutron-star magnetic fields are enormous by
terrestrial scales, but still far below the theoretical
upper limit. An order-of-magnitude estimate of this limit
can be obtained by equating the gravitational energy of the
star to its electromagnetic energy \cite{ChandraFermi}. For
neutron stars, such estimate gives  the limiting field
$B_\mathrm{max}\sim10^{18}$\,--\,$10^{19}$~G
\cite{LaiShapiro91}. Numerical simulations of hydrostatic
equilibrium of magnetized neutron stars show that
$B_\mathrm{max}\lesssim10^{18}$~G
\cite{Bocquet,Cardall,KiuchiKotake,FriebenRezzolla12}. 
Still stronger magnetic fields imply so intense electric
currents that their interaction would disrupt the star. Note
in passing that the highest magnetic field that can be
accommodated in quantum electrodynamics (QED) is, by order of
magnitude, $[\mel^2
c^3/(e\hbar)]\exp(\pi^{3/2}/\sqrt{\alphaf})\approx10^{42}$~G
\cite{ShabadUsov06}, where $\alphaf= e^2/(\hbar
c)\approx1/137$ is the fine structure constant, and $\hbar$
is the Planck constant divided by $2\pi$.

We will see below that magnetic fields $B\gtrsim10^{11}$~G
strongly affect the most important characteristics of
neutron-star envelopes. These effects are particularly
pronounced at radiating surfaces and in atmospheres, which are
the main subject of the present review.

\subsection{General Relativity effects}
\label{sect:GR}

The significance of the GR effects
for a star is quantified by the compactness parameter 
\beq
  \xg=r_g/R,
\label{x_g}
\eeq
where
\beq
  r_g=2GM/c^2\approx 2.95\,M/M_\odot \textrm{ km}
\label{r_g}
\eeq
is the Schwarzschild radius, and $G$ is the gravitational
constant. The compactness parameter of a typical neutron
star lies between 1/5 and 1/2, that is not small (for
comparison, the Sun has $\xg=4.24\times10^{-6}$).  Hence,
the GR effects are not negligible. Two important
consequences follow: first, the quantitative theory of
neutron stars must be wholly  relativistic; second,
observations of neutron stars open up a unique opportunity
for measuring the GR effects and verification of the GR
predictions.

\begin{figure*}
\begin{center}
\includegraphics[width=.2\textwidth]{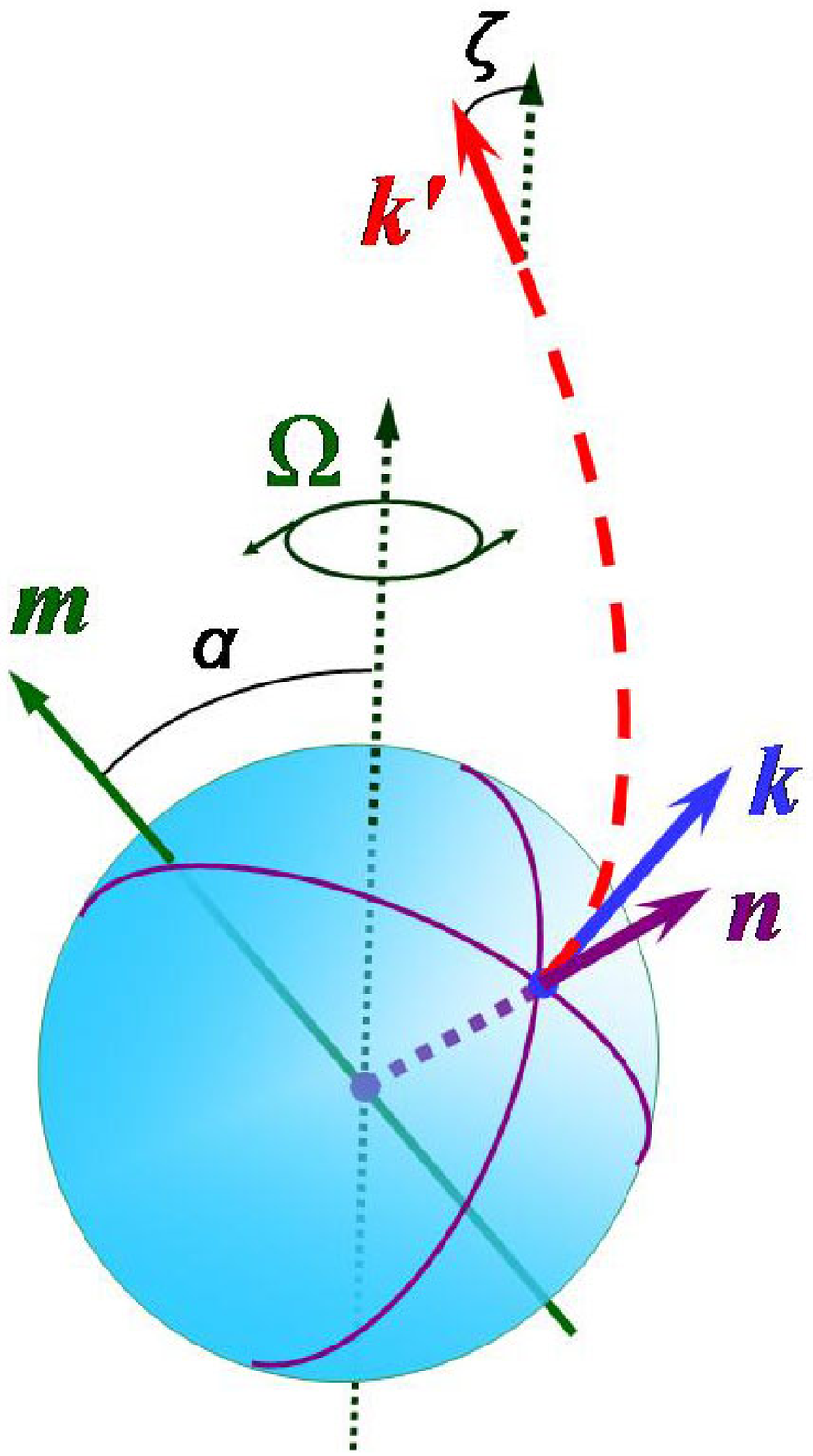}
\hspace*{.1\textwidth}
\includegraphics[width=.26\textwidth]{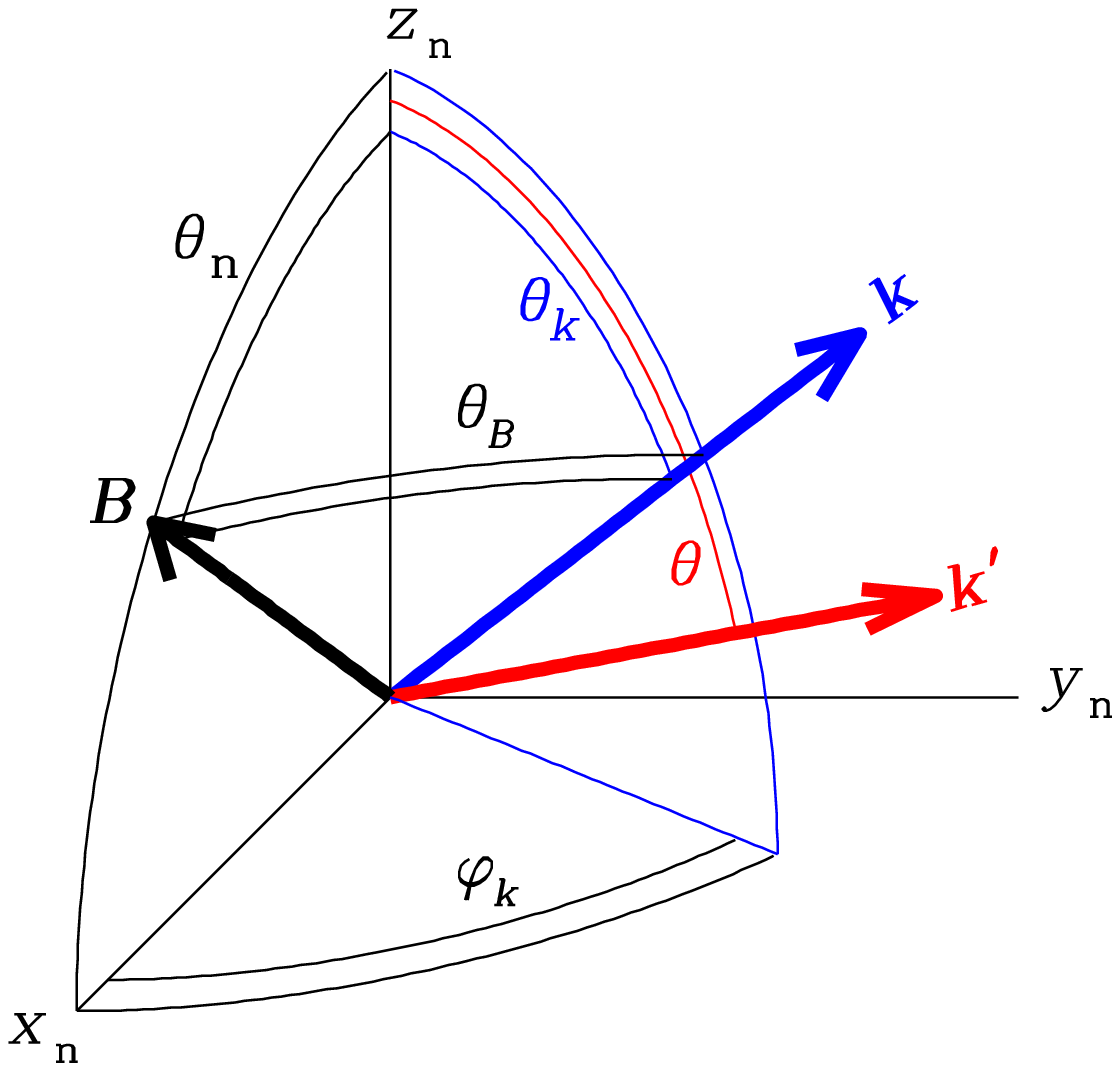}
\hspace*{.05\textwidth}
\includegraphics[width=.24\textwidth]{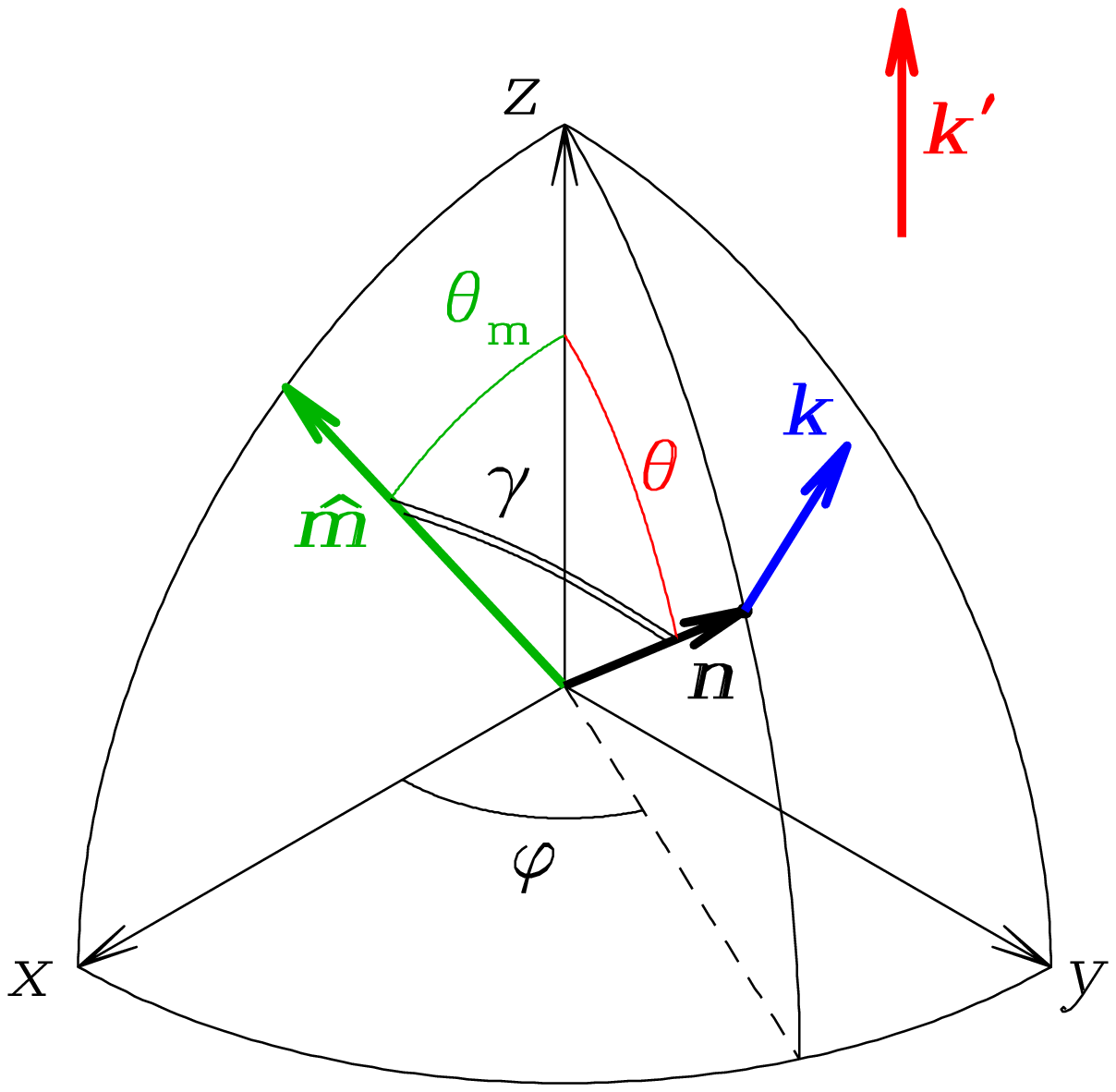}
\caption{
\emph{Left panel}: an illustration of the
gravitational light-bending near a neutron star;
$\mathbf{n}$ is the normal to the surface at a radiating
point, $\bm{k}$ is the wave vector of an emitted ray in the
local reference frame, $\bm{k'}$ is the wave vector in the
observer's reference frame. In addition, the stellar
rotation vector $\mathbf{\Omega}$ and magnetic moment
$\bm{m}$ are shown. The angles formed by the rotation axis
with the magnetic moment ($\alpha$) and with the line of
sight ($\zeta$) are indicated. 
\emph{Middle panel}:  wave vectors $\bm{k}$, $\bm{k'}$, and
the magnetic field vector $\bm{B}$ in the local reference
frame $(x_\mathrm{n} y_\mathrm{n} z_\mathrm{n})$ with the
$z$-axis along $\mathbf{n}$ and the $x$-axis along the
projection of $\bm{B}$ on the surface; $\theta_\mathrm{n}$
is the angle between $\bm{B}$ and $\mathbf{n}$, $\theta_k$
and $\theta$ are the angles between the wave vectors and the
normal, $\theta_B$ is the angle between the ray and
the magnetic field, and $\varphi_k$ is the azimuth.
\emph{Right panel}: vectors $\mathbf{n}$, $\bm{k}$,
$\bm{k'}$, and $\hat{\mathbf{m}}\equiv\bm{m}/|\bm{m}|$ in
the coordinate system $(xyz)$ with the $z$-axis along the
line of sight and the $x$-axis along the projection of
$\bm{m}$ on the picture plane; $\theta_\mathrm{m}$ is the
angle between $\bm{m}$ and the line of sight, $\gamma$ is
the angle between $\mathbf{n}$ and $\bm{m}$, and $\varphi$
is the azimuth.
}
\label{fig:bending}
\end{center}
\end{figure*}

In GR, gravity at the
stellar surface  is determined by the equation
\beq 
   g=\frac{GM}{R^{2}\,\sqrt{1-\xg}} \approx
      \frac{1.328\times10^{14}
        }{
          \sqrt{1-\xg}}
          \,\frac{M/M_\odot}{
                \,R_6^{2}}\textrm{ cm s}^{-2}.
\eeq
Stellar hydrostatic equilibrium is governed by the
Tolman-Oppenheimer-Volkoff equation (corrections due to the
rotation and magnetic fields are negligible for the majority
of neutron stars):
\beq
   \frac{\dd P}{\dd r} =
    - \left( 1 + \frac{P}{\rho c^2} \right)\,
      \left( 1 + \frac{4\pi r^3 P}{M_r c^2} \right)\,
      \left( 1 - \frac{2GM_r}{rc^2} \right)^{-1/2},
\label{TOV}
\eeq
where $r$ is the radial coordinate measured from the stellar
center, and $M_r$ is the mass inside a sphere of
radius $r$.

The photon frequency, which equals $\omega$ in the local
inertial reference frame, undergoes a redshift
to a smaller frequency
$\omega_\infty$ in the remote observer's reference frame. 
Therefore a thermal spectrum with effective temperature
$\Teff$, measured by the remote observer,
corresponds to a lower effective temperature
\beq
   \Teff^\infty = \Teff / (1+z_g),
\label{Tinfty}
\eeq
where
\beq
 z_g \equiv \omega/\omega_\infty -1  = (1-\xg)^{-1/2} -1
\label{z_g}
\eeq
is the redshift parameter. Here and hereafter the symbol
$\infty$ indicates that the given quantity is measured at a
large distance from the star and can differ from its value
near the surface. 

Along with the radius $R$ that is determined by the
equatorial length $2\pi R$ in the local reference frame, one
often considers an \emph{apparent radius} for a remote
observer, 
\beq
   R_\infty = R \,(1+z_g).
\label{Rinfty}
\eeq
With decreasing $R$, $z_g$ increases so that the
apparent radius has a minimum, $\min R_\infty
\approx12$\,--\,14 km (\cite{NSB1}, Chapt.~6).

The apparent
photon luminosity $L_\mathrm{ph}^\infty$ and the luminosity
in the stellar reference frame $L_\mathrm{ph}$
are determined by the Stefan-Boltzmann law
\beq
  L_\mathrm{ph}^\infty=4\pi\sSB\,R_\infty^2
          \,(\Teff^\infty)^4,
\quad
  L_\mathrm{ph} = 4\pi \sSB\,R^2 \Teff^4
\label{LSB}
\eeq
with $\sSB = \pi^2\kB^4/(60\hbar^3 c^2)$.
According to (\ref{Tinfty})\,--\,(\ref{Rinfty}), they
are interrelated as
\beq
   L_\mathrm{ph}^\infty = (1-\xg)\, L_\mathrm{ph} =
   L_\mathrm{ph}/(1+z_g)^2.
\label{Linfty}
\eeq
In the absence of the perfect spherical symmetry, it is
convenient to define a local effective surface
temperature $\Ts$ by the relation
\beq
   F_\mathrm{ph}(\theta,\varphi) = \sSB\Ts^4,
\label{Fgamma}
\eeq
where $F_\mathrm{ph}$ is the local radial
flux density at the surface point, determined by the polar
angle ($\theta$) and azimuth ($\varphi$) in the spherical
coordinate system. Then
\beq
   L_\mathrm{ph} =
\int_0^{\pi}\sin\theta\,\dd\theta
\int_0^{2\pi}\dd\varphi\,R^2 F_\mathrm{ph}(\theta,\varphi)\,.
\label{Lintegral}
\eeq
The same relation connects the apparent luminosity
$L_\mathrm{ph}^\infty$ (\ref{Linfty}) with the apparent flux
$F_\mathrm{ph}^\infty=\sSB\,(\Ts^\infty)^4$ in the remote
system, in accord with the relation
$\Ts^\infty = \Ts/(1+z_\mathrm{g})$ analogous to
(\ref{Tinfty}).

The expressions (\ref{Tinfty}), (\ref{Rinfty}) and
(\ref{Linfty}) agree with the concepts of the light ray
bending and time dilation near a massive body. If the angle
between the wave vector $\bm{k}$ and the normal to the
surface $\mathbf{n}$ at the emission point is $\theta_k$,
then the observer receives a photon whose wave vector $\bm{k'}$
makes an angle $\theta > \theta_k$ with $\mathbf{n}$
(Fig.~\ref{fig:bending}). The rigorous theory of the
influence of the light bending near a star on its observed
spectrum has been developed in \cite{PechenickFC83} and cast
in a convenient form in \cite{Page95,PavlovZavlin00}. The
simple approximation \cite{Beloborodov02}
\beq
  \cos\theta_k = \xg + (1-\xg)\cos\theta
\label{Beloborodov}
\eeq
is applicable at $\xg<0.5$ with an error within a few
percent. At $\cos\theta_k < \xg$, \req{Beloborodov} gives
$\theta > \pi/2$, as if the observer looked behind the
neutron-star horizon. In particular, for a star with a
dipole magnetic field and a sufficiently large inclination
angle $\theta_\mathrm{m}$ of the dipole moment vector
$\bm{m}$ to the line of site, the observer can see the two
opposite magnetic poles at once. Clearly, such effects
should be taken into account while comparing theoretical
neutron-star radiation models with observations.

Let $I_\omega$ be the specific intensity per unit circular
frequency
(if $I_\nu$ is the specific intensity per unit frequency,
then
$I_\omega=I_\nu/(2\pi)$;
see~\cite{Zheleznyakov}). A contribution to the observed
radiation flux density from a small piece of the surface
$\dd\mathcal{A}$ in the circular frequency interval
$[\omega,\omega+\dd\omega]$ equals
\cite{PoutanenGierlinski03,PoutanenBeloborodov06}
\beq
   \dd F_{\omega_\infty}^\infty =
          I_\omega(\bm{k})\,\cos\theta_k
               \left| \frac{\dd\cos\theta_k}{\dd\cos\theta}
                    \right|\,
                   \frac{\dd\mathcal{A}}{D^2}\,
                   (1-\xg)\,\dd\omega,
\label{dF}
\eeq
where $\dd\omega=(1+z_g)\,\dd\omega_\infty$. Here and
hereafter we assume that the rotational velocity of the
patch $\dd\mathcal{A}$ is much smaller than the speed of
light.  If this condition is not satisfied, then the
right-hand side of \req{dF} should be multiplied by
$(\cos\tilde\theta_k/\cos\theta_k)^4$, where
$\tilde\theta_k$ is the angle between the surface normal and
the wave vector in the reference frame, comoving with the
patch $\dd\mathcal{A}$ at the moment of radiation
\cite{PoutanenGierlinski03,PoutanenBeloborodov06}. For
a spherical star, Eqs.~(\ref{Beloborodov}), (\ref{dF}) give
\beq
   F_{\omega_\infty}^\infty = 
   (1-\xg)^{3/2}\frac{R^2}{D^2}
        \int I_\omega(\bm{k};\theta,\varphi) \,
       \cos\theta_k \sin\theta
           \,\dd\theta\,\dd\varphi,
\label{Fintegral}
\eeq
where the integration is restricted by 
the condition $\cos\theta_k>0$.

The magnetic field is also distorted by
the space curvature in the GR. For the uniform and dipole
fields, this distortion is described by Ginzburg \& Ozernoi
\cite{GinzburgOzernoi}. In the dipole field case, the
magnetic vector is
\beq
 \bm{B} = B_\textrm{p}\,(\mathbf{n}\cdot\hat{\mathbf{m}})\,\mathbf{n}
 + B_\textrm{eq}\,\big[(\mathbf{n}\cdot\hat{\mathbf{m}})\,\mathbf{n}
  - \hat{\mathbf{m}}\big],
\label{eq:dipole}
\eeq
where  $\hat{\mathbf{m}}=\bm{m}/|\bm{m}|$ is the magnetic
axis direction,
$B_\textrm{eq}$ and $B_\textrm{p}$ are the equatorial and
polar field strengths, respectively, and their ratio equals
\beq
  \frac{B_\textrm{eq}}{B_\textrm{p}}
   = \frac{\xg^2/2 - (1-\xg) \ln(1-\xg) - \xg
        }{
         [\ln(1-\xg)+\xg + \xg^2/2] \sqrt{1-\xg}}.
\label{eq:dipole2}
\eeq
In the limit of flat geometry ($\xg\to0$)
$B_\textrm{eq} \to B_\textrm{p}/2$, but in general
$B_\textrm{eq} / B_\textrm{p} > 1/2 + \xg/8$.

Muslimov \& Tsygan \cite{MuslimovTsygan86} obtained
expansions of the components of a poloidal magnetic field
vector $\bm{B}$
over the scalar spherical harmonics near a static neutron
star beyond the dipole approximation. Equations
(\ref{eq:dipole}) and (\ref{eq:dipole2}) are a particular
case of this expansion. Petri \cite{Petri13} developed a
technique of expansion of electromagnetic fields around a
rotating magnetized star over vector spherical harmonics,
which allows one to find a solution of the Maxwell equations
in the GR for an arbitrary multipole component of the
magnetic field. In this case, the solutions for a
nonrotating star in the GR \cite{MuslimovTsygan86} and for a
rotating dipole in the flat geometry \cite{Deutsch55} are
reproduced as particular cases.

\subsection{Measuring masses and radii by ther\-mal
spectrum}

Information on the mass and radius of a neutron star can be
obtained from its thermal spectrum. To begin with, let us consider
the perfect blackbody radiation whose spectrum is described by
the Planck function\footnote{%
$\mathcal{B}_{\omega,T}$ is the specific intensity of
nonpolarized blackbody radiation related to the circular
frequency (see \cite{Zheleznyakov}).}
\beq
\mathcal{B}_{\omega,T} = 
\frac{\hbar\omega^3}{4\pi^3c^2}\,
\frac{1}{\exp[\hbar\omega/\kB T]-1},
\label{Bomega}
\eeq
and neglect interstellar absorption and
nonuniformity of the surface temperature distribution.
The position of the spectral maximum 
$\hbar\omega_\mathrm{max}=2.8\kB T$
gives us the effective
temperature $\Teff^\infty$, and the measured intensity gives
the total flux density $F_\mathrm{bol}$ that reaches
the observer. If the star is located at distance $D$, then
its apparent photon luminosity is $L_\mathrm{ph}^\infty=4\pi
D^2 F_\mathrm{bol}$, and \req{LSB} yields $R_\infty$.

In reality, comparison of theoretical and measured spectra
depends on a larger number of parameters. First, the
spectrum is modified by absorption in the interstellar
matter. The effect of the interstellar gas on the X-ray part
of the spectrum is approximately described by factor
$\exp[-(\NH/10^{21}\mbox{~cm}^{-2})\,
(\hbar\omega/0.16\mbox{~keV})^{-8/3}]$, where $\NH$
is the hydrogen column density on the line of sight
\cite{WilmsAMC00}. Thus one can evaluate  $\NH$
from an analysis of the spectrum. If $D$ is unknown, one can
try to evaluate it assuming a typical interstellar gas
density for the given Galaxy region and using $D$ as a
fitting parameter.

Second, the temperature distribution can be nonuniform over
the stellar surface. For example, at contrast to the cold
poles of the Earth, the pulsars have heated regions near
their magnetic poles, ``hot polar caps.'' The polar caps
of accreting neutron stars with strong magnetic fields are
heated by matter flow from a companion star through an
accretion disk and accretion column (see
\cite{Romanova-ea12,LaPalombara-ea12} and references
therein). The polar caps of isolated pulsars and magnetars
are heated by the current of charged particles, created in
the magnetosphere and accelerated by the electric field
along the magnetic field lines (see the reviews
\cite{Beskin99,HardingLai06,BeskinIF13}, papers
\cite{MedinLai10,Beloborodov13}, and references therein).
The thermal spectrum of such neutron stars is sometimes
represented as consisting of two components, one of them
being related to the heated region and the other to the rest
of the surface, each with its own value of the effective
temperature and effective apparent radius of the emitting
area (e.g., \cite{Burwitz-ea03}). Besides,  variable
strength and direction of the magnetic field over the
surface affect the thermal conductivity of the envelope.
Hence, the temperature $\Ts$ of a cooling neutron star
outside the polar regions is also nonuniform (see, e.g.,
\cite{PYCG03,PonsMG09}).

Finally, a star is not a perfect blackbody, therefore its
radiation spectrum differs from the Planck function.
Spectral modeling is a complex task, which includes solving
equations of hydrostatic equilibrium, energy balance, and
radiative transfer (below we will consider it in more
detail). Coefficients of these equations depend on chemical
composition of the atmosphere, effective temperature,
gravity, and magnetic field. Making different assumptions
about the chemical composition, $M$, $R$, $\Teff$, and $B$
values, and about distributions of $\Ts$ and $\bm{B}$ over
the surface, one obtains different model spectra. Comparison
of these spectra with the observed spectrum yields an
evaluation of acceptable values of the parameters. With the known
shape of the spectrum, one can calculate $F_\mathrm{bol}$
and evaluate $R_\infty$ using \req{LSB}. 
Identification of spectral features may provide $z_g$. A
simultaneous evaluation of $z_g$ and $R_\infty$ allows one
to calculate $M$ from Eqs.~(\ref{x_g}), (\ref{r_g}),
(\ref{z_g}), and (\ref{Rinfty}). This method of mass and radius
evaluation requires a reliable theoretical description of
the envelopes that affect the surface temperature and
radiation spectrum. 

\subsection{Neutron-star envelopes}
\label{sect:envelopes}

Not only the superdense core of a neutron star, but also the
envelopes are mostly under
conditions unavailable in the laboratory. By the terrestrial
standards, they are characterized by superhigh pressures,
densities, temperatures, and magnetic fields. The envelopes
differ by their composition, phase state, and their role in
the evolution and properties of the star.

In the deepest envelopes, just above the core of a neutron
star, matter forms a neutron liquid with immersed atomic
nuclei and electrons. In these layers, the neutrons and
electrons are strongly degenerate, and the nuclei are
neutron-rich, that is, their neutron number can be several
times larger than the proton number, so that only the huge
pressure keeps such nuclei together. Electrostatic
interaction of the nuclei is so strong that they are
arranged in a crystalline lattice, which forms the solid
stellar crust. There can be a mantle between the crust and
the core (though not all of the modern models of the dense
nuclear matter predict its existence). Atomic nuclei in the
mantle take exotic shapes of extended cylinders or planes
\cite{Lorenz93}. Such matter behaves like liquid crystals
\cite{PethickPotekhin98}.

The neutron-star crust is divided into the inner and outer
parts. The outer crust is characterized by the absence of
free neutrons. The boundary lies at the critical
neutron-drip density $\rho_\mathrm{nd}$. According to
current estimates \cite{PearsonGC11}, 
$\rho_\mathrm{nd}=4.3\times10^{11}$~\gcc. With decreasing
ion density $\nion$, their electrostatic interaction
weakens, and finally a Coulomb liquid becomes
thermodynamically stable instead of the crystal. The
position of the melting boundary, which can be called the
bottom of the neutron-star ocean, depends on temperature and
chemical composition of the envelope. If all the ions in the
Coulomb liquid have the same charge $Ze$ and mass
$\mion= Am_\mathrm{u}$, where
$m_\mathrm{u}=1.66\times10^{-24}$~g is the atomic mass
unit, and if the magnetic field is not too strong, then ion
dynamics is determined only by the Coulomb coupling constant
$\Gami$, that is the typical electrostatic to thermal energy
ratio for the ions:
\beq
   \Gami = \frac{(Ze)^2}{\aion\kB T}
      = \frac{22.75\,Z^2}{T_6}
        \,\left(\frac{\rho_6}{A}\right)^{1/3},
\eeq
where $\aion=(4\pi\nion/3)^{-1/3}$, $T_6\equiv T/(10^6$~K)
and $\rho_6\equiv\rho/(10^6$~\gcc). Given the strong
degeneracy, the electrons are often considered as a uniform
negatively charged background. In this model, the melting
occurs at $\Gami=175$ \cite{PC00}. However, the ion-electron
interaction and quantizing magnetic field can shift the
melting point by tens percent \cite{PC00,PC13}.

The strong gravity drives rapid separation of chemical
elements
\cite{AlcockIllarionov,HameuryHB83,BBC02,ChangBA10,BeznogovYak13}.
Results of Refs.~\cite{BBC02,ChangBA10,BeznogovYak13}
can be combined to find that the characteristic
sedimentation time for the impurity
ions with mass and charge numbers $A'$ and $Z'$ (that is the
time at which the ions pass the pressure scale height $P/\rho g$)
in the neutron-star ocean is
\beq
   t_\mathrm{sed}\approx
 \frac{46\,Z^{2.9}{(Z')}^{0.3}A^{-1.8}}{A'-AZ'/Z+
  \Delta_T+\Delta_\mathrm{C}}\,
\frac{\rho_6^{1.3}}{g_{14}^2 T_6^{0.3}} \mbox{~~days},
\label{sed}
\eeq
where $g_{14}\equiv g/(10^{14}$ cm~s$^{-2})\sim$1\,--\,3,
$\Delta_T$ is a thermal correction to the ideal degenerate
plasma model \cite{HameuryHB83,ChangBA10}, and
$\Delta_\mathrm{C}$ is an electrostatic (Coulomb) correction
\cite{ChangBA10,BeznogovYak13}. The Coulomb correction
$\Delta_\mathrm{C}\sim10^{-3}-10^{-2}$ dominates in strongly
degenerate neutron-star envelopes (at $\rho\gtrsim10^3$
\gcc), and at smaller densities
$\Delta_T\gtrsim\Delta_\mathrm{C}$. Ions with larger $A/Z$
ratios settle faster, while among ions with equal $A/Z$ the
heavier ones settle down faster
\cite{HameuryHB83,ChangBA10,BeznogovYak13}. It follows from
(\ref{sed}) that $t_\mathrm{sed}$ is small compared with the
known neutron-star ages, therefore neutron-star envelopes
consist of chemically pure layers separated by  transition
bands of diffusive mixing.

Especially important is the thermal blanketing envelope that
governs the flux density $F_\mathrm{ph}$ radiated by a cooling
star with a given internal temperature $T_\mathrm{int}$.
$F_\mathrm{ph}$ is mainly regulated by the thermal
conductivity in the ``sensitivity strip''
\cite{GPE83,YakovlevPethick}, which plays the role of a
``bottleneck'' for the heat leakage. Position of this strip
depends on the stellar parameters $M$, $R$,
$T_\mathrm{int}$, magnetic field, and chemical composition of
the envelope. Since the heat transport across the magnetic
field is hampered, the depth of the sensitivity strip can be
different at different places of a star with a strong
magnetic field: it lies deeper at the places where the
magnetic field is more inclined to the surface \cite{VP01}.
As a rule, the sensitivity strip embraces the deepest layer
of the ocean and the upper part of the crust and lies in
the interval $\rho\sim10^5$\,--\,$10^9$ \gcc.

\subsection{Atmosphere}
\label{sect:atm-gen}

With decreasing density, the ion electrostatic energy and
electron Fermi energy eventually become smaller than the
kinetic ion energy. Then the degenerate Coulomb liquid gives
way to a nondegenerate gas. The outer gaseous envelope of
a star constitutes the atmosphere. In this paper, we will
consider models of quasistationary atmospheres. They 
describe stellar radiation only in the
absence of intense accretion, since otherwise it is formed
mainly by an accretion disk or by flows of infalling matter.

It is important that the sensitivity strip, mentioned in
\S\,\ref{sect:envelopes}, always lies at large optical
depths. Therefore radiative transfer in the atmosphere
almost does not affect the full thermal flux, so that one
can model a spectrum while keeping $F_\mathrm{ph}$
determined and $\Ts$ from a simplified model of heat
transport in the atmosphere. Usually such model is based on
the Eddington approximation (e.g., \cite{Sobolev}). Shibanov
\etal{} \cite{Shibanov_ea98} verified the high accuracy of
this approximation for determination of the full thermal
flux from neutron stars with strong magnetic fields.

Atmospheres of ordinary stars are divided into the lower
part called photosphere, where radiative transfer dominates,
and the the upper atmosphere, whose temperature is 
determined by processes other than the radiative transfer.
Usually the upper atmosphere of neutron stars is thought to
be absent or negligible. Therefore one does not discriminate
between the notions of atmosphere and photosphere for the
neutron stars. In this respect let us note that vacuum
polarization in superstrong magnetic fields (see
\S\,\ref{sect:vacpol}) makes magnetosphere birefringent, so
that the magnetosphere, being thermally decoupled from
radiation propagating from the star to the observer, can
still affect this radiation. Thus the magnetosphere can play
the role of an upper atmosphere of a magnetar.

Geometric depth of an atmosphere is several millimeters in
relatively cold neutron stars and centimeters in relatively
hot ones. These scales can be easily obtained from a simple
estimate: as well as for the ordinary stars, a typical depth
of a neutron-star photosphere is by order of magnitude
slightly larger than the barometric height scale, the latter
being equal to $\kB T/(\mion
g)\approx(0.83/A)\,(T_6/g_{14})$~cm. The
photosphere depth to the neutron-star radius ratio is only
$\sim10^{-6}$ (for comparison, for ordinary stars this ratio
is $\sim10^{-3}$), which allows one to calculate local
spectra neglecting the surface curvature.

The presence of atoms, molecules, and ions with bound states
significantly changes the electromagnetic absorption
coefficients in the atmosphere, thereby affecting the
observed spectra. A question arises, whether  the processes
of particle creation and acceleration near the surface of
the pulsars let them to have a partially ionized atmosphere.
According to canonical pulsar models
\cite{ManchesterTaylor,Beskin99,Michel04}, the magnetosphere
is divided in the regions of open and closed field lines,
the closed-lines region being filled up by charged particles
so that the electric field of the magnetosphere charge in
the comoving (rotating) reference frame
cancels the electric field arising from the rotation of the
magnetized star. The photosphere that lies below this part
of the magnetosphere is stationary and electroneutral.

At contrast, there is a strong electric field near the
surface in the open-line region. This field accelerates the
charged particles almost to the speed of light. It is not
obvious that these processes do not affect the
photosphere, therefore quantitative estimates are needed.
Let us define the column density
\beq
  \ycol = \int_r^\infty (1+z_g)\,\rho(r)\dd r,
\label{ycol}
\eeq
where the factor $(1+z_g)$ takes account of the relativistic
scale change in the gravitational field. According to
\cite{Tsai74}, in the absence of a strong magnetic field,
ultrarelativistic electrons lose their energy mostly to
bremsstrahlung at the depth where
$\ycol\sim60\mbox{~g~cm}^{-2}$. As noted by Bogdanov \etal{}
\cite{BogdanovRG07}, such column density is orders of
magnitude larger than the typical density of a nonmagnetic
neutron-star photosphere. Therefore, the effect of the
accelerated particles reduces to an additional deep heating.

The situation changes in a strong magnetic field. Electron
oscillations driven by the electromagnetic wave are hindered
in the directions perpendicular to the magnetic field, which
thus decreases the coefficients of electromagnetic wave
absorption and scattering by the electrons and atoms
(\S\,\ref{sect:opac}). Therefore the strong magnetic field
``clarifies'' the plasma, that is, the same mean (Rosseland
\cite{Rosseland24,Mihalas}) optical depth $\tau_\mathrm{R}$
is reached at a larger density. For a typical neutron star
with $B\gtrsim 10^{11}$~G, the condition 
$\tau_\mathrm{R}=3/2$ that is required to have  $T(r)=\Teff$
in the Eddington approximation, is fulfilled at the density
\cite{PCY07}
\beq
   \rho\sim B_{12} ~~\gcc,
\label{rhoB12}
\eeq
where $B_{12}\equiv B/(10^{12}$~G). Thus the density of the
layer where the spectrum is formed increases with
growing $B$. At the same time the main mechanism of electron
and positron deceleration changes, which is related to
Landau quantization (\S\,\ref{sect:QLandau}). In the strong
magnetic field, the most effective deceleration mechanism is the
magneto-Coulomb interaction, which makes the charged
particles colliding with plasma ions to jump to excited
Landau levels with subsequent de-excitation through
synchrotron radiation \cite{KotovKB86}. The magneto-Coulomb
deceleration length is inversely proportional to $B$. An
estimate \cite{KotovKB86} of the characteristic depth of the
magneto-Coulomb deceleration of ultrarelativistic electrons
in the neutron-star atmosphere can be written as
\beq
   \ycol \approx
           \big[(\gammar/700)\,Z^2\,A^{-3}
             \,B_{12}^{-2}\big]^{0.43}\,T_6
                 \mbox{~g~cm}^{-2},
\label{magnetocoulomb}
\eeq
$\gammar\sim10^3$\,--\,$10^8$ being the Lorentz
factor. One can easily see from
(\ref{rhoB12}) and (\ref{magnetocoulomb}) that at
$B\gtrsim3\times10^{12}$~G the electrons are decelerated by
emitting high-energy photons in an optically thin layer. In
this case, the magneto-Coulomb radiation constitutes a
nonthermal supplement to the thermal photospheric
spectrum of the polar cap. 

At the intermediate magnetic fields  $10^{11}$~G~$\lesssim B
\lesssim 3\times10^{12}$~G, the braking of the accelerated
particles occurs in the photosphere. Such polar caps require
special photosphere models, where the equations of
ionization, energy, and radiative balance would take the
braking of charged particles into account.

The photospheres can have different chemical compositions.
Before the early 1990s, it was commonly believed that the
outer layers of a neutron star consist of iron, as it is the
most stable chemical element remaining after the supernova
burst that gives birth to a neutron star
\cite{Shklovsky-SNe}. Nevertheless, the outer envelopes of
an isolated neutron star may contain hydrogen and helium
because of accretion of interstellar matter
\cite{Shvartsman71,Blaes-ea92}. Even if the star is in the
ejector regime \cite{Lipunov}, that is, its rotating
magnetosphere throws away the infalling plasma, a small
fraction of the plasma still leaks to the surface (see
\cite{Romanova-ea12} and references therein). Because of the
rapid separation of ions in the strong gravitational field
(\S\,\ref{sect:envelopes}), an accreted atmosphere can
consist entirely of hydrogen. In the absence of magnetic
field, hydrogen completely fills the photosphere if its
column density exceeds $\ycol\gtrsim0.1\mbox{~g~cm}^{-2}$.
In the field  $B\sim10^{14}$~G this happens at
$\ycol\gtrsim10^3\mbox{~g~cm}^{-2}$. Even in the latter case
an accreting mass of $\sim10^{-17}M_\odot$ would suffice.
But if the accretion occurred at the early stage of the
stellar life, when its surface temperature was higher than a
few MK, then hydrogen could diffuse into deeper and hotter
regions where it would be burnt in thermonuclear reactions
\cite{ChiuSalpeter64}, leaving helium on the surface
\cite{ChangBildsten03}. The same might happen to helium
\cite{ChiuSalpeter64}, and then the surface would be left
with carbon \cite{Rosen68,ChangBA10}. Besides, a mechanism
of spallation of heavy chemical elements into lighter ones
operates in pulsars due to the collisions of the accelerated
particles in the open field line regions, which produces
lithium, beryllium, and boron isotopes \cite{Jones78}.
Therefore, only an analysis of observations can elucidate
chemical composition of a neutron star atmosphere.

The Coulomb liquid may turn into the gaseous phase abruptly.
This possibility arises in the situation of a first-order
phase transition between the condensed matter and the
nondegenerate plasma (see~\S\,\ref{sect:cond}). Then the
gaseous layer may be optically thin.
In the latter case, a neutron star is called naked
\cite{TurollaZD04}, because its spectrum is formed at a
solid or liquid surface uncovered by an atmosphere.

Although many researchers studied neutron-star atmospheres
for tens of years, many unsolved problems still persist,
especially when strong magnetic fields and incomplete
ionization are present. The state of the art of these
studies will be considered below. 

\section{Neutron stars with thermal spectra}
\label{sect:NSthermal}

In general, a neutron-star spectrum includes contributions
caused by different processes beside the thermal emission:
for example, processes in pulsar magnetospheres, pulsar
nebulae, accretion disk, etc. A small part of such spectra
allow one to separate the thermal component from the other
contributions (see \cite{Zavlin09}, for review).
Fortunately, their number constantly increases. Let us list
their main classes.

\subsection{X-ray transients}
\label{sect:XT}

The X-ray binary systems where a neutron star accretes
matter from a less massive star (a Main Sequence star or a
white dwarf) are called low-mass X-ray binaries (LMXBs). In
some of the LMXBs, periods of intense accretion alternate with
longer (usually of months, and sometimes years) ``periods of
quiescence,'' when accretion stops and the remaining X-ray
radiation comes from the heated surface of the neutron star.
During the last decade, such  soft X-ray transients (SXTs)
in quiescence (qLMXBs) yield ever increasing amount of
valuable information on the neutron stars. 

Compression of the crust under the weight of newly accreted
matter results in deep crustal heating, driven by exothermic
nuclear transformations \cite{HZ90,HZ08}.  These
transformations occur in a nonequilibrium layer, whose
formation was first studied by Bisnovatyi-Kogan and
Chechetkin \cite{BisnoChech74}. In the review of the same
authors \cite{BisnoChech79}, this problem is exposed in more
detail with applications to different real objects. For a
given theoretical model of a neutron star, one can calculate
the heating curve \cite{Yak-ea04}, that is the dependence of
the equilibrium accretion-free effective temperature $T_0$
on the accretion rate averaged over a large preceding 
period of time. Comparing the heating curves with a measured
$T_0$ value, one can draw conclusions on parameters of a
given neutron star and properties of its matter. Such
analysis has provided restrictions on the mass and
composition of the core of the neutron star in SXT SAX
J1808.4--3658 \cite{Yak-ea04}. In
\cite{LevenfishHaensel07,Ho11}, a possibility to constrain
critical temperatures of proton and neutron superfluidities
in the stellar core was demonstrated. Prospects of
application of such analysis to various classes of X-ray
transients are discussed in  \cite{WijnandsDP13}.

The SXTs that have recently turned into quiescence allow one
to probe the state of the neutron-star crust by the decline
of $\Teff$. Brown \etal~\cite{BBR98} suggested that during
this decline the radiation is fed by the heat that was
deposited in the crust in the preceding active period. In
2001, SXT KS 1731--260, which was discovered in 1989 by
Sunyaev's group \cite{Sunyaev-ea90}, turned from the active
state into quiescence~\cite{Wijnands-ea01}. Subsequent
observations have provided the cooling rate of the surface
of the neutron star in this SXT. In 2007, Shternin
\etal~\cite{Shternin-ea07} analyzed the 5-year cooling of KS
1731--260 and obtained constraints to the heat conductivity
in the neutron-star crust. In particular, they showed that
the hypothesis on an amorphous state of the crust
\cite{Jones04}  is incompatible with the observed cooling
rate, which means that the crust has a regular crystalline
structure.

\begin{figure}[t]
\begin{center}
\includegraphics[width=.85\linewidth]{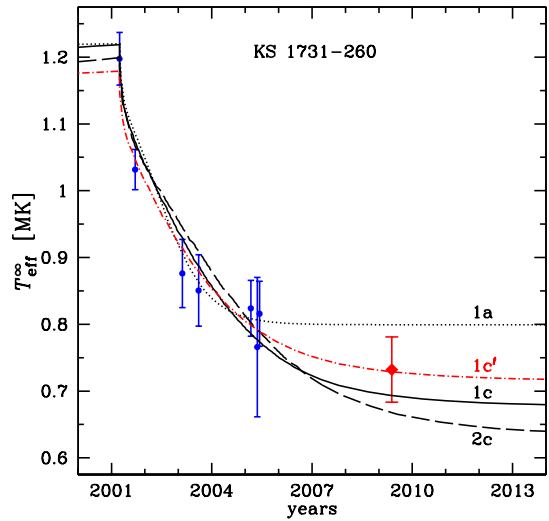}
\caption{
Theoretical cooling curves for different neutron-star models
compared with observations of KS 1731--260. The
observational data are from Table~1 of
Ref.~\cite{Cackett-ea10}. The blue dots correspond to the
observations used in \cite{Shternin-ea07}, and the red
diamond is the new observation. The $1\sigma$-errorbars are
plotted. For the cooling curves, we use the numerical data
and notations from Ref.~\cite{Shternin-ea07}:  1a --
$M=1.6\,M_\odot$, $T_0=0.8$ MK, $E_{44}=2.6$; 1c --
$M=1.6\,M_\odot$, $T_0=0.67$ MK, $E_{44}=2.4$; 2c --
$M=1.4\,M_\odot$, $T_0=0.63$ MK, $E_{44}=2.4$. The model 1a,
unlike the other three models, assumes an accreted envelope
and a moderate (in terms of \cite{Shternin-ea07}) neutron
superfluidity in the crust. The curve marked $\mathrm{1c}'$
was not shown in \cite{Shternin-ea07}. It corresponds to
$M=1.65\,M_\odot$, $T_0=0.7$ MK, $E_{44}=2$. 
\label{fig:ks1731}}
\end{center}
\end{figure}

Figure~\ref{fig:ks1731} shows theoretical cooling curves
compared to observations of KS 1731--260. The theoretical
models differs in assumptions on the neutron-star mass,
composition of its heat-blanketing envelope, neutron
superfluidity in the crust, heat $E_\mathrm{tot}$ deposited
in the crust in the preceding accretion period
($E_{44}\equiv E_\mathrm{tot}/10^{44}$ erg), and the
equilibrium effective temperature $T_0$. The models 1a, 1c,
and 2c were among others described and discussed in
\cite{Shternin-ea07}. At that time when only the first 7
observations had been available, it was believed that the
thermal relaxation of the crust was over, and $T_0=0.8$ MK
\cite{Cackett-ea06}, which corresponds to the curve 1a in
Fig.~\ref{fig:ks1731}. Shternin \etal~\cite{Shternin-ea07}
were the first to call this paradigm in question. They
demonstrated that the available observations could be
described by the curves 1c ($T_0=6.7\times10^5$ K) and 2c
($T_0=6.3\times10^5$ K) as well. In 2009, new
observations of KS 1731--260 were performed, which confirmed
that the cooling continues \cite{Cackett-ea10}. The whole
set of observations is best described by the model 
$\mathrm{1c}'$ (the dot-dashed line in
Fig.~\ref{fig:ks1731}), which only slightly differs from the
model 1c and assumes $T_0=0.7$~MK.

In 2008, a cooling curve of SXT MXB 1659--29 was constructed
for crustal thermal-relaxation stage, which had been
observed during 6 years \cite{Cackett-ea08}. This curve
generally agreed with the theory. In 2012, however, the
spectrum suddenly changed, as if the temperature abruptly
dropped  \cite{Cackett-ea13}.  However, the spectral
evolution driven by the cooling has already had to reach an
equilibrium. The observed change of the spectrum can be
explained by a change of the line-of-sight hydrogen column
density. The cause of this change remains unclear.
Indications to variations of $\NH$ were also found
in the cooling qLMXB EXO 0748--676 \cite{DiazTrigo-ea11}.

Several other qLMXBs have recently turned into quiescence
and show signs of thermal relaxation of the neutron-star
crust. A luminosity decline was even seen during a single
8-hour observation of SXT XTE J1709--267 after the end of
the active phase of accretion \cite{DegenaarWM13}. Analyses
of observations of some qLMXBs (XTE J1701--462
\cite{Fridriksson-ea11,PageReddy13}, EXO 0748--676
\cite{TurlioneAP13}) confirm the  conclusions of Shternin
\etal~\cite{Shternin-ea07} on the crystalline structure of
the crust and  give additional information on the heating
and composition of the crust of accreting neutron stars
\cite{DegenaarWM13,Degenaar-ea13,TurlioneAP13}. In
\S\,\ref{sect:qLMXB} we will discuss the interpretation of
the observed qLMXB spectra that underlies such analysis.

Transiently accreting X-ray pulsars Aql X-1, SAX
J1808.4--3658, and IGR J00291+5734 reveal similar
properties, but an analysis of their spectral evolution is
strongly impeded by the possible presence of a nonthermal
component and hot polar caps
(see~\cite{Heinke-ea09,CotiZelati-ea14}, and
references therein). Their X-ray luminosities in
quiescence vary nonmonotonically, as well as those of
qLMXBs Cen X-4 \cite{Cackett-ea10b} and EXO
1745--248 \cite{DegenaarWijnands12}. The variations of
thermal flux that do not conform to the thermal-relaxation
scenario may be caused by an accretion on the neutron star,
which slows down but does not stop in quiescence
\cite{Rutledge-ea02a,TurlioneAP13,CotiZelati-ea14}.

\subsection{Radio pulsars}

There are several normal pulsars whose spectra clearly
reveal a thermal component: these are relatively young (of
the age $t_*\lesssim10^5$ years) pulsars J1119--6127, 
B1706--44, and Vela, and middle-aged ($t_*\sim10^6$ years)
pulsars  B0656+14, B1055$-$52, and Geminga. The spectra of
the latter three objects, dubbed ``Three Musketeers''
\cite{BeckerTrumper}, are described by a three-component
model, which includes a power-law spectrum of magnetospheric
origin, a thermal spectrum of hot polar caps, and a thermal
spectrum of the rest of the surface \cite{Zavlin09}. In most
works the thermal components of pulsar spectra is
interpreted with the blackbody model, and less often a model
of the fully ionized H atmosphere with a predefined surface
gravity. We will see that both are physically ungrounded.
Only recently, in Ref.~\cite{Ng-ea12}, the X-ray radiation
of PSR J1119--6127 was interpreted using a H atmosphere model
with allowance for the incomplete
ionization. This result will be described in
\S\,\ref{sect:1119}.

A convenient characteristic of the slowdown of pulsar
rotation is the loss rate of the rotational kinetic energy
$\dot{E}_\mathrm{rot}=-I\Omega\dot{\Omega}$ of a standard
rotator with the moment of inertia
$I=10^{45}\mbox{~g~cm}^2$, typical of neutron stars, where
$\Omega=2\pi/\mathcal{P}$ is the angular frequency of the rotation,
and $\dot{\Omega}$ is its time derivative
(see~\cite{ShapiroTeukolsky}). As follows from observations,
spectra of millisecond pulsars with
$\dot{E}_\mathrm{rot}>10^{35}\mbox{~erg~s}^{-1}$ are mainly
nonthermal. However, millisecond pulsars PSR J0030+0451,
J0437\,--\,4715, J1024--0719, and J2124--3358, with
$\dot{E}_\mathrm{rot}\lesssim10^{34}\mbox{~erg~s}^{-1}$ show
a thermal spectral component on the nonthermal background.
In \S\,\ref{sect:msPSR} we will consider interpretation of
this thermal component based on photosphere models.

\subsection{Bursters}
\label{sect:bursters}

Accreting neutron stars in close binary systems, which
produce X-ray bursts with intervals from hours to days, are
called bursters. The theory of the bursters were formulated
in \cite{vanParadijs78} (see also review
\cite{StrohmayerBildsten}). 

During intervals between the bursts, a burster's atmosphere
does not essentially differ from an atmosphere of a cooling
neutron star. In such periods, the bulk of the observed
X-ray radiation arises from transformation of gravitational
energy of the accreting matter into thermal energy.  The
matter, mostly consisting of hydrogen and helium, piles up
on the surface and sooner or later (usually during several
hours or days) reaches such densities and temperatures that
a thermonuclear burst is triggered, which is observed from
the Earth as a Type~I X-ray burst.\footnote{Some binaries
show Type~II X-ray bursts, which recur more frequently than
the Type~I bursts, typically every several minutes or
seconds. They may be caused by gravitational instabilities
of accreting matter, rather than by thermonuclear reactions
\cite{HoffmanML78}.} Some of such bursts last over a minute
and are called long X-ray bursts. They arise in the periods
when the accretion rate is not high, so that the luminosity
$L_\mathrm{ph}$ before the burst does not exceed several
percent of the Eddington limit $L_\mathrm{Edd}$
(\S\,\ref{sect:RTE}). In this case, the inner part of the
accretion disk is a hot ($\kB T\sim20-30$~keV) flow of
matter with an optical thickness about unity. It almost does
not affect the burst, nor screen it \cite{Suleimanov-ea11}.
As we will see in \S\,\ref{sect:burstatm},  the observed
spectrum of a burster, its evolution during a long burst,
and subsequent relaxation are successfully interpreted with
nonmagnetic atmosphere models.

But if the accretion rate is higher, so that 
$L_\mathrm{ph}\gtrsim0.1L_\mathrm{Edd}$, then the accretion
disk is relatively cool and optically thick down to the
neutron-star surface. In this case, the disk can strongly
shield the burst and reprocess its radiation
\cite{BaskoSunyaev76,LapidusST85}, while at the
surface a boundary spreading layer is formed. The theory of
such layer is developed in
\cite{InogamovSunyaev99,InogamovSunyaev10}. The spreading
layer spoils the spectrum so that its usual decomposition
becomes ambiguous and needs to be modified, as described in
\cite{GilfanovRM03,RevnivtsevSP13}.

\subsection{Radio quiet neutron stars}

The discovery of radio quiet\footnote{This term is rather
relative, because some of such objects have revealed radio
emission \cite{MalofeevMT07,Teplykh-ea11}.} neutron stars,
whose X-ray spectra are apparently purely thermal, has
become an important milestone in astrophysics. The radio
quiet neutron stars include central compact objects in
supernova remnants (CCOs) \cite{DeLuca08,Harding13} and
X-ray dim isolated neutron stars (XDINSs)
\cite{Haberl07,PopovProkhorov,Mereghetti08,Turolla09,Harding13}.

Exactly seven XDINSs are known since 2001, and they are
dubbed ``Magnificent Seven'' \cite{PopovProkhorov}.
Observations have provided stringent upper limits
($\lesssim0.1$~mJy) to their radio emission
\cite{Kondratiev-ea09}. XDINSs have longer periods
($>3$ s) than the majority of pulsars, and their
magnetic field estimations by Eq.~(\ref{PPdot}) give, as a
rule, rather high values $B \sim (10^{13}$\,--\,$10^{14})$~G
\cite{Mereghetti08,Kaplan-vanKerkwijk11}.  It is possible
that XDINSs are descendant of magnetars
\cite{Mereghetti08,Mereghetti13,Turolla09}. 

About ten CCOs are known to date
\cite{HalpernGotthelf,Harding13}. Pulsations have been found
in radiation of three of them. The periods of these
pulsations are rather small (0.1 s to 0.42 s) and very
stable. This indicates that CCOs have relatively weak
magnetic field $B \sim 10^{11}$~G, at
contrast to XDINSs. For this reason they are sometimes
called ``antimagnetars''
\cite{HalpernGotthelf,Harding13,GotthelfHA13}. Large
amplitudes of the pulsations of some CCOs indicate strongly
nonuniform surface temperature distribution. To explain it,
some authors hypothesized that a superstrong magnetic field
might be hidden in the neutron-star crust
\cite{ShabaltasLai12}.

The X-ray source 1RXS J141256.0+792204, which was discovered
in 2008 and dubbed Calvera, initially was considered as a
possible eighth object with the properties of the
``Magnificent Seven'' \cite{Rutledge-ea08}. However,
subsequent observations suggest that its properties are
closer to the CCOs. In 2013, observations of Calvera at the
orbital observatory \textit{Chandra} provided the period
derivative $\dot{\mathcal{P}}$ \cite{HalpernBG13}. According to
\req{PPdot}, its value corresponds to
$B\approx4.4\times10^{11}$~G. The authors \cite{HalpernBG13}
characterize Calvera as an ``orphaned CCO,'' whose magnetic
field is emerging through supernova debris. Calvera is also
unique in that it is the only energetic pulsar that emits
virtually no radio nor gamma radiation, which places
constraints on models for particle acceleration in
magnetospheres \cite{HalpernBG13}.

\subsection{Neutron stars with absorption lines in
their thermal spectra}

CCO 1E 1207.4--5209 has been the first neutron star whose
thermal spectrum was found to possess features resembling
two broad absorption lines \cite{Sanwal-ea02}.  The third
and fourth spectral lines were reported \cite{Bignami-ea03},
but their statistical significance was called in question
\cite{MoriCH}. It is possible that the complex shape of CCO
PSR J0821--4300 may also be due to an absorption line
\cite{GotthelfHA13}.

Features, which are possibly related to resonant absorption,
are also found in spectra of four XDINSs: RX J0720.4--3125
\cite{Haberl-ea04b,Hambaryan-ea09}, RX J1308.6+2127
(RBS1223) \cite{Schwope-ea07}, 1RXS J$214303.7+065419$
(RBS1774)
\cite{Cropper-ea07,Kaplan-vanKerkwijk09,Schwope-ea09} and RX
J1605.3+3249 \cite{vanKerkwijk-ea04}. Possible absorption
features were also reported in spectra of two more XDINSs,
RX J0806.4--4123 and RX J0420.0--5022 \cite{Haberl-ea04a},
but a confident identification is hampered by uncertainties
related to ambiguous spectral background subtraction
\cite{Kaplan-vanKerkwijk11}. Only the ``Walter star'' RX
J1856.5\,--\,3754 that was discovered the first of the
``Magnificent Seven'' \cite{Walter96} has a smooth spectrum
without any features in the X-ray range \cite{Burwitz-ea01}.

An absorption line has been recently found in the spectrum
of SGR 0418+5729 \cite{Tiengo-ea13}. Its energy  varies from
$<1$~keV to $\sim4$~keV with the rotational phase. The
authors interpret it as a proton cyclotron line associated
with a highly nonuniform magnetic-field distribution between
$\sim2\times10^{14}$~G and $\sim10^{15}$~G. The  discrepancy
with the estimate $B\sim6\times10^{12}$~G according to
\req{PPdot} \cite{Rea-ea13} the authors \cite{Tiengo-ea13}
explain by an absence of a large-scale dipolar component of
the superstrong magnetic field (which can be, e.g.,
contained in spots). They reject the electron-cyclotron
interpretation on the grounds that it would imply
$B\sim(1-5)\times10^{11}$~G, again at odds with the estimate
\cite{Rea-ea13}. Note that the latter contradiction can be
resolved in the models
\cite{Marsden-ea01,ErtanAXP,BisnoIkhsanov14,Truemper-ea13}
that involve a residual accretion torque
(\S\,\ref{sect:intro}). There is also no discrepancy if the
line has a magnetospheric rather than photospheric origin.
Similar puzzling lines had been previously observed in
gamma-ray bursts of magnetars
\cite{StrohmayerIbrahim00,Ibrahim-ea02,GavriilKW02}.

Unlike the radio quiet neutron stars, spectra of the
ordinary pulsars were until recently successfully described
by a sum of smooth thermal and nonthermal spectral models.
The first exception is the radio pulsar PSR J1740+1000, in
whose X-ray spectrum is found to possess absorption features
\cite{Kargaltsev-ea12}. This discovery fills the gap
between the spectra of pulsars and radio quiet neutron stars
and shows that similar spectral features can be pertinent
to different neutron-star classes.

Currently there is no unambiguous and incontestable
theoretical interpretation of the features in neutron-star
spectra. There were more or less successful attempts to
interpret spectra of some of them.
In \S\,\ref{sect:obs} we will consider the interpretations
that are based on magnetic neutron-star atmosphere models.

\section{Nonmagnetic atmospheres}
\label{sect:NMA}

\subsection{Which atmosphere can be treated as nonmagnetic?}
\label{sect:weakB}

The main results of atmosphere modeling are the outgoing
radiation spectra. Zavlin \etal~\cite{ZavlinPS96} formulated
the conditions that allow calculation of a neutron-star
spectrum without account of the magnetic field. In the
theory of stellar atmospheres, interaction of 
electromagnetic radiation with matter is conventionally
described with the use of opacities $\opac$, that is
absorption and scattering cross sections counted
per unit mass of the medium. Opacities of fully
ionized atmospheres do not depend on magnetic field at
the frequencies $\omega$ that are much larger than the
electron cyclotron frequency $\omc$, which
corresponds to the energy $
   \hbar\omc \approx
        11.577\,B_{12}\mbox{~keV}.
$
On this ground, Zavlin \etal~\cite{ZavlinPS96} concluded
that for the energies $\hbar\omega\sim(1$\,--\,10) $\kB T$
that correspond to the maximum of a thermal spectrum one can
neglect the magnetic-field effects on opacities, if
\beq
 B\ll (\mel c / \hbar e)\,\kB T \sim 10^{10}\,
    T_6\mbox{~G}.  
\label{weakB1} 
\eeq 
Strictly speaking, the estimate (\ref{weakB1}) is very
relative. If the atmosphere contains an appreciable
fraction of atoms or ions in bound states, then even a weak
magnetic field changes the opacities by spectral
line splitting (the Zeeman and Paschen-Back effects).
Besides, magnetic field polarizes radiation in plasmas 
\cite{Ginzburg}. The 
Faraday and Hanle effects that are related to the polarization serve as
useful tools for studies of the stellar atmospheres and
magnetic fields, especially the Sun (see
\cite{DonatiLandstreet09}, for a review). But the bulk of
neutron-star thermal radiation is emitted in X-rays, whose
polarimetry only begins to develop, therefore one
usually neglects such fine effects for the neutron stars.

Magnetic field drastically affects opacities of partially
ionized photospheres, if the electron cyclotron frequency
$\hbar\omc$ is comparable to or larger than the electron
binding energies $E_\mathrm{b}$. Because of the high density
of neutron-star photospheres, highly excited states do not
survive as they have relatively large sizes and low binding
energies (the disappearance of bound states with increasing
density is called pressure ionization). For low-lying
electron levels of atoms and positive atomic ions in the
absence of a strong magnetic field, the binding energy can be
estimated as $E_\mathrm{b}\sim (Z+1)^2$~Ry, where $Z$ is the
charge of the ion, and $\mbox{Ry}=\mel
e^4/(2\hbar^2)=13.6057$~eV is the Rydberg constant in energy
units. Consequently the condition $\hbar\omc\ll
E_\mathrm{b}$ is fulfilled at
\beq
  B\ll
    B_0\,(Z+1)^2/2,
\label{weakB2}
\eeq
where
\beq
   B_0 = \frac{\mel^2\,c\,e^3}{\hbar^3} =
   2.3505\times10^9\mbox{~G}
\label{B0}
\eeq
is the atomic unit of magnetic field. The conditions
(\ref{weakB1}) and (\ref{weakB2}) are fulfilled for most
millisecond pulsars and accreting neutron stars.

\subsection{Radiative transfer}
\label{sect:RTE}

A nonmagnetic photosphere of a neutron star does not
essentially differ from photospheres of the ordinary stars.
However, quantitative differences can give rise to specific 
problems: for instance, the strong gravity results in high
density, therefore the plasma nonideality that is
usually neglected in stellar atmospheres can become
significant. Nevertheless, the spectrum
that is formed in a nonmagnetic neutron-star photosphere can
be calculated using the conventional methods that are
described in the classical monograph by Mihalas
\cite{Mihalas}. For stationary neutron-star atmospheres,
thanks to their small thickness, the approximation of
plane-parallel locally uniform layer is quite accurate. The
local uniformity means that the specific intensity at a
given point of the surface can be calculated neglecting the
nonuniformity of the flux distribution over the surface,
that is, the nonuniformity of 
$\Ts$.

Almost all models of neutron-star photospheres assume the
radiative and local thermodynamic equilibrium (LTE; see
\cite{Ivanov-perenos} for a discussion of this and
alternative approximations). Under these conditions, it is
sufficient to solve a system of three basic equations:
equations of radiative transfer, hydrostatic equilibrium,
and energy balance.

The first equation can be written in a
plane-parallel layer as (see, e.g., \cite{Chandra-perenos})
\beq
 \cos\theta_k \frac{\dd I_\omega(\khat)}{\dd \ycol} =
    \opac_\omega I_\omega -
      \int_{(4\pi)}\!\!
          \opac_\omega^\mathrm{s}(\khat',\khat)
         I_\omega(\khat') \dd\khat'
     - \opac_\omega^\mathrm{a}
     \mathcal{B}_{\omega,T},
\label{RTEgen}
\eeq
where $\khat$ is the unit vector along $\bm{k}$, 
$\opac_\omega=\opac_\omega^\mathrm{a}+
\int_{(4\pi)}\opac_\omega^\mathrm{s}(\khat',\khat)\,\dd\khat'/(4\pi)$
is the total opacity, $\opac_\omega^\mathrm{a}$ and
$\opac_\omega^\mathrm{s}(\khat',\khat)$ are its components
due to, respectively, the true absorption and the scattering
that changes the ray direction from $\khat'$ to $\khat$, and
$\dd\khat' = \sin\theta_{k'}\dd\theta_{k'}\dd\varphi_{k'}$
is a solid angle element. Most studies of the neutron-star
photospheres neglect the dependence of
$\opac_\omega^\mathrm{s}$ on $\khat'$ and $\khat$. As shown
in \cite{Haakonsen-ea12}, the inaccuracy that is introduced
by this simplification does not exceed 0.3\% for the thermal
spectral flux of a neutron star at $\hbar\omega<1$ keV and
reaches a
few percent at higher energies.

For simplicity, in \req{RTEgen} we have neglected
polarization of radiation and a change of frequency at the
scattering. In general, the radiative transfer equation
includes an integral of  $I_\omega$ not only over angles,
but also over frequencies, and contains, with account of
polarization, a vector of Stokes parameters instead of
$I_\omega$, while the scattering cross section is replaced
by a matrix. A detailed derivation of the transfer
equations for polarized radiation is given, e.g., in
\cite{Chandra-perenos}, and solutions of the radiative
transfer equation with frequency redistribution are studied
in \cite{Ivanov-perenos}.

The condition of hydrostatic equilibrium follows from
\req{TOV}. Given that $|R-r|\ll R$,
$|M-M_r|\ll M$, and $P\ll\rho c^2$ in the photosphere,
we have
\beq
   \frac{\dd P}{\dd \ycol} = g - g_\mathrm{rad},
\label{barometric}
\eeq
where (see, e.g., \cite{SuleimanovPouW12})
\bea
   g_\mathrm{rad} &=&
    \frac{1}{c}\,\frac{\dd}{\dd\ycol}
   \int_0^\infty \dd\omega
    \int_{(4\pi)} \dd\khat\,\cos^2\theta_k\,
    I_\omega(\khat)
\nonumber\\&\approx&
    \frac{2\pi}{c} 
   \int_0^\infty \dd\omega\,\opac_\omega
    \int_0^\pi
    \cos\theta_k\, I_\omega(\khat)
       \sin\theta_k\,\dd\theta_k.
\hspace*{2em}\label{grad}
\eea
The last approximate equality becomes exact for the isotropic
scattering.
The quantity $g_\mathrm{rad}$ takes account of the radiation
pressure that counteracts gravity. It becomes appreciable at
 $\Teff\gtrsim10^7$~K. Therefore, $g_\mathrm{rad}$ is usually
dropped in calculations of the spectra of the cooler
isolated neutron stars, but included in the models of
relatively hot bursters. Radiative flux of the bursters
amply increases during the bursts, thus increasing
$g_\mathrm{rad}$. The critical value of $g_\mathrm{rad}$
corresponds to the limit of stability, beyond which matter
inevitably flows away under the pressure of light. In a hot
nonmagnetic atmosphere, where
the Thomson scattering dominate, the instability
appears when the luminosity  $L_\mathrm{ph}$ exceeds the
Eddington limit
\bea\hspace*{-2em}
L_\mathrm{Edd} &=& 4\pi c\,(1+z_g)\, {G M m_\mathrm{p}}/{\sigmaT}
\nonumber\\\hspace*{-2em}
& \approx&
1.26\times10^{38}\,(1+z_g)\,({M}/{M_\odot}) ~~ \textrm{erg~s}^{-1}, 
\label{Edd}
\eea
where $m_\mathrm{p}$ is the proton mass, and
\beq
   \sigmaT=\frac{8\pi}{3}\left(\frac{e^2}{\mel c^2}\right)^2
\label{Thomson}
\eeq
is the Thomson cross section, A temperature-dependent
relativistic correction to $\sigmaT$ \cite{Pacz83} increases
$L_\mathrm{Edd}$ approximately by 7\% at typical
temperatures $\sim3\times10^7$~K at the bursters
luminosity maximum \cite{Suleimanov-ea11,SuleimanovPouW12}.

Finally, the energy balance equation in the stationary state
expresses the fact that the energy acquired by an elementary
volume equals the lost energy. The radiative equilibrium
assumes that the energy transport through the photosphere is
purely radiative, that is, one neglects electron heat
conduction and convection, as well as other sources and
leaks of heat. Under these conditions, the energy balance
equation reduces to
\beq
   \int_0^\infty\dd\omega \int_{(4\pi)} I_\omega(\khat)
      \,\cos\theta_k\,\dd\khat
      = F_\mathrm{ph},
\label{enbal}
\eeq
where $F_\mathrm{ph}$ is the local flux at the surface that
is related to $\Ts$
according to Eq.~(\ref{Fgamma}).

Radiation is almost isotropic at large optical
depth 
\beq
   \tau_\omega =
\int_r^\infty \opac_\omega(r') \, \dd \ycol(r'),
\eeq
therefore one may
restrict to the first two terms of the intensity expansion in
spherical functions:
\beq
   I_\omega(\khat) = J_\omega +
   \frac{3}{4\pi}\bm{F}_\omega\cdot\khat.
\eeq
Here,
$
   J_\omega = \frac{1}{4\pi}
    \int_{(4\pi)}
    I_\omega(\khat)\,\dd\khat
$
is the mean intensity, averaged over all directions, and
$
   \bm{F}_\omega = \int_{(4\pi)}
   I_\omega(\khat)\khat\,\dd\khat
$
is the diffusive flux vector. Then integro-differential
equation (\ref{RTEgen}) reduces to a diffusion-type equation
for $J_\omega$. If scattering is isotropic, then in the
plane-parallel locally-uniform approximation the stationary
diffusion equation has the form
\beq
   \frac{\dd^2}{\dd\tau_\omega^2} 
       \,\frac{J_\omega}{3}
   =
     \frac{\opac_\omega^\mathrm{s}}{\opac_\omega}\,
     (J_\omega - \mathcal{B}_{\omega,T})
\label{diffusion}
\eeq
(see \cite{Ishimaru} for derivation of the diffusion
equation from the radiative transfer equation in a more
general case). Sometimes the diffusion approximation is
applied to the entire atmosphere, rather than only to its
deep layers. In this case, one has to replace
$J_\omega/3$ on the
left-hand side of Eq.~(\ref{diffusion}) 
by $f_\omega J_\omega$, where
$f_\omega(\tau_\omega)$ is the so called Eddington factor
\cite{Mihalas},
which is determined by iterations of the
radiative-transfer and energy-balance equations with
account of the boundary conditions (see \cite{ZavlinPS96} for details).

In modeling bursters atmospheres, one usually employs
\req{diffusion} with the Eddington factor on the left-hand
side and an additional term on the right-hand side, a
differential Kompaneets operator \cite{Kompaneets}
acting on $J_\omega$ (see, e.g.,
\cite{ZavlinShibanov91,GrebenevSunyaev02,SuleimanovPoutanen06,SuleimanovPW11}).
The Kompaneets operator describes, in the diffusion
approximation, the photon frequency redistribution due to the
Compton effect, which cannot be neglected at the high
temperatures typical of the bursters.

In order to close the system of equations of radiative
transfer and hydrostatic balance, one needs the EOS
and opacities $\opac_\omega^\mathrm{s,a}$ for all
densities and temperatures encountered in the photosphere.
In turn, in order to determine the EOS and opacities, it is
necessary to find ionization distribution for the chemical
elements that compose the photosphere. The basis for
solution of these problems is provided by quantum mechanics
of all particle types that give a significant contribution
to the EOS or opacities. In the nonmagnetic neutron-star
photospheres, these particles are only the electrons and
atomic ions, because molecules do not survive the typical
temperatures $T\gtrsim3\times10^5$~K.

We will not consider in detail the calculations of the EOS
and opacities in the absence of a strong magnetic field,
because they do not basically differ from the ones for the
ordinary stellar atmospheres, which have been thoroughly
considered, e.g., in the review \cite{Carson76}. Detailed
databases have been developed for them (see
\cite{RogersIglesias98}, for review), the most suitable of
which for the neutron-star photospheres are \textsc{OPAL}
\cite{OPAL} and \textsc{OP} \cite{OP}.\footnote{\raggedright
The \textsc{OPAL} opacities are included in the
\textsc{MESA} project \cite{MESA}, and the database
\textsc{OP} is available at
http://cdsweb.u-strasbg.fr/topbase/TheOP.html} In the
particular cases where the neutron-star atmosphere consists
of hydrogen or helium, all binding energies are smaller than
$\kB T$, therefore the approximation of an ideal gas of
electrons and atomic nuclei is applicable.

Systematic studies of neutron-star photospheres of different
chemical compositions, from hydrogen to iron, started from
the work by Romani \cite{Romani87}. In the subsequent
quarter of century, the nonmagnetic neutron-star
photospheres have been studied in many works (see
\cite{Zavlin09} for a review). Databases of neutron-star
hydrogen photosphere model spectra have been published
\cite{ZavlinPS96,GaensickeBR02,Heinke-ea06},\footnote{Models
\textsc{NSA}, \textsc{NSAGRAV}, and \textsc{NSATMOS}
in the database \textit{XSPEC}
\cite{XSPEC}.} and a numerical code for their calculation
has been released
\cite{Haakonsen-ea12}.\footnote{https://github.com/McPHAC/}
A publicly available database of model spectra for the
carbon photospheres has been recently published
\cite{Suleimanov-ea14}.\footnote{Model \textsc{CARBATM} in the
database \textit{XSPEC} \cite{XSPEC}.} In addition, model
spectra were calculated for neutron-star photospheres
composed of helium, nitrogen, oxygen, iron (e.g.,
\cite{RajagopalRomani96,Pons-ea02,Heinke-ea06,HoHeinke09}),
and mixtures of different elements
\cite{GaensickeBR02,Pons-ea02}.

\subsection{Atmospheres of bursters}
\label{sect:burstatm}

Burster spectra were calculated by many authors
(see, e.g.,  \cite{Suleimanov-ea11}, for references),
starting from the pioneering works
\cite{LondonHT84,LondonTH86,LapidusST85} (see, e.g., 
\cite{Suleimanov-ea11}, for references). These calculations
as well as observations show that the X-ray spectra of
bursters at high luminosities are close to so called diluted
blackbody spectrum
\beq
   F_\omega \approx w \mathcal{B}_{\omega,\Tbb},
\label{diluted}
\eeq
where $\mathcal{B}_{\omega,T}$ is the Planck function
(\ref{Bomega}),
the parameter $\Tbb$ is
called color temperature, normalization $w$ is
a dilution factor, and
the ratio
$f_\mathrm{c}=\Tbb/\Teff$ (typically $\sim3/2$) 
is called color correction
\cite{LapidusST86,LondonTH86,Suleimanov-ea11}. The
apparent color temperature $\Tbb^\infty$ is related to
$\Tbb$ by the relation analogous to (\ref{Tinfty}).

If the luminosity reaches the Eddington limit during a
thermonuclear burst, then the photosphere radius $\Rph$
first increases, and goes back to the initial value $R$ at
the relaxation stage \cite{Pacz83}. Based on this model,
Kaminker \etal~\cite{Kaminker-ea89} suggested a method of
analysis of the Eddington bursts of the bursters and for the
first time applied it to obtaining constraints of the
parameters of the burster MXB 1728--34. Subsequently this
method was amended and modernized by other authors (see
\cite{Suleimanov-ea11}, for references).

According to
\req{LSB}, the bolometric flux equals
$F_\mathrm{bol}=L_\mathrm{ph}^\infty/(4\pi
D^2)=\sSB(\Teff^\infty)^4 (\Rph^\infty/D)^2$. But the
approximation (\ref{diluted}) implies $F_\mathrm{bol}= w
\sSB(\Tbb^\infty)^4(\Rph^\infty/D)^2$. Therefore, at the
late stage of a long burst, when $\Rph=R={}$constant, $w\propto
f_\mathrm{c}^{-4}$. 
On the other hand, the dependence of $f_\mathrm{c}$ on
$L_\mathrm{ph}$ can be obtained from numerical calculations.
This possibility lies in the basis of the method of 
studying bursters that was implemented in the series of
papers by Suleimanov \etal{} 
\cite{SuleimanovPoutanen06,Suleimanov-ea11}. The
calculations show that $f_\mathrm{c}$ mainly depends on the
ratio $l_\mathrm{ph}=L_\mathrm{ph}/L_\mathrm{Edd}$, and also
on gravity $g$ and chemical composition of the photosphere
(mostly on the helium-to-hydrogen fractional abundance, and
to a less extent on the content of heavier elements). Having
approximated the observed spectral normalizations
$f_\mathrm{c}^{-4}(l_\mathrm{ph})$ by the results of
theoretical calculations, one finds the chemical composition
that provides an agreement of the theory with observations.
For this selected composition, one finds the color
correction that corresponds to the observed one at different
values of $g$,  and thus obtains a curve of allowable values
in the $(M,R)$-plane. The point at this curve that
satisfies the condition  $F_\mathrm{bol}=l_\mathrm{ph}
F_\mathrm{Edd}$,  $F_\mathrm{Edd}=L_\mathrm{Edd}/[4\pi D^2
(1+z_g)^2]$ being the bolometric flux that corresponds to
the Eddington luminosity \req{Edd}, gives an estimate of the
mass and radius of the neutron star, if the distance $D$ is
known. If $D$ is unknown, then this analysis allows one to
obtain restrictions on joint values of $M$, $R$, and $D$.

This method was successfully applied to analyzing
the long bursts of bursters 4U 1724--307
\cite{Suleimanov-ea11} and GS 1826--24 \cite{ZamfirCG12}. In
both cases, there was a marked agreement of the observed and
calculated dependences $f_\mathrm{c}(l_\mathrm{ph})$. In
\cite{ZamfirCG12}, the authors have also simulated the light
curves, that is, the time dependences of  $F_\mathrm{bol}$.
As well as in an earlier work \cite{Heger-ea07}, they
managed to find the chemical composition of the atmosphere
and the accretion rate that give an agreement of the
theoretical light curve of each burst and of the intervals between
the bursts with observations. Thus they
obtained an absolute calibration of the luminosity. A
comparison of the theoretical and observed dependences gives
an estimate of the ratio $f_\mathrm{c}/(1+z_g)$, which does
not depend on the distance $D$, thus providing
additional constraints to the neutron-star mass and radius
\cite{Suleimanov-ea11,ZamfirCG12}. A possible anisotropy of
the emission, which modifies the total flux (e.g., because
of screening and reflection of a part of radiation by an
accretion disk) is equivalent to a multiplication of $D$ by
a constant factor, therefore it does not affect the
$D$-independent estimates \cite{ZamfirCG12}.

In \cite{OzelBG10,Ozel13,GuverOzel13} the authors used a
simplified analysis of spectra of bursters, ignoring the
dependence $f_\mathrm{c}(l_\mathrm{ph})$, but only assuming
that the Eddington luminosity is reached at the ``touchdown
point,'' determined by the maximum of the color temperature.
This assumption is inaccurate, therefore such simplified
analysis fails: it gives considerably lower $R$ values, than
the method described above. In addition, the authors of
\cite{OzelBG10,Ozel13,GuverOzel13} analyzed the ``short''
bursts, for which the theory fails to describe the
dependence $f_\mathrm{c}(l_\mathrm{ph})$, and the usual
separation of spectral components becomes ambiguous (see
\S\,\ref{sect:bursters}). Therefore, the
simplified estimates of neutron-star parameters
\cite{OzelBG10,Ozel13,GuverOzel13} are unreliable (see the
discussion in \cite{Suleimanov-ea11}).

We must note that the current results for the bursters still
do leave some open questions. First, the estimates for two
different sources in \cite{Suleimanov-ea11} and
\cite{ZamfirCG12} are hard to conciliate: in the case of the
H atmosphere model, the former estimate indicates a
relatively large neutron-star radius, thus a stiff EOS,
whereas the latter gives a constraint, which implies a soft
EOS. Second, a good agreement between the theory and
observations has been achieved only for a restricted
decaying part of the lightcurves. Third, there is a lack of
explanation to different normalizations of spectra for the
bursts that have different recurrence times. In
\cite{ZamfirCG12}, the authors discuss these uncertainties
and possible prospects of their resolution with the aid of
future observations.

\subsection{Photospheres of isolated neutron stars}
\label{sect:INS}

\begin{figure}[t]
\includegraphics[width=\linewidth]{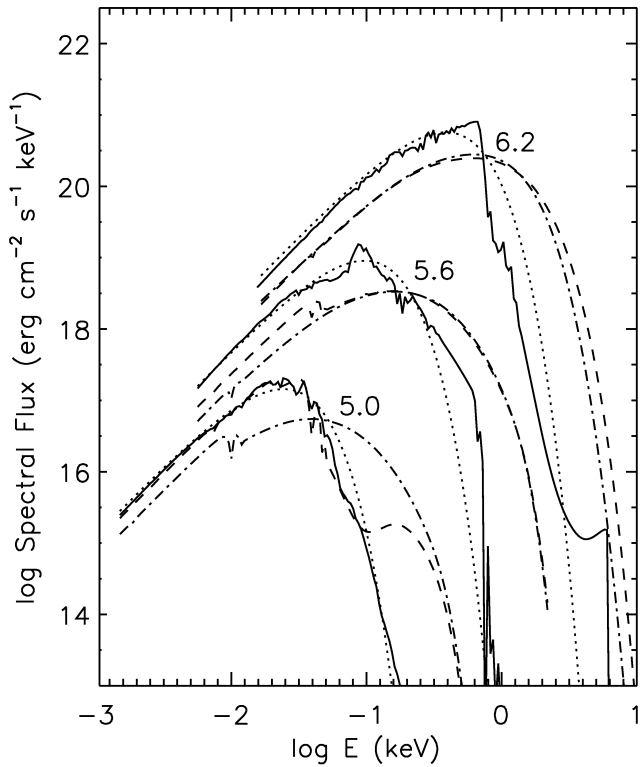}
\caption{
Radiation energy flux densities as functions of photon
energy $E=\hbar\omega$ for a photosphere composed of iron
(solid lines), helium (dashed lines), and hydrogen
(dot-dashed lines) as compared to the blackbody spectrum
(dotted curves) at $g_{14}=2.43$ for different values of
effective temperature (numbers at the curves correspond to 
$\log\,\Teff$ [K]). (Fig.~3 from \cite{Pons-ea02}, courtesy
of J.~Pons and \copyright AAS.)
\label{fig:Pons02f3}}
\end{figure}

Nonmagnetic atmospheres of isolated neutron stars differ
from accreting neutron stars atmospheres, first of all, by a
lower effective temperature
$\Ts\sim3\times(10^5$\,--\,$10^6$)~K, and may be also by
chemical composition. Examples of spectra of such
atmospheres are given in Fig.~\ref{fig:Pons02f3}. 

If there was absolutely no accretion on a neutron star, then
the atmosphere should consist of iron. A spectrum of such
atmosphere has the maximum in the same wavelength range as
the blackbody spectrum, but contains many features caused by
bound-bound transitions and photoionization
\cite{Romani87,RajagopalRomani96,WernerDeetjen00,GaensickeBR02}.
Absorption lines and photoionization edges are smeared with
increasing $g$, because the photosphere becomes denser, thus
increasing the effects leading to line broadening
\cite{Sobelman} (for example, fluctuating microfields in the
plasma \cite{PGC02}).

If the atmosphere consists of hydrogen and
helium, the spectrum is smooth, but shifted to higher
energies compared to the blackbody spectrum at the same
effective temperature \cite{Romani87,ZavlinPS96}. As shown
by Zavlin \etal~\cite{ZavlinPS96}, this shift is caused by
the decrease of light-element opacities according to the
law  $\opac_\omega\propto\omega^{-3}$ at $\hbar\omega>\kB
T\sim 0.1$ keV, which makes photons with larger
energies to come from deeper and hotter photosphere layers.
Zavlin \etal~\cite{ZavlinPS96} payed attention also to the
polar diagrams of radiation coming from the atmosphere.
Unlike the blackbody radiation, it is strongly anisotropic
($I_\omega(\khat)$ quickly decreases at large angles
$\theta_k$), and the shape of the polar diagram depends on
the frequency $\omega$ and on the chemical composition of
the atmosphere.

Suleimanov \& Werner \cite{SuleimanovWerner07} have taken
account of the Compton effect on the spectra of isolated
neutron stars, using the same technique as for the bursters.
They have shown that this effect results in a decrease of
the high-energy flux at $\hbar\omega\gg1$ keV for the
hydrogen and helium atmospheres. It becomes considerable at
high effective temperatures $\Ts > 10^6$~K, where the
spectral maximum shifts to the energies $E \gtrsim 1$ keV.
This effect makes the spectra of hot hydrogen and helium
atmospheres closer to the blackbody spectrum with color
correction $f_\mathrm{c}\approx1.6$\,--\,1.9. 

Papers \cite{WernerDeetjen00,RauchSW08} stand apart, being
the only ones where non-LTE calculations were done for a
spectrum of an iron neutron-star atmosphere. At
$T=2\times10^5$~K, the difference from the LTE model is
about 10\% for the flux in the lines and much less in the
continuum \cite{WernerDeetjen00,GaensickeBR02}. As noted in 
\cite{GaensickeBR02}, the difference may be larger at higher
temperatures, which turned out to be the case indeed in
\cite{RauchSW08}.

Pons \etal~\cite{Pons-ea02} performed a thorough study in
attempt to describe the observed spectrum of the Walter star
RX J1856.5\,--\,3754 by the nonmagnetic atmosphere models
with various chemical compositions. It turned out that the
hydrogen atmosphere model that reproduces the X-ray part of
the spectrum predicts approximately 30 times larger optical
luminosity than observed, whereas an iron-atmosphere model
corresponds to a too small radius. This demonstrates once
again that a neutron-star radius estimate strongly depends
on the assumptions on its atmosphere. Satisfactory results
have been obtained for a chemical composition corresponding
to the ashes of thermonuclear burning of matter that was
accreted on the star at the early stage of its life. This
model, as well as other models of atmospheres composed of
elements heavier than helium, predicted absorption lines in
the X-ray spectrum. However, subsequent deep
X-ray observations with space observatories \textit{Chandra}
\cite{Drake-ea02} and \textit{XMM-Newton}
\cite{Burwitz-ea03} have not found such lines.

The failure of the interpretation of the Walter star
spectrum with nonmagnetic atmosphere models can be explained
by the presence of a strong magnetic
field. The field is indicated by a nearby nebula
glowing in the H$\alpha$ line
\cite{vanKerkwijkKulkarni01}. Such nebulae are found near
pulsars, which ionize interstellar hydrogen by shock waves
arising from hypersonic pulsar magnetosphere interaction
with interstellar medium
\cite{KaspiRH06,GaenslerSlane06}. Doubts had initially been cast on the
pulsar analogy by the absence of observed pulsations of
radiation of this star, but soon such pulsations were
discovered \cite{TiengoMereghetti07}. Interpretation of the
Walter star spectrum with magnetic atmosphere models will be
considered in \S\,\ref{sect:1856}.

The first successful interpretation of an isolated neutron
star spectrum based on a nonmagnetic atmosphere model was
done in  \cite{HoHeinke09}. The authors showed that the
observed X-ray spectrum of the CCO in Cassiopeia A supernova
remnant, which appeared around 1680, is well described by a
carbon atmosphere model with the effective temperature
$\Teff\sim2\times10^6$~K. Subsequent observations revealed
that $\Teff$ appreciably decreases with time
\cite{HeinkeHo10}, which was explained by the heat-carrying
neutrino emission outburst caused by the superfluid
transition of neutrons \cite{Page-ea11,Shternin-ea11}. At
$t_*\approx330$ yrs this agrees with the cooling theory
\cite{YakovlevPethick}. An independent analysis 
\cite{Posselt-ea13} confirmed the decrease of the registered
flux, but the authors stressed that the statistical
significance of this result is not high and that the same
observational data allow other interpretations.
Recently, a spectrum of one more CCO, residing in supernova
remnant HESS J1731--347, was also satisfactorily described
by a nonmagnetic carbon atmosphere model
\cite{Klochkov-ea13}.

\subsection{Atmospheres of neutron stars in qLMXBs}
\label{sect:qLMXB}

Many SXTs reside in globular clusters, whose distances are
known with accuracies of 5\,--\,10\%. This  reduces a major
uncertainty that hampers the spectral analysis. As we noted
in \S\,\ref{sect:NSthermal}, spectra of SXTs in quiescence,
called qLMXBs, are probably determined by  neutron-star
thermal emission. In early works, these spectra were
interpreted with the Planck function, which overestimated
the effective temperature and underestimated the effective
radius of emitting area. However, Rutledge \etal{}
\cite{Rutledge-ea99,Rutledge-ea02a,Rutledge-ea02b} found
that the nonmagnetic hydrogen atmosphere model provides an
explanation to the SXT spectra as caused by radiation from
the entire neutron-star surface with acceptable values of
the temperature and radius.

Currently tens qLMXBs in globular clusters are known (they
are listed in  \cite{Heinke-ea03,Guillot-ea09}), and the use
of hydrogen atmosphere models for their spectral analysis
has become customary. For instance, the analysis of the
cooling of KS 1731--260 and the other similar objects
that was discussed in \S\,\ref{sect:XT} was based on the
measurements of the effective temperature  $\Teff$ with the
use of the models \cite{ZavlinPS96} and NSATMOS
\cite{Heinke-ea06}. 

In many works (including
\cite{Cackett-ea06,Cackett-ea08,Cackett-ea10}),
the neutron-star mass and radius were a priori fixed to
$M=1.4\,M_\odot$ and $R=10$~km, which entrain
$g_{14}=2.43$. It was shown in \cite{Heinke-ea06}, that
such fixing of $g$ may strongly bias estimates of the
neutron-star parameters (which means, in particular, that
the estimates of $\Teff$ for KS 1731--260 and MXB 1659--29,
quoted in \S\,\ref{sect:XT}, are unreliable). An analysis of
thermal spectrum of qLMXB X7 in the globular cluster 47 Tuc,
free of such fixing, gave a 90\%-confidence area of $M$
and $R$ estimates, which agrees with relatively stiff EOSs
of supranuclear matter \cite{Heinke-ea06}. However, the
estimates that were obtained in \cite{Guillot-ea13} by an
analogous analysis for five qLMXBs in globular clusters,
although widely scattered, generally better agree with soft
EOSs. In \cite{Servillat-ea12,Catuneanu-ea13}, thermal
spectra of two qLMXBs were analyzed using hydrogen and
helium atmosphere models. It turned out that the former
model leads to low estimates of $M$ and $R$, compatible with
the soft EOSs, while the latter yields high values, which
require a stiff EOS of superdense matter. Thus, despite the
progress achieved in recent years, the estimates of
neutron-star masses and radii based on the qLMXBs spectral
analysis are not yet definitive.

\subsection{Photospheres of millisecond pulsars}
\label{sect:msPSR}

Magnetic fields of most millisecond pulsars satisfy the
weak-field criteria formulated in 
\S\,\ref{sect:weakB}. Nevertheless, magnetic field does play
certain role, because the open field line areas (``polar
caps'') may be heated by deceleration of fast particles
(see~\S\,\ref{sect:atm-gen}). Therefore, one should take
nonuniform temperature distribution into account, while
calculating the integral spectrum.

Models of rotating neutron stars with hot spots were
presented in many publications (e.g., 
\cite{PoutanenGierlinski03,Cadeau-ea07,GarasyovDK11}, and
references therein), however most of them
used the blackbody radiation model. This model
is acceptable for a preliminary qualitative description of
the spectra and light curves of the millisecond pulsars, but a
detailed quantitative analysis must take the photosphere
into account. Let us consider results of such analyses.

The nearest and the brightest of the four millisecond
pulsars with observed thermal radiation is PSR
J0437\,--\,4715. It belongs to a binary system with a
6-billion-year-old white dwarf. The low effective
temperature of the white dwarf ($\sim4000$~K), as well as
the brightness of the pulsar and a relatively low intensity
of its nonthermal emission favor the analysis of the thermal
spectrum. Recently, the pulsar's thermal radiation has been
extracted from the white-dwarf radiation even in the
ultraviolet range \cite{Durant-ea12}, although the maximum
of the pulsar thermal radiation lies at X-rays. Zavlin \&
Pavlov \cite{ZavlinPavlov98} showed that the thermal X-ray
spectrum of PSR J0437\,--\,4715 can be explained by emission
of two hot polar caps with hydrogen photospheres and a
nonuniform temperature distribution, which was presented by
the authors as a steplike function with a higher value
$T\approx(1$\,--\,$2)\times10^6$~K in the central circle of
radius $0.2$\,--\,$0.4$~km and a lower value
$T\approx(3$\,--\,$5)\times10^5$~K in the surrounding broad
ring of radius about several kilometers.

Subsequent observations of the binary system J0437\,--\,4715
in spectral ranges from infrared to hard X-rays and their
analysis in 
\cite{Zavlin-ea02,BogdanovRG07,BogdanovGR08}
have generally confirmed the qualitative conclusions of
\cite{ZavlinPavlov98}. In particular, Bogdanov \etal{}
\cite{BogdanovRG07,BogdanovGR08} reproduced not
only the spectrum, but also the light curve of this pulsar
at X-rays, using the model of a hydrogen atmosphere with a
steplike temperature distribution, supplemented with a
power-law component. These authors have also explained
\cite{BogdanovGR06} the power-law spectral component by the
Compton scattering of thermal polar-cap photons on energetic
electrons in the magnetosphere or in the pulsar wind. Thus
all the spectral components may have thermal
origin. Finally, Bogdanov \cite{Bogdanov13} reanalyzed the
phase-resolved X-ray spectrum of PSR J0437\,--\,4715 using
the value  $M=(1.76\pm0.20)\,M_\odot$ obtained from radio
observations \cite{Verbiest-ea08}, the distance of 156.3 pc
measured by radio parallax \cite{Deller-ea08}, a nonmagnetic
hydrogen atmosphere model NSATMOS \cite{Heinke-ea06}, and a
three-level distribution of $\Teff$ around the polar caps.
As a result, he came to the conclusion that the radius of a
neutron star of such mass cannot be smaller than 11 km,
which favors the stiff equations of state of supranuclear
matter.

The presence of a hydrogen atmosphere helps one to explain
not only the spectrum but also the relatively large
pulsed fraction (30\,--\,50\%) in thermal radiation of this
and the three other millisecond pulsars with observed thermal
components of radiation (PSR J0030+0451,
J2124--3358, and J1024--0719). According to 
\cite{BogdanovGR08,Zavlin09}, such strong pulsations may
indicate that all similar pulsars have hydrogen atmospheres.
The measured spectra and light curves of all the four pulsars
agree with this assumption
\cite{BogdanovGR08}.

\section{Matter in strong magnetic fields}

The conditions of \S\,\ref{sect:weakB} are not satisfied for
most of the known isolated neutron stars, therefore magnetic
fields drastically affect radiative transfer in their
atmospheres. Before going on to magnetized atmosphere models, it
is useful to consider the magnetic-field effects on their
constituent matter.

\subsection{Landau quantization}
\label{sect:QLandau}

Motion of charged particles in a magnetic field is quantized
in Landau levels \cite{Landau30}. It means that only
longitudinal (parallel to $\bm{B}$) momentum of the particle can
change continuously. Motion of a classical charged particle
across magnetic field is restricted to circular orbits,
corresponding to a set of discrete quantum states,
analogous to the states of a
two-dimensional oscillator.

The complete theoretical description of the quantum
mechanics of free electrons in a magnetic field is given in
monograph \cite{SokTer}. It is convenient to characterize
magnetic field by its strength in
relativistic units, $b$, and in atomic units, $\gamma$:
\bea 
 b&=&
{\hbar\omc}/({\mel c^2}) = B/B_\mathrm{QED} =
{B_{12}}/{44.14} \,,
\label{magpar}
\\
\gamma&=& B/B_0 = 425.44\,B_{12}.
\eea
We have already dealt with the atomic unit $B_0$ in
\S\,\ref{sect:weakB}. The relativistic unit $B_\mathrm{QED}
= \mel^2 c^3 / (e\hbar) = B_0/\alphaf^2$ is the critical
(Schwinger) field, above which specific QED effects
become pronounced. In astrophysics, the
magnetic field is called \emph{strong}, if $\gamma\gg1$, and
\emph{superstrong}, if $b\gtrsim1$.

In the nonrelativistic theory, the distance between Landau
levels equals the cyclotron energy $\hbar\omc$. In the
relativistic theory, Landau level energies equal $E_N=\mel
c^2 \,(\sqrt{1+2bN}-1)$ ($N=0,1,2,\ldots$). The wave
functions that describe an electron in a magnetic field have
a characteristic transverse scale $\sim\am=(\hbar
c/eB)^{1/2}=\aB/\sqrt{\gamma}$, where $\aB$ is the Bohr
radius. The momentum projection on the magnetic field
remains a good quantum number, therefore we have the Maxwell
distribution for longitudinal momenta at thermodynamic
equilibrium. For transverse motion, however, we have the
discrete Boltzmann distribution over $N$.

In practice, the Landau quantization becomes important when
the electron cyclotron energy $\hbar\omc$ is at least
comparable to both the electron Fermi energy $\EF$ and the
characteristic thermal energy $\kB T$. If $\hbar\omc$ is
appreciably larger than both these energies, then most
electrons reside on the ground Landau level in thermodynamic
equilibrium, and the field is called strongly quantizing.
For it to be the case, simultaneous conditions 
$\rho<\rho_B$ and $\zete\gg1$ must be fulfilled, where
\bea
  \rho_B &=&
 \frac{\mion}{\pi^2\sqrt2\,\am^3\,Z}
  = 7045 \,\frac{A}{Z}
       \,B_{12}^{3/2}\text{ \gcc},
\label{rho_B}
\\
   \zete &=& \frac{\hbar\omc}{\kB T} = 134.34\,
   \frac{B_{12}}{T_6} .
\label{zeta_e}
\eea
In the neutron-star atmospheres, these conditions are
satisfied, as a rule, at  $B\gtrsim10^{11}$~G. In the
opposite limit $\zete\ll1$, the Landau quantization can be
neglected. Note that in the magnetospheres, which have lower
densities, electrons can condensate on the lowest Landau
level even at $B\sim10^8$~G because of the violation of the
LTE conditions (\S\,\ref{sect:LTE}).

Ions in a neutron-star atmosphere can be treated as
nondegenerate and nonrelativistic particles. The parameter
$\zete$ is replaced for them by 
\beq
   \zeti = \hbar\omci/\kB T = 0.0737\,(Z/A)
           B_{12}/T_6.
\label{omci}
\eeq
Here, $\omci=ZeB/(\mion c)$ is the ion cyclotron frequency,
and $\hbar\omci=6.35(Z/A)B_{12}$~eV is the ion cyclotron
energy. In magnetar atmospheres, where $B_{12}\gtrsim100$
and $T_6\lesssim10$, the parameter $\zeti$ is not small,
therefore the Landau quantization of ion motion should be
taken into account.

\subsection{Interaction with radiation}
\label{sect:crosssect}

The general expression for a differential cross
section of absorption of a plane electromagnetic wave by a
quantum-mechanical system can be written as (e.g., \cite{Armstrong})
\beq
   \dd\sigma=\frac{4\pi^2}{\omega c}
   \left|\bm{e}\cdot\langle f| \bm{j}_\mathrm{eff} | i \rangle
   \right|^2\,\delta(E_f-E_i-\hbar\omega)\,\dd\nu_f,
\label{dsigma}
\eeq
where$|i\rangle$ and $|f\rangle$ are, respectively, the
initial and final states of the system, $\dd\nu_f$ is the
number of final states in the considered energy interval $\dd E_f$,
$\bm{e}$ is the electromagnetic polarization vector, 
$\bm{j}_\mathrm{eff}=\sum_i q_i
\mathrm{e}^{\mathrm{i}\bm{k}_i}\dot{\bm{r}}_i$ is the
effective electric-current operator, and 
$\dot{\bm{r}}_i$ is the velocity operator acting on a
particle with charge $q_i$. While calculating the matrix
elements 
$\langle f| \bm{j}_\mathrm{eff} | i \rangle$, it is
important to remember that $\dot{\bm{r}}_i$ is not
proportional to the canonical momentum $\bm{p}$ in a magnetic
field. For the system ``electron+proton'' interacting with
radiation in a constant magnetic field, these matrix
elements are derived analytically in \cite{PP97}.

In the dipole approximation, the cross section of photon
interaction with a plasma particle can be expanded in three
components corresponding to the longitudinal, right, and
left polarizations with respect to the magnetic field (e.g.,
\cite{Ginzburg,VenturaNM79}):
\beq
   \sigma(\omega,\theta_B) = 
     \sum_{\alpha=-1}^1 \sigma_\alpha\low(\omega)
      \,\,|e_\alpha(\omega,\theta_B)|^2.
\label{sumalpha}
\eeq
Here, $\omega$ is the photon frequency, $\theta_B$ is the
angle between $\bm{k}$ and $\bm{B}$
(Fig.~\ref{fig:bending}), and
$
e_0\equiv e_z$
and
$e_{\pm1}\equiv {(e_x\pm\mathrm{i}e_y)}/{\sqrt{2}}
$
are the components of the expansion of the electromagnetic
polarization vector $\bm{e}$ in a cyclic basis in the
coordinate system with the $z$-axis along $\bm{B}$.
Representation (\ref{sumalpha}) is convenient because
$\sigma_\alpha$ do not depend on $\theta_B$.

Scattering cross-sections in neutron-star photospheres are
well known
\cite{Ventura79,KaminkerPS82,Mesz}.
For $\alpha=-1$, the photon-electron scattering has a
resonance at the cyclotron frequency $\omc$. Outside a
narrow (about the Doppler width) frequency interval around
$\omc$, the cross sections for the basic polarizations
$\alpha=0,\pm1$ are written as
\beq
    \sigma_\alpha^\mathrm{s,e} =
          \frac{\omega^2}{(\omega+\alpha\omc)^2
             +\nu_{\mathrm{e},\alpha}^2}\, \sigma_\mathrm{T},
\label{sigma-se}
\eeq
where $\sigma_\mathrm{T}$ is the nonmagnetic Thomson cross
section,
\req{Thomson}, and the effective damping factors
$\nu_{\mathrm{e},\alpha}$ are equal to the half of the total
rate of spontaneous and collisional decay of the electron
state with energy $\hbar\omega$ (see \cite{PL07}). The ion
cross section looks analogously,
\beq
    \sigma_\alpha^\mathrm{s,i} =
      \left( \frac{\mel}{\mion}\right)^2
          \frac{\omega^2\,Z^4}{(\omega-\alpha\omci)^2
             +\nu_{\mathrm{i},\alpha}^2}\, \sigma_\mathrm{T}.
\label{sigma-sp}
\eeq
Unlike the nonmagnetic case, in superstrong fields one cannot
neglect the scattering on ions, since 
$\sigma_{+1}^\mathrm{s,i}$
has a resonance at frequency $\omci$.

In the absence of magnetic field, absorption of a photon by
a free electron is possible only at interaction with a third
particle, which takes the difference of the total
electron-photon momentum before and after the absorption. In
a quantizing magnetic field, in addition, also electron
transitions between the Landau levels are possible. In the
nonrelativistic theory, such transitions occur between the
equidistant neighboring levels at the frequency $\omc$,
which corresponds to the dipole approximation. In the
relativistic theory, the multipole expansion leads to an
appearance of cyclotron harmonics \cite{Zheleznyakov}.
Absorption cross-sections at these harmonics were derived in
\cite{PavlovSY80} in the Born approximation without
allowance for the magnetic quantization of electron motion,
and represented in a compact form in 
\cite{SuleimanovPW12}. 

Allowance for the quantization of electron motion leads to
the appearance of cyclotron harmonics in the nonrelativistic
theory as well. In \cite{PavlovPanov}, also in the Born
approximation, photon-electron absorption cross-sections
were derived for an electron, which moves in a magnetic field and
interacts with a nonmoving point charge. This model is
applicable at $\omega \gg \omci$. In the superstrong field
of magnetars, the latter condition is unacceptable, therefore one
should consider absorption of a photon by the system of
finite-mass charged particles, which yields \cite{PC03,P10}
\beq
   \sigma_\alpha^\mathrm{ff}(\omega)
   =
          \frac{4\pi e^2
          }{ 
     \mel c} \,
   \frac{\omega^2\,\nu_{\alpha}^{\mathrm{ff}}(\omega)
          }{
          (\omega+\alpha\omc)^2 (\omega-\alpha\omci)^2
             +\omega^2 \tilde\nu_\alpha^2(\omega)},
\label{sigma-fit0}
\eeq
where $\nu_{\alpha}^{\mathrm{ff}}$ is an effective
photoabsorption collision
frequency, and $\tilde\nu_\alpha$ is an effective frequency
including also other collisions. We see from
(\ref{sigma-fit0}) that
$\sigma_{-1}^\mathrm{ff}$ and
$\sigma_{+1}^\mathrm{ff}$ have a resonance at the
frequencies $\omc$ and $\omci$, respectively. 
Expressions of the effective collision frequencies
$\nu_{\alpha}^{\mathrm{ff}}$ and
$\tilde\nu_\alpha$ in
the electron-proton plasma are given in \cite{PC03}.
One can write
\beq
      \nu_{\alpha}^{\mathrm{ff}}(\omega) =
        \frac{4}{3}\,\sqrt{\frac{2\pi}{\mel T}}\,
          \frac{n_\mathrm{e}\, e^4}{\hbar \omega}
 \Lambda_{\alpha}^{\mathrm{ff}},
\label{nu-ff}
\eeq
where
$\Lambda_{\alpha}^{\mathrm{ff}}=(\pi/\sqrt3)g_{\alpha}^{\mathrm{ff}}$
is a Coulomb logarithm and $g_{\alpha}^{\mathrm{ff}}$ is a
Gaunt factor, and
$g_{-1}^{\mathrm{ff}}=g_{+1}^{\mathrm{ff}}$.
Without the magnetic field, the Gaunt factor is a smooth
function of $\omega$. A calculation with allowance for the
Landau quantization shows, however,
that $\nu_{\alpha}^{\mathrm{ff}}(\omega)$ has peaks at the
multiples of the electron and ion cyclotron frequencies for
all polarizations
$\alpha$. 

\begin{figure}[t]
\includegraphics[width=\linewidth]{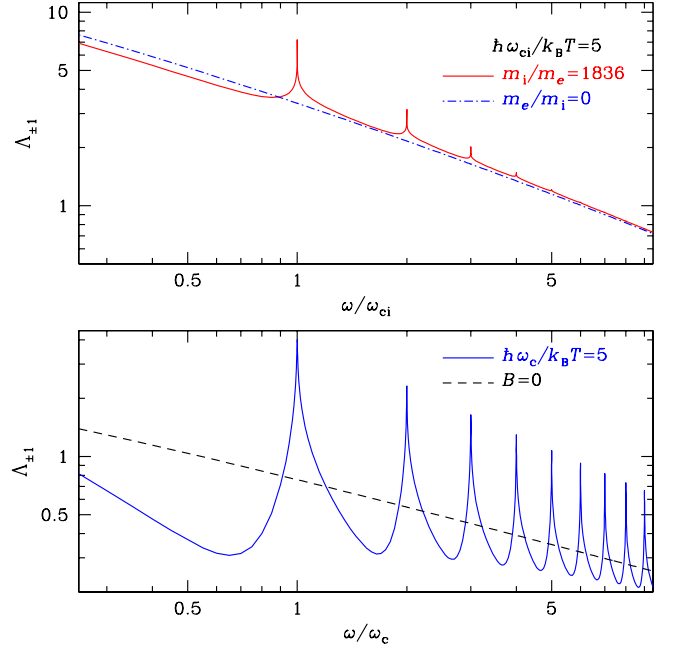}
\caption{
Electron (lower panel) and proton (upper panel) cyclotron
harmonics of the Coulomb logarithm for free-free absorption
at $\hbar\omc=5\kB T$ and
$\hbar\omci=5\kB T$, respectively, for the photon
polarization across the magnetic field. Solid lines show the
result of the accurate calculation of
$\Lambda_1(\omega)=\Lambda_{-1}(\omega)$ in the Born
approximation, and dot-dashed line in the upper panel shows the
infinite-proton-mass approximation (in the lower panel it
effectively coincides with the accurate result). For
comparison, the dashed line in the lower panel shows the
nonmagnetic Coulomb logarithm.
\label{fig:Lambda_ff}}
\end{figure}

Free-free absorption in a hydrogen plasma with account of
both (electron and ion) types of the cyclotron harmonics has
been first calculated in  \cite{PC03}, a detailed
consideration is given in \cite{P10}, and a generalization
to the case of arbitrary hydrogenlike ions and a discussion
of non-Born corrections are presented in \cite{PL07}. If
$\omci/\omega\to0$, then the results of
Ref.~\cite{PavlovPanov} for the electron photoabsorption are
reproduced, but one should keep in mind that the ion
cyclotron harmonics cannot be obtained by a simple scaling
of the electron ones. Such scaling was used in neutron-star
atmosphere models starting from the work \cite{Pavlov-ea95}
until the publication \cite{PC03}, where it was shown to be
qualitatively wrong. One can see it in
Fig.~\ref{fig:Lambda_ff}, where the electron and ion
cyclotron harmonics are shown at equal scales. In spite of
the choice of the same cyclotron frequency to temperature
ratio, the cyclotron peaks in the upper panel are much
weaker than in the lower panel. Physical reasons and
consequences of this fact are discussed in detail in
\cite{P10}. It has been also demonstrated \cite{P10} that
the ion cyclotron harmonics are so weak that they can be
neglected in the neutron-star atmospheres.

\subsection{Atoms}
\label{sect:atoms}

As first noticed in \cite{CLR70}, atoms with bound states
should be much more abundant at $\gamma\gg1$ than at
$\gamma\lesssim1$ in a neutron-star atmosphere at the same
temperature. This difference is caused by the
magnetically-induced increase of
binding energies and decrease of sizes of atoms in so-called
tightly-bound states, which are
characterized by electron-charge
concentration at short distances to the nucleus. Therefore
it is important to take account of the bound states and
bound-bound transitions in a strong magnetic field even for
light-element atmospheres, which would be almost fully
ionized in the nonmagnetic case.

Pioneering works by Loudon, Hasegawa and Howard 
\cite{Loudon,HasegawaHoward61}\footnote{The papers
\cite{Loudon,HasegawaHoward61} and some of the works cited
below were devoted to the Mott exciton in a magnetized
solid, which is equivalent to the problem of a hydrogen atom
in a strong magnetic field.} were at the origin of numerous
studies of atoms in strong magnetic fields. In most of these
studies the authors used the model of an atom with an
infinitely heavy (fixed in space) nucleus. Their results are
summarized in a number of reviews (e.g.,
\cite{Garstang77,Ruder-ea}). The model of an infinitely
massive nucleus is too crude to describe the atoms in the
strongly magnetized neutron-star atmospheres, but it is a
convenient first approximation. Therefore, in this section
we keep to this model, and postpone going beyond its frames
to  \S\,\ref{sect:motion}.

According to the Thomas-Fermi model, a typical size of an
atom with a large nuclear charge $\Znuc\gg1$ is proportional
to $\gamma^{-2/5}$ in the interval $\Znuc^{4/3} \ll \gamma \ll
\Znuc^3$ \cite{Kadomtsev70}. At $\gamma\gtrsim\Znuc^3$, the
usual Thomas-Fermi model becomes inapplicable for
an atom \cite{LiebSY92}. In particular,
it cannot describe the difference of the transverse and
longitudinal atomic sizes, which becomes huge
in such strong fields. In this field range, however, a good
starting approximation is provided by so called adiabatic
approximation, which presents each electron orbital as a
product of a Landau function \cite{SokTer}, describing free
electron motion in the plane transverse to the field, and a
function describing a one-dimensional motion of the electron
along magnetic field lines in the field of an effective
potential, similar to the Coulomb potential truncated at
zero \cite{HainesRoberts}. At $\gamma\gg\Znuc^3$, all
electron shells of the atom are strongly compressed in the
directions transverse to the field. In the ground
state, atomic sizes along and transverse to $\bm{B}$,
respectively, can be estimated as
\cite{KadomtsevKudryavtsev}
\beq
l_\perp\approx \sqrt{2\Znuc-1}\,\am,
\quad
l_\|\approx\frac{\Znuc^{-1}\aB
}{
\ln[\sqrt{\gamma}/(\Znuc\sqrt{2\Znuc-1})]}.
\eeq
In this case, the binding energy $E^{(0)}$ of the ground
state increases with increasing $\bm{B}$ approximately as 
$(\ln \gamma)^2$. Here and hereafter, the superscript
(0) indicates the approximation of a nonmoving nucleus.
At $\Znuc\gg1$ and $\gamma/\Znuc^3\to\infty$, the
asymptotic estimate reads
$E^{(0)}\sim -\Znuc\hbar^2/(\mel l_\|^2)$
\cite{KadomtsevKudryavtsev}. However, this asymptote is
never reached in practice (see~\S\,\ref{sect:QED}).

Particularly many works were devoted to the simplest atom in
magnetic field, the H atom. Since the electron resides on
the ground Landau level $N=0$ in the hydrogen atom at
$B>10^9$~G, its spin being directed opposite to the field, a
bound state is determined by quantum numbers $s$ and $\nu$,
where $s = 0,1,2,\ldots$ corresponds to the electron
orbital-momentum projection on the magnetic-field direction,
$-\hbar s$, and $\nu=0,1,2,\ldots$ in the adiabatic
approximation is equal to the number of wave-function nodes
along this direction. The tightly-bound atomic states are
characterized by the value $\nu=0$, while all non-zero
values of $\nu$ correspond to loosely-bound states.

\begin{figure}[t]
\includegraphics[width=\linewidth]{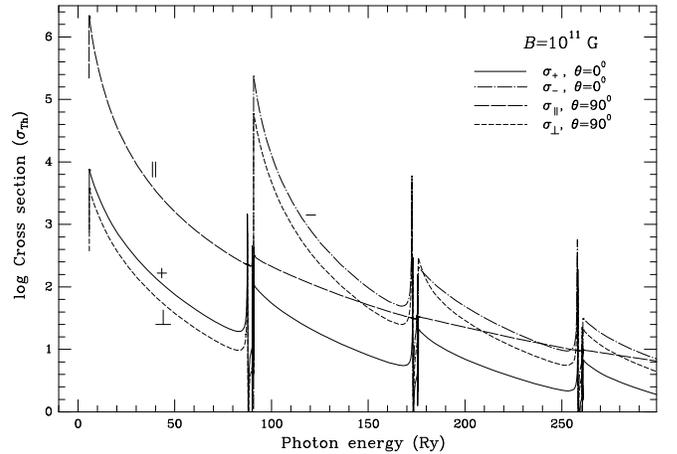}
\caption{
Logarithm of photoionization cross-section, normalized to the
Thomson cross section (\ref{Thomson}),
$\log(\sigma/\sigma_\mathrm{T})$, as function of photon energy
$\hbar\omega$ for the ground state of the hydrogen atom at
rest in magnetic field $B=10^{11}$~G. The curves labelled
by ``+'', ``$-$'', and ``$\|$'' display the cross sections
for circular and longitudinal polarizations $\alpha=+1$,
$-1$, and 0, respectively, and the curve labelled ``$\perp$''
is for radiation polarized perpendicular to  $\bm{B}$. The
wave vector $\bm{k}$ is directed along  $\bm{B}$ for
$\alpha\pm1$ and perpendicular to  $\bm{B}$ for the other
two cases. (Fig.~4 from \cite{PPV97}, reproduced with the
permission of \copyright ESO.)
\label{fig:ppv4}}
\end{figure}

Calculations of the hydrogen-atom properties beyond the
adiabatic approximation were performed by various methods
(variational, discrete-mesh, etc.). At $\gamma\gg1$, the
most natural method of calculations is the expansion of the
wave function over the Landau orbitals, which constitute a
complete orthogonal functional basis in the plane
perpendicular to the magnetic field \cite{SimolaVirtamo}.
Such calculations were done in
\cite{SimolaVirtamo,Rosner84,Forster84} for the bound states
and in \cite{PPV97} also for the continuum states, which
allowed one to obtain the oscillator strengths as well as
photoionization cross-sections. Examples of such
cross-sections are presented in Fig.~\ref{fig:ppv4} for the
hydrogen atom at rest in strong magnetic fields with account
of the finite proton mass. The broad peaks correspond to
transitions to excited Landau levels $N>0$, while the narrow
peaks and dips near corresponding partial thresholds with
$\hbar\omega\approx N\hbar\omc$ are due to resonances
related to autoionization of metastable states \cite{PPV97}.

Analytical expressions for atomic characteristics are best
suited for astrophysical modeling. However, the asymptotic
estimates at $\gamma\gg1$ do not provide the desirable
accuracy. For example, the binding energy of the
ground-state hydrogen
atom at rest, $E^{(0)}_{s\nu}$ at
$s=\nu=0$, when calculated in frames of the nonrelativistic
quantum mechanics, goes to $(\ln\gamma)^2$\,Ry in the limit
$\gamma\to\infty$ \cite{Loudon,HainesRoberts}, but this
estimate is in error by a factor over 2 at any $B$ values
that are encountered in the neutron stars. With account of
two further terms of the asymptotic expansion
\cite{HasegawaHoward61}  
$E^{(0)}_{00}\sim\ln^2(\tilde\gamma/\ln^2\tilde\gamma)$~Ry, 
where $\tilde\gamma\approx0.28\gamma$. But even this
estimate differs from accurate results by 40\,--\,80~\% at $B
\sim 10^{12}$\,--\,$10^{14}$~G. A possible way of solution
to this problem consists in constructing analytical
approximations to the results of numerical calculations. In
\cite{P98} we gave accurate fitting formulae for many bound
states of the hydrogen atom at
$B\lesssim10^{14}$~G. The energy levels in the infinite-mass
approximation have been recently revisited by Popov and
Karnakov 
\cite{PopovKarnakov12}, who obtained analytical expressions,
applicable at $B\gtrsim10^{11}$~G. Here we will give
another
approximation for the tightly-bound levels, valid at
\emph{any} $B$. Temporarily ignoring corrections for vacuum
polarization (\S\,\ref{sect:QED}) and finite nuclear mass
(\S\,\ref{sect:motion}), we present the binding energy as
\beq
\frac{E_{s,0}^{(0)}}{\mbox{Ry}} = \frac{(1+s)^{-2}+
       (1+s)\, x/a_1 + a_3 x^3 + a_4 x^4 + a_6 x^6
      }{
       1 + a_2 x^2 + a_5 x^3 + a_6 x^4},
\label{Hfit0}
\eeq
where $x=\ln(1+a_1\gamma)$. Here, $a_i$ are numerical
parameters, which we approximate as functions of $s$:
\bea
     a_1 &=& {(0.862+2.5\,s^2)
           }/{
            (1+0.018\,s^3)} \,,
\nonumber\\
        a_2 &=& 0.275 +
          0.1763\,\delta_{s,0} + s^{2.5}/6 \,,
\nonumber\\
      a_3 &=& 0.2775+0.0202\,s^{2.5},
\qquad
\nonumber\\
      a_4 &=& {0.3157}/{(1+2s)^2}
           - 0.26\,\delta_{s,0},
\nonumber\\
      a_5 &=& 0.0431,
\nonumber\\
      a_6 &=& {2.075\times10^{-3}}/{
        (1+7s^2)^{0.1}}
          + 1.062\times10^{-4}\,s^{2.5}\,.
\nonumber
\eea
Approximation (\ref{Hfit0}) accurately reproduces the Zeeman
shift of the lowest sublevel of each multiplet in the
weak-field limit and the correct asymptote in the
strong-field limit. Its inexactness is confined within 3\%
for $s<30$ at $\gamma>1$ and for $s<5$ at any $\gamma$, and
within 0.3\% for $s=0$ at any $\gamma$.

Binding energies of the loosely-bound states
($\nu\geqslant1$) can be evaluated at $\gamma\gtrsim1$ as
\beq
   E_{s,\nu}^{(0)} = \frac{\mbox{1~Ry}}{(n+\delta)^2},
\label{Hlike}
\eeq
where
\bea&&\hspace*{-2em}
     n=\frac{\nu+1}{2},
\quad
\delta\approx\frac{1+s/2}{1+2\sqrt\gamma+0.077\gamma}
\quad
\mbox{for odd $\nu$;}
\nonumber\\&&\hspace*{-2em}
     n=\frac{\nu}{2},
\quad
      \delta\approx \frac{1+s/8}{
          0.6+1.28\ln(1+0.7\gamma^{1/3})}
\quad
\mbox{for even $\nu$.}
\nonumber\eea
At
$\gamma\to\infty$, energies (\ref{Hlike}) tend to those of a
field-free H atom ($n^{-2}$~Ry),
therefore the loosely-bound states are often called
``hydrogenlike'' (this picture is broken by 
vacuum polarization, \S\,\ref{sect:QED}).

In the approximation of an infinite nuclear mass, the energy
of any one-electron ion is related to the hydrogen atom
energy as $E(\Znuc,B)=\Znuc^2\,E(1,B/\Znuc^2)$
\cite{SurmelianOConnel74}. Thus one sees, in particular,
that the adiabatic approximation for the single-electron
ions is applicable at  $\gamma\gg \Znuc^2$, which is a
weaker condition than for many-electron atoms. Analogous
similarity relations exist also for the cross sections of
radiative transitions \cite{Wunner-ea82}. However, they are
violated if one takes motion across the magnetic field into
account. Even for an atom at rest, the account of the finite
nuclear mass can be important at $s\neq0$. These effects
will be considered in \S\,\ref{sect:motion}.

Binding energies and oscillator strengths of many-electron
atoms were successfully calculated with the use of different
methods: variational (e.g., \cite{AlHujajSchmelcher04} and
references therein), density-functional
\cite{Jones85,RelovskyRuder,Braun02}, Monte Carlo
\cite{JonesOC97,Buecheler-ea07}, and the Hartree-Fock method
\cite{Fock,Froese}. In the simplest version of the
Hartree-Fock method
\cite{Ruder-ea,MillerNeuhauser91,EngelKW09}, the
wave-function basis is constructed from the one-electron
wave functions in the adiabatic approximation. This method
is reliable for calculations of the energies, oscillator
strengths, and photoionization cross sections of the helium
atom \cite{MedinLP08}. But for many-electron atoms the
condition of applicability of the adiabatic approximation
$\gamma\gg\Znuc^3$ is too restrictive. It is overcome in the
mesh Hartree-Fock method, where each one-electron orbital is
numerically determined as a function of the longitudinal
($z$) and radial coordinates on a two-dimensional mesh
\cite{Ivanov91} (see also \cite{IvanovSchmelcher00}, and
references therein), and in the ``twice self-consistent''
method \cite{SchimeczekEW12}, where a transverse part of
each orbital is presented as a linear superposition of the
Landau functions with numerically optimized coefficients.
These works gave a number of important results but were not
realized in astrophysical applications. In practice, the
optimal method for modeling neutron-star atmospheres
containing atoms and ions of elements with  $2<
\Znuc\lesssim10$ proves to be the method by Mori and Hailey
\cite{MoriHailey02}, where corrections to the adiabatic
Hartree approximation are treated by perturbation. The
latter method can provide an acceptable accuracy 
at moderate computational expenses.

\subsection{Molecules and molecular ions}

Molecular properties in strong magnetic fields have been
studied during almost 40 years, but remain insufficiently
known. Known the best are the properties of diatomic
molecules oriented along the field, especially the H$_2$
molecule (see \cite{SchmelcherDC01}, and references
therein). Lai \cite{Lai01} obtained approximate expressions
for its binding energy at $\gamma\gtrsim10^3$, which grows
approximately at the same rate $\propto(\ln\gamma)^2$ as the
atomic binding energy. In such strong fields, the ground
state of this molecule is the state where the spins of both
electrons are opposite to the magnetic field and the
molecular axis is parallel to it, unlike the weak fields
where the ground state is $^1\Sigma_g$. In moderate fields,
the behavior of the molecular terms is quite nontrivial. If
the molecular axis is parallel to $\bm{B}$, then the states
$^1\Sigma_g$ and $^3\Pi_u$ are metastable at
$0.18<\gamma<12.3$, and decay into the channel $^3\Sigma_u$
\cite{DetmerSC98}. It turns out, however, that the molecular
orientation along $\bm{B}$ is not optimal in such fields:
for example, at $\gamma=1$ the triplet state of the molecule
oriented perpendicular to the field has the lowest energy,
and at $\gamma=10$ the ground state is inclined at
37$^\circ$ to $\bm{B}$ \cite{Kubo07}.

The ion H$_2$$^+$ is well studied, including its arbitrary
orientations in a magnetic field (e.g., \cite{KS96}, and
references therein). An analysis by Khersonskii
\cite{Khers87b} shows that the abundance of {H$_2$}$^+$ is
very small in neutron-star atmospheres, therefore the
these ions are unlikely to affect the observed
spectra.

Strong magnetic fields stabilize the molecule He$_2$ and its
ions He$_2$$^{+}$, He$_2$$^{2+}$, and He$_2$$^{3+}$, which
do not exist in the absence of the field. Mori and Heyl
\cite{MoriHeyl} have performed the most complete study of
their binding energies in neutron-star atmospheres. The ions
HeH$^{++}$, H$_3$$^{++}$, and other exotic molecular ions,
which become stable in the strong magnetic fields, were also
considered (see \cite{Turbiner07,TurbinerLVG10}, and
references therein). Having evaluated the ionization
equilibrium by the Khersonskii's method \cite{Khers87b}, one
can easily see that the abundance of such ions is extremely
small at the densities, temperatures, and magnetic fields
characteristic of the neutron stars. Therefore, such ions do
not affect the thermal spectrum.

There are rather few results on molecules composed of
atoms heavier than He. Let us note the paper 
\cite{MedinLai06a}, where the authors applied the
density-functional method to calculations of binding
energies of various molecules from H$_n$ to Fe$_n$ with $n$
from 1 through 8 at $B$ from $10^{12}$~G to
$2\times10^{15}$~G. The earlier studies of heavy molecules
in strong magnetic fields are discussed in the review by Lai
\cite{Lai01}. All these studies assumed the model of
infinitely massive atomic nuclei.

\subsection{Relativistic effects}
\label{sect:QED}

One can encounter the statement that the use of the
nonrelativistic quantum mechanics for calculation of atomic
and molecular structure is justified only at $B <
B_\mathrm{QED}$. However, a treatment of the hydrogen atom
in strong magnetic fields based on the Dirac
equation
\cite{LindgrenVirtamo79,ChenGoldman92,NakashimaNakatsuji10}
has not revealed any significant differences from the
solution to the same problem based on the Schr\"odinger
equation. The reasons for that are clear. One can always
expand a wave function over a complete basis of
two-dimensional functions, such as the set of the Landau
functions for all electrons. The Landau functions have the
same form in the relativistic and nonrelativistic theories
\cite{SokTer}. Coefficients of such expansion are functions
of $z$ corresponding to the electron motion along $\bm{B}$.
This motion is nonrelativistic for the bound electrons,
because the maximal binding energy is much smaller than the
electron rest energy $\mel c^2=511$~keV. Therefore, a system
of equations for the functions of $z$ in question can be solved
in the nonrelativistic approximation, which thus provides
the accurate wave function.

Nevertheless, there is a specific relativistic effect, which
is non-negligible in superstrong fields. As noted by
Heisenberg and Euler \cite{HeisenbergEuler}, the virtual
electron-positron pairs that appear in an electromagnetic
field according to the Dirac theory, modify the Maxwell
equations. This effect is called vacuum polarization. To
date it has not been observed, but it was studied in many
theoretical works, reviewed in detail by Schubert
\cite{Schubert}. A strong electromagnetic field creates a
nonzero space charge by acting on the virtual pairs. Such
charge, in particular, screens the Coulomb interaction
between an electron and an atomic nucleus at distances
comparable to the Compton wavelength
$\lambda_\mathrm{C}=2\pi\hbar/(\mel c)=2\pi\alphaf\aB$. 
Shabad and Usov \cite{ShabadUsov07,ShabadUsov08} noted that
this screening affects the even atomic levels in superstrong
magnetic fields, which squeeze the atom so that its size
becomes comparable to $\lambda_\mathrm{C}$. As a result,
instead of the unlimited growth of the binding energies of
the tightly-bound states that is predicted by the
nonrelativistic theory for unlimited increase of $B$, these
energies ultimately level off. For the same reason, the
double degeneracy of the loosely-bound states that follows
from \req{Hlike} at $\gamma\to\infty$, does not realize.

Machet and Vysotsky \cite{MachetVysotsky11} have thoroughly
studied this effect, confirmed the qualitative conclusions
of Shabad and Usov, and obtained more accurate quantitative
estimates. In particular, according to their results (see
also \cite{PopovKarnakov12}), the effect of the vacuum
polarization on the electron binding energies in a nonmoving
Coulomb potential can be simulated by replacing the
parameter $\gamma$ to $\gamma^\ast = \gamma/[1+
\alphaf^3\gamma/(3\pi)]$. As a result, the binding energy of
the hydrogen atom cannot exceed 1.71 keV at any $B$.

\subsection{The effects of finite nuclear mass}
\label{sect:motion}

An overwhelming majority of studies of atoms in strong
magnetic fields assumed the nuclei to be infinitely massive
(fixed in space). For magnetic neutron-star atmospheres,
this approximation is very serious and often an
undesirable simplification.

Let us start with an atom with a nonmoving center of mass.
The nucleus of a finite mass, as any charged particle,
undergoes circular oscillations in the plane perpendicular
to $\bm{B}$. In the atom, these oscillations cannot be
separated from the electron oscillations, therefore the
longitudinal projections of the orbital moments of the
electrons and the nucleus are not conserved separately.
Only their difference is conserved. Different atomic quantum
numbers correspond to different oscillation energies of the
atomic nucleus, multiple of its cyclotron energy. As a
result, the energy of every level gets an addition, which is
non-negligible if the parameter $\gamma$ is not small
compared to the nucleus-to-electron mass ratio. For the
hydrogen atom and hydrogenlike ions, $\hbar s$ in
\req{Hfit0} now corresponds to the difference of
longitudinal projections of orbital moments of the atomic
nucleus and the electron, and the sum $N+s$ plays role of a
nuclear Landau number, $N$ being the electron Landau number.
For the bound states in strong magnetic fields, $N=0$,
therefore the nuclear oscillatory addition to the energy
equals $s\hbar\omci$. Thus the binding energy of a hydrogen
atom at rest is
\beq
 E_{s\nu}
    = E_{s\nu}^{(0)}(\gamma^\ast) - \hbar\omci s, 
\label{E0}
\eeq
where $\gamma^\ast  = \gamma/(1+4.123\times10^{-8}\,\gamma)$
according to \S\,\ref{sect:QED}. It follows that the number
of $s$ values is limited for the bound states. In
particular, one can easily check using Eqs.~(\ref{Hfit0})
and (\ref{E0}) that all bound states have zero
moment-to-field projection ($s=0$) at $B>6\times10^{13}$~G.

The account of the finite nuclear mass is more complicated
for multielectron atoms. Al-Hujaj and Schmelcher
\cite{AlHujajSchmelcher03} have shown that the contribution
of the nuclear motion to the binding energy of a non-moving atom 
equals
$\hbar\omci S(1+\delta(\gamma))$, where ($-S$) is the total
magnetic quantum number and $|\delta(\gamma)|\ll1$.

\begin{figure*}
\begin{center}
\includegraphics[width=.319\textwidth]{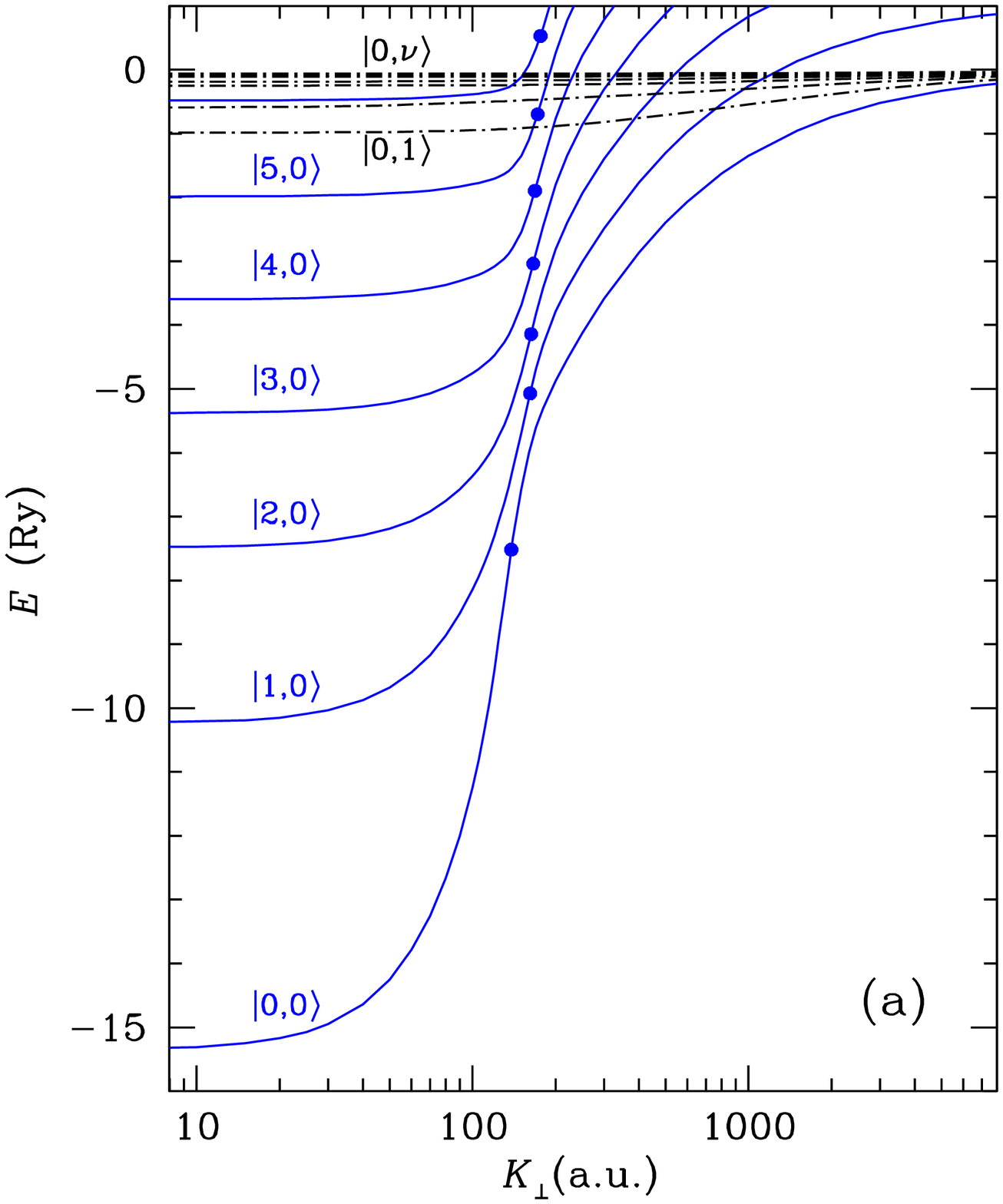}
\,
\includegraphics[width=.336\textwidth]{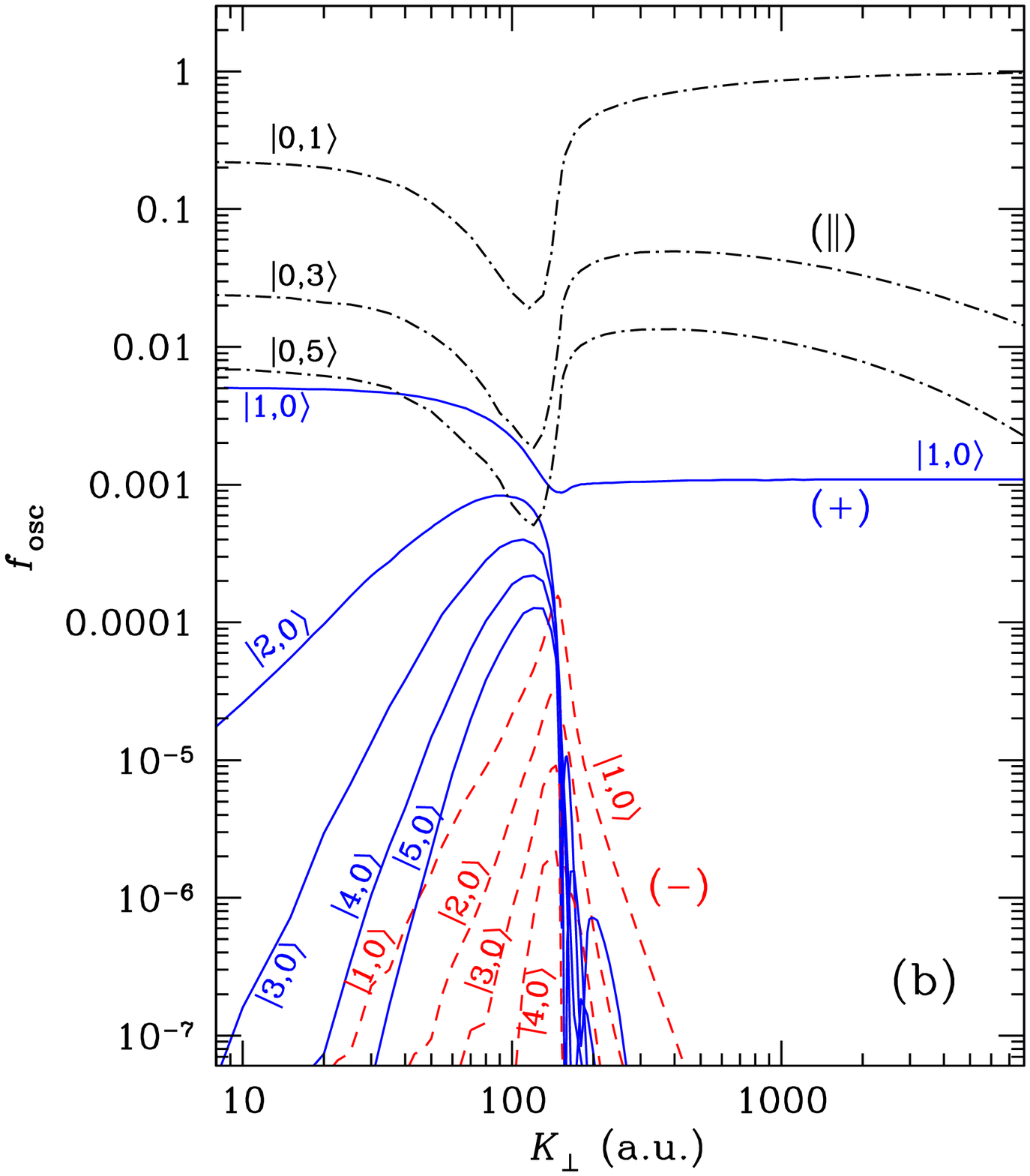}
\,
\includegraphics[width=.314\textwidth]{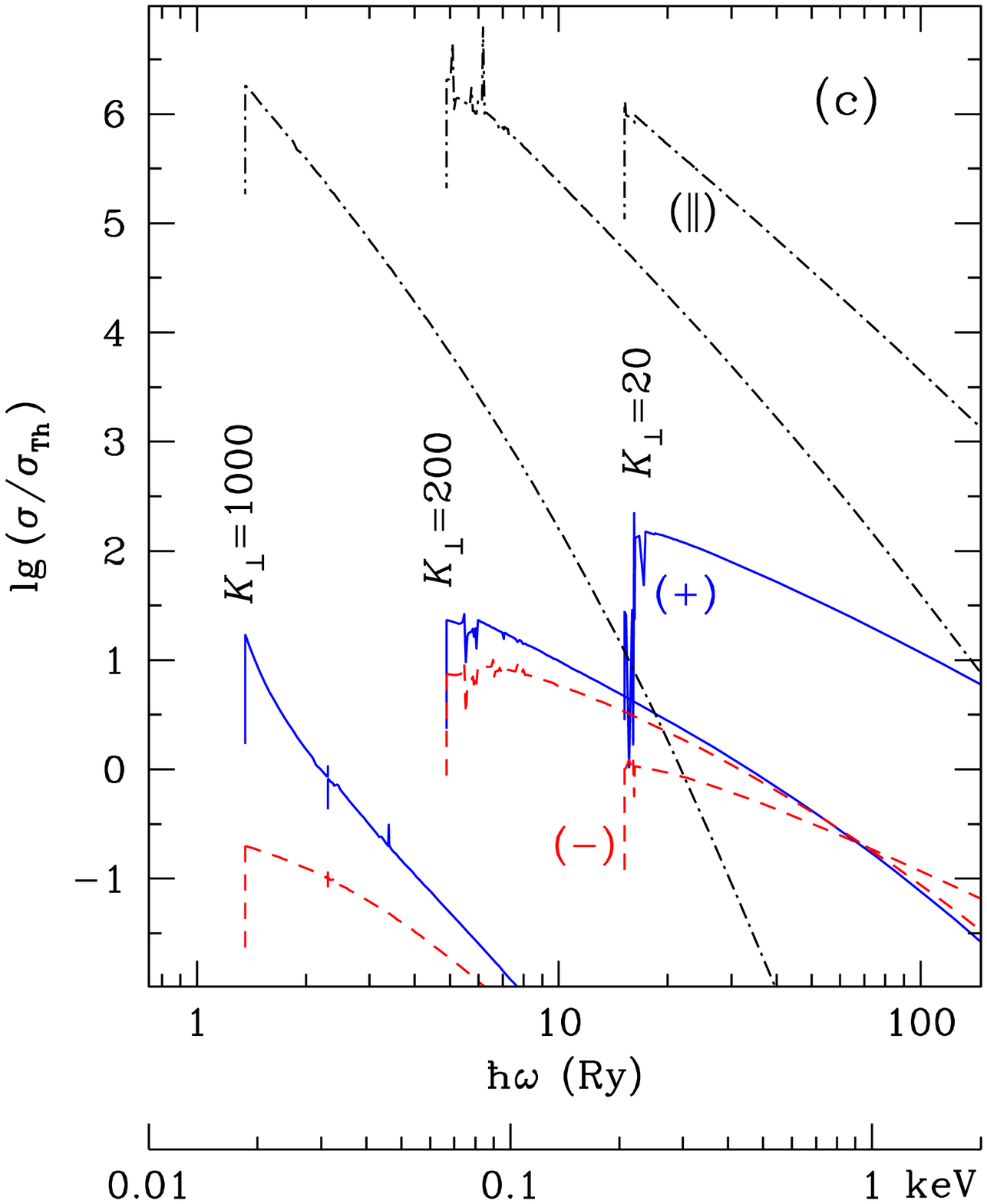}
\caption{
(a) energies, (b) oscillator strengths, and (c)
photoionization cross-sections for a hydrogen atom moving in
magnetic field $B=2.35\times10^{12}$~G. Energies of states
$|s,0\rangle$ (solid curves) and $|0,\nu\rangle$ (dot-dashed
curves) are shown as functions of the transverse
pseudomomentum $\Kp$ (in atomic units). The heavy dots
on the solid curves are the inflection points at
$\Kp=\Kc$.  The  $\Kp$-dependence of
oscillator strengths (b) is shown for transitions from the
ground state to the states $|s,0\rangle$ under influence of
radiation with polarization $\alpha=+1$ (solid curves) and
$\alpha=-1$ (dashed curves), and also for transitions into
states $|0,\nu\rangle$ for $\alpha=0$ (dot-dashed curves).
Cross sections of photoionization (c) under the influence of
radiation with $\alpha=+1$ (solid curves), $\alpha=-1$
(dashed curves), and $\alpha=0$ (dot-dashed curves) are
shown for the ground state as functions of the photon energy
in Ry (the upper x-scale) and keV (the lower x-scale) at
$\Kp=20$~a.u.{} (the right curve),  $\Kp=200$~a.u.{}
(the middle curve), and $\Kp=1000$~a.u.{} (the left curve
of every type).
}
\label{fig:Hatom}
\end{center}
\end{figure*}

The astrophysical simulations require an account of finite
temperatures, hence thermal motion of particles. The theory
of motion of a system of point charges in a constant
magnetic field is reviewed in \cite{JHY83,BayeVincke90}. The
canonical momentum $\bm{P}$ is not conserved in this motion,
but a pseudomomentum
$\bm{K}=\bm{P}+(1/2c)\,\bm{B}\times\sum_i q_i \bm{r_i}$ 
is conserved. The
pseudomomentum of a single charged particle has a one-to-one
correspondence to the position of the guiding center
in the $(xy)$ plane, perpendicular to the
magnetic field, while a pseudomomentum of an atom or ion
equals the sum of pseudomomenta of its constituent
particles. If the system is electrically neutral as a whole,
then all the components of  $\bm{K}$ are good quantum
numbers. For a charged system (an ion), $K^2$ is a good
quantum number, while $K_x$ and $K_y$ do not commute.
The specific effects related to collective motion of a
system of charged particles are especially important in a
neutron-star atmosphere at $\gamma\gg1$. In particular, so
called decentered states may become populated, where an
electron is localized mostly in a ``magnetic well'' aside
from the Coulomb center.

For a hydrogen atom,
$
   \bm{K} = \bm{P} + ({e}/{2c})\,\bm{B}\times \bm{R},
$
where the vector $\bm{R}$ connects the electron to the
proton. The studies
of this particular case were initiated in the pioneering
works  \cite{GorkovDzyal,Burkova,IPM84}. Numerical
calculations of the energy spectrum of the hydrogen atom with
account of the effects of motion across a strong magnetic
field were performed in  \cite{VDB92,P94}. Probabilities of
various radiative transitions were studied in a
series of papers ended with \cite{PP97}.

Figure~\ref{fig:Hatom} shows the
energies, oscillator strengths, and photoionization
cross-sections of a hydrogen atom moving in a magnetic field
with $\gamma=1000$. The negative energies in
Fig.~\ref{fig:Hatom}a correspond to bound states. The
reference point is taken to be the sum of the zero-point
oscillation energies of free electron and proton,
$(\hbar\omc+\hbar\omci)/2$. At small transverse
pseudomomenta $\Kp$, the energies of low levels in
Fig.~\ref{fig:Hatom}a exceed the binding energy of the
field-free hydrogen atom (1~Ry) by an order of magnitude.
However, the total energy increases with increasing
$\Kp$, and it can become positive for the states with
$s\neq0$ due to the term $\hbar\omci s$ in \req{E0}. Such
states are metastable. In essence, they are continuum
resonances. Note that the transverse atomic velocity equals
$\partial E/\partial \bm{K}$, therefore it is maximal at the
inflection points ($\Kp=\Kc$) on the curves
in Fig.~\ref{fig:Hatom}a and decreases with further increase
of  $\Kp$ \cite{P94}, while the average
electron-proton distance continues to increase. The atom
goes into the decentered state, where the electron and
proton are localized near their guiding centers, separated
by distance $r_* = (\aB^2/\hbar)\Kp/\gamma$.

The dependences of the binding energies on $\Kp$ are
approximately described at $\Kp\ll \Kc$ and $\Kp\gg \Kc$,
respectively, by expressions
\bea\hspace*{-2em}
   E_{s\nu}^{(<)} &\!\!=&\!\! E_{s\nu}^{(0)}
         - \Kp^2/(2m_\mathrm{eff}) - \hbar\omci s,
\label{Ecentred}
\\\hspace*{-2em}
   E_{s\nu}^{(>)} &\!\!\!\!\!=&\!\! \!\!\!
         \frac{\mbox{2\,Ry}}{\sqrt{
       \hat{r}_*^{2\phantom{/}}
             + (2\nu+1)\, \hat{r}_*^{3/2} 
     + \ldots }
                }
          - \hbar\omci s\,,
\label{Edecentred}
\eea
where $\hat{r}_* \equiv r_*/\aB$ and $m_\mathrm{eff}$ is an
effective ``transverse mass.'' The latter is expressed
through the values of $E_{s\nu}^{(0)}$ for the given and
neighboring levels by the perturbation theory
\cite{VB88,PM93}. However, for excited states even a small
inaccuracy in $E_{s\nu}^{(0)}$ may lead to a fatal error in
$m_\mathrm{eff}$. Therefore in practice it is more
convenient to use the approximation
$m_\mathrm{eff}\approx
m_\mathrm{a}\,[1+(\gamma/\gamma_{s\nu})^{p_{s\nu}}]$, where
$\gamma_{s\nu}$ and $p_{s\nu}$ are dimensionless parameters,
and $m_\mathrm{a}$ is the true mass of the atom. For the
tightly bound levels,
$\gamma_{s0}\approx6\times10^3/(1+2s)^2$ and
$p_{s\nu}\approx0.9$. At $B\lesssim10^{13}$~G we can
approximately describe the energies of the states with
$\nu=0$ at arbitrary $\Kp$, if we replace the ellipsis under
the square root in
\req{Edecentred} by the expression 
$\hat{r}_*/(5+3s)+\big(\mbox{2~Ry}/E_{s\nu}^{(0)}\big)^2$,
and replace the inflection point $\Kc$ by intersection of
$E_{s\nu}^{(<)}(\Kp)$ with $E_{s\nu}^{(>)}(\Kp)$. At
stronger fields or for $\nu\neq0$, the transition between
the centered and decentered states smears, and one has
to resort to more complex  fitting formulae~\cite{P98}.

Figure~\ref{fig:Hatom}b shows oscillator strengths for the
main dipole-allowed transitions from the ground state to
excited discrete levels as functions of $\Kp$. Since the
atomic wave-functions are symmetric with respect to the
$z$-inversion for the states with even $\nu$, and
antisymmetric for odd $\nu$, only the transitions that
change the parity of $\nu$ are allowed for the polarization
along the field ($\alpha=0$), and only those preserving the
parity for the orthogonal polarizations
($\alpha=\pm1$). For the atom at rest, in the dipole
approximation, due to the conservation of the $z$-projection
of the total angular momentum of the system,
absorption of a photon with polarization $\alpha=0,\pm1$ results in
the change of $s$ by $\alpha$. This
selection rule for a non-moving atom manifests itself in
vanishing oscillator strengths at $\Kp\to0$ for
$s\neq\alpha$. In an appropriate
coordinate system \cite{Burkova,P94}, the symmetry is
restored at $\Kp\to\infty$, therefore the transition
with $s=\alpha$ is the only one that survives also in the
limit of large pseudomomenta. But in the intermediate region
of $\Kp$, where the transverse atomic velocity is not
small, the cylindrical symmetry is broken, so that
transitions to other levels are allowed. Thus the corresponding
oscillator strengths in Fig.~\ref{fig:Hatom}b have maxima at
$\Kp\approx \Kc$. Analytical approximations for
these oscillator strengths are given in \cite{P98}.

Figure~\ref{fig:Hatom}c shows photoionization cross-sections
for hydrogen in the ground state as functions of photon
energy at three values of $\Kp$. The leftward shift of
the ionization threshold with increasing $\Kp$
corresponds to the decrease of the binding energy that is
shown in Fig.~\ref{fig:Hatom}a, while the peaks and dips on
the curves are caused by resonances at transitions to
metastable states  $|s,\nu;K\rangle$ with positive energies
(see \cite{PP97}, for a detailed discussion).

Quantum-mechanical calculations of the characteristics of
the He$^+$ ion that moves in a strong magnetic field are
performed in \cite{BPV97,PB05}. The basic difference from
the case of a neutral atom is that the the ion
motion is restricted by the field in the transverse plane,
therefore the values of $K^2$ are quantized
\cite{JHY83,BayeVincke90}. Clearly, the similarity relations
for the ions with nonmoving nuclei
(\S\,\ref{sect:atoms}) do not hold anymore.

Currently there is no detailed calculation of binding
energies, oscillator strengths, and photoionization
cross-sections for atoms and ions other than H and He$^+$,
arbitrarily moving in a strong magnetic field. For such
species one usually neglects the decentered states and uses
a perturbation theory with respect to $\Kp$
\cite{VB88,PM93}. Such approach was realized, e.g., in
\cite{MoriHailey02,MedinLP08}. It can be sufficient for
simulations of relatively cool atmospheres of moderately
magnetized neutron stars. Detailed conditions of
applicability of the perturbation theory \cite{VB88,PM93}
require calculations, but a rough order-of-magnitude
estimate can be obtained by requiring that the mean Lorentz
force acting on a bound electron because of the atomic
thermal motion should be small compared to the Coulomb
forces. As a result, for an atom with mass
$m_\mathrm{a}=Am_\mathrm{u}$ we get the condition $\kB
T/E_\mathrm{b}\ll m_\mathrm{a}/(\gamma
\mel)\approx4A/B_{12}$, where $E_\mathrm{b}$ is the atomic
ionization energy. If $B\lesssim10^{13}$~G and
$T\lesssim10^6$~K, it is well satisfied for
low-lying levels of carbon and heavier atoms.

\subsection{Equation of state}
\label{sect:EOS}

Theoretical description of thermodynamics of partially
ionized plasmas can be based on either ``physical'' of
``chemical'' models (see, e.g., a discussion and references
in \cite{Dappen92,Rogers00}). In the chemical model of
plasmas, bound states (atoms, molecules, ions) are treated
as separate members of the thermodynamic ensemble, while in
the physical model the only members of the ensemble are
atomic nuclei and electrons. Each of the models can be
thermodynamically self-consistent, but the physical model is
more relevant from the microscopic point of view, because it
does not require a distinction of electrons bound
to a given nucleus. Such a distinction becomes very ambiguous
at high densities, where several nuclei can attract the same
electron with comparable forces. On the other hand,
calculations in frames of the physical model are technically more
complicated. As a rule, they are based on a
diagram expansion, which requires an increase of the number
of terms with the density increase. For this reason, even
the most advanced equation of state for nonmagnetic
photospheres that is based on the physical model
\cite{OPAL-EOS} still restricts to the domain
$\rho\lesssim 10\, T_6^3$ \gcc.

Studies of magnetic neutron-star photospheres, as a rule,
are based on the chemical plasma model. In this case, the
ionization equilibrium is evaluated by minimizing the
Helmholtz free energy $F$ given by 
\beq
   F= F_\mathrm{id}^{(e)} + F_\mathrm{id}^\mathrm{(i)} +
      F_\mathrm{int} +
      F_\mathrm{ex},
\label{F}
\eeq
where $F_\mathrm{id}^{(e)}$ and $F_\mathrm{id}^{(i)}$
describe the ideal electron and ion gases, $F_\mathrm{int}$
includes internal degrees of freedom for bound states, and
$F_\mathrm{ex}$ is a nonideal component. All thermodynamic
functions that are required for modeling a photosphere
with a given chemical composition are expressed through
derivatives of $F$ over $\rho$ and $T$ \cite{LaLi-SP1}.

According to the Bohr-van Leeuwen theorem,\footnote{This
theorem was proved by different methods in PhD theses by
Niels Bohr in 1911 and H.-J.~van Leeuwen in 1919, and
published by the latter in 1921 \cite{vanLeeuwen}.}
magnetic field does not affect thermodynamics of classical
charged particles. The situation differs in the quantum
mechanics. The importance of the quantum effects depends on
the parameters $\zete$ (\ref{zeta_e}) and $\zeti$
(\ref{omci}).

We use the equality \cite{LaLi-SP1}
$
   F_\mathrm{id}^{(e)}/V =
   \mu_\mathrm{e} n_\mathrm{e}  - P_\mathrm{id}^{(e)}
$
where $V$ is the volume of the system, and
$\mu_\mathrm{e}$, $n_\mathrm{e}$, and $P_\mathrm{id}^\mathrm{(e)}$ 
are, respectively, the chemical potential, number density, and
pressure in the ideal electron gas model. The equation of
state is determined by a relation between these quantities,
which can be found from relations (e.g.,
 \cite{NSB1,PC13})
\beq
   \Bigg\{
   \begin{array}{l}
       n_\mathrm{e} \\ P_\mathrm{id}^{(e)}
   \end{array}
   \Bigg\}
         =
   \sum_{N,\sigma}
    \frac{(1+2bN)^{1/4}}{\pi^{3/2}\am^2\lambda_\mathrm{e}}
   \Bigg\{
   \begin{array}{l}
     {\partial I_{1/2}(\chi_N,\tau_N)}/{\partial \chi_N}
     \\
        \kB T\,I_{1/2}^{\phantom{I}}(\chi_N,\tau_N)
   \end{array}
   \Bigg\},
\label{nPe}
\eeq
where $\lambda_\mathrm{e} = [{2\pi\hbar^2}/({ \mel \kB T})]^{1/2}$
is the thermal de Broglie wavelength,
$\tau_N = \kB T/(\mel c^2 \sqrt{1+2bN})$,
$\chi_N = \mu_\mathrm{e} / (\kB T) + \tau_0^{-1} - \tau_N^{-1}$,
\beq
   I_{1/2}(\chi_N,\tau_N) \equiv \int_0^\infty
  \frac{ \sqrt{x \,(1+\tau_N x/2)}
    }{ \exp(x-\chi_N)+1 }\,{\dd}x
\label{I_nu}
\eeq
is the Fermi-Dirac integral, and the summation is done over
all $N$ and all values of spin projections on the magnetic
field, $\hbar\sigma/2$, so that $\sigma=\pm1$ for positive
$N$ and $\sigma=-1$ at $N=0$.

In a strongly quantizing magnetic field, it is sufficient to
retain only the term with $N=0$ in the sums (\ref{nPe}). In
this case, the electron Fermi momentum  equals $\pF =
2\pi^2\am^2\hbar n_\mathrm{e}$. Therefore, with increasing
$n_\mathrm{e}$ at a fixed $B$, the degenerate electrons
begin to fill the first Landau level when $n_\mathrm{e}$
reaches $n_B=(\pi^2\sqrt2\,\am^3)^{-1}$. This value just
corresponds to the density $\rho_B$ in \req{rho_B}. The
ratio of the Fermi momentum $\pF$ in the strongly quantizing
field to its nonmagnetic value $\hbar(3\pi^2
n_\mathrm{e})^{1/3}$ equals
$[{4\rho^2}/{(3\rho_B^2)}]^{1/3}$. Therefore, the Fermi
energy at a given density $\rho<\sqrt{3/4}\,\rho_B$ becomes
smaller with increasing $B$, that is, a strongly quantizing
magnetic field relieves the electron-gas degeneracy. For
this reason, strongly magnetized neutron-star photospheres
remain mostly nondegenerate, as it were in the absence of
the field, despite their densities are orders of
magnitude higher than the nonmagnetic photosphere densities.

The free energy of nondegenerate nonrelativistic ions is
given by
\bea
 \frac{F_\mathrm{id}^\mathrm{(i)}}{\Nion \kB T}
  &=& 
      \ln\left(2\pi \frac{\nion \lambdi\am^2}{Z}\right)
    + \ln\left( 1- \mathrm{e}^{- \zeti}\right) -1
\nonumber\\&& +
      \frac{\zeti}{2} +  \ln\Bigg(\frac{
     \sinh[\gfact\,\zeti \Mspin/4] }{ \sinh(\gfact\,\zeti/ 4)
       }
     \Bigg),
\label{Fp}
\eea
where $\lambdi = [2\pi\hbar^2 / (\mion \kB T)]^{1/2}$ is the
thermal de Broglie wavelength for the ions, $\sion$ is the
spin number, and $\gfact$ is the spin-related g-factor (for
instance, $\sion=1/2$ and $\gfact=5.5857$ for the proton). All the
terms in  (\ref{Fp}) have clear physical meanings. At
$\zeti\to0$, the first and second terms give together $\ln(
\nion \lambdi^3)$, which corresponds to the
three-dimensional Boltzmann gas. The first term corresponds
to the one-dimensional Boltzmann gas model at $\zeti\gg1$.
The second-last term in (\ref{Fp}) gives the total energy
$\Nion\hbar\omci/2$ of zero-point oscillations transverse to
the magnetic field. Finally, the last term represents the
energy of magnetic moments in a magnetic field.

The nonideal free-energy part $F_\mathrm{ex}$ contains the
Coulomb and exchange contributions of the electrons and the
ions, and the electron-ion polarization energy. In the
case of incomplete ionization $F_\mathrm{ex}$ 
includes also interactions of ions and electrons with
atoms and molecules. In turn, the interaction between the
ions is described differently depending on the phase state
of matter. The terms that constitute $F_\mathrm{ex}$ depend
on magnetic field only if it quantizes the motion of these
interacting particles. Here we will not discuss these terms
but address an interested reader to the paper \cite{PC13}
and references therein. This nonideality is
negligible in the neutron-star atmospheres, but it
determines the formation of a condensed surface, which will
be considered in \S\,\ref{sect:cond}.

\subsection{Ionization equilibrium}
\label{sect:Saha}

For photosphere simulations, it is necessary to determine
the fractions of different bound states, because they affect
the spectral features that are caused by bound-bound and
bound-free transitions. Solution to this problem is laborious
and ambiguous. The principal difficulty in the chemical
plasma model, namely the necessity to distinguish the bound
and free electrons and ``attribute'' the bound electrons to
certain nuclei, becomes especially acute at high densities,
where the atomic sizes cannot be anymore neglected with
respect to their distances. Current approaches to the
solution of this problem are based, as a rule, on the
concept of so called occupation probabilities of
quantum states. For example, consider electrons in
thermodynamic equilibrium with ions of the $Z$th chemical
element, and let $j$ be the ionization degree of every ion
(i.e., the number of lacking electrons), $\kappa$ is its
quantum state, and $E_{j,\kappa}$ and $g_\kappa^{(j)}$ are,
respectively, its binding energy and statistical weight. An
occupation probability $w_{j,\kappa}$ is an additional
statistical weight of the given state under the condition of
plasma nonideality, that is under interaction of the ion
$(Z,j,\kappa)$ with surrounding particles, with respect to
its weight without such interactions.\footnote{This ratio is
not necessarily less than unity, thus the term
``probability'' is not quite correct, but we adhere to the
traditional terminology.} As first noted by
Fermi~\cite{Fermi24}, occupation probabilities
$w_{j,\kappa}$ cannot be arbitrary but should be consistent
with $F_\mathrm{ex}$. Minimizing $F$ with account of the
Landau quantization leads to a system of
ionization-equilibrium equations for $n_j\equiv \sum_\kappa
n_{j,\kappa}$ \cite{Khers87a,RRM97}
\bea\hspace*{-2em}
 \frac{n_j}{n_{j+1}}
      &=& n_\mathrm{e} \lambda_\mathrm{e}^3\,\,
      \frac{\sinh(\zeta_j/2)}{\zeta_j}
         \,\frac{\zeta_{j+1}}{\sinh(\zeta_{j+1}/2)}
\nonumber\\&&\times
          \,\frac{\tanh(\zete/2)}{\zete} \,
          \frac{\mathcal{Z}_{\mathrm{int},j}}{
            \mathcal{Z}_{\mathrm{int},j+1}}\,
           \exp\left(\frac{E_{j,\mathrm{ion}}}{\kB T}\right),
\label{Saha-m}
\eea
where $\mathcal{Z}_{\mathrm{int},j}=\sum_\kappa
g_\kappa^{(j)}\,w_{j,\kappa}\,
\exp\left[({E_{j,\kappa}-E_{j,\mathrm{gr.st}} })/({ {\kB}
T})\right]$ is internal partition function for the $j$th
ion type,  $E_{j,\mathrm{gr.st}}$ is its ground-state
binding energy,
$E_{j,\mathrm{ion}}=E_{j,\mathrm{gr.st}}-E_{j+1,\mathrm{gr.st}}$
is its ionization energy, and $\zeta_j$ is the magnetic
quantization parameter (\ref{omci}). Equation (\ref{Saha-m})
differs from the usual Saha equation, first, by the terms
with $\zete$ and $\zeta_j$, representing partition
functions for distributions of free electrons and ions over
the Landau levels, and second, by the occupation
probabilities $w_{j,\kappa}$ in the expressions for the
partition functions $\mathcal{Z}_{\mathrm{int},j}$.

There were many attempts to find such approximation for the
occupation probabilities that best reproduced the real
plasma EOS. They were discussed, for example, by Hummer and
Mihalas \cite{HummerMihalas}, who proposed an approximation
based on the Inglis-Teller criterion \cite{IT} for
dissolution of spectral lines because of their smearing due
to the Stark shifts in plasma microfields. However, the
translation of the spectroscopic criterion to thermodynamics
is not well grounded. It is necessary to clearly distinguish
between the disappearance of spectral lines of an atom and
the complete destruction of this atom with increasing
pressure, as was stressed, e.g., in
\cite{EckerKroell,Rogers86,StehleJacquemot}. In order to
take this difference into account, in  \cite{P96b} we
introduced a concept of optical occupation probabilities
$\tilde{w}_{j,\kappa}$, which resemble the Hummer-Mihalas
occupation probabilities and should be used for calculation
of spectral opacities, but differ from the thermodynamic
occupation probabilities $w_{j,\kappa}$ that are used in
the EOS calculations.

Equation (\ref{Saha-m}) was applied to modeling partially
ionized atmospheres of neutron stars, composed of iron,
oxygen, and neon \cite{Miller92,RRM97,MoriHailey06,MoriHo}.
The effects related to the finite nuclear masses
(\S\,\ref{sect:motion}) were either ignored or treated in
the first order of the perturbation theory. Since
quantum-mechanical characteristics of an atom in a strong
magnetic field depend on the transverse pseudomomentum
$\Kp$, the atomic distribution over $\Kp$ cannot be
written in a closed form, and only the distribution over
longitudinal momenta $K_z$ remains Maxwellian. The first
complete account of these effects has been taken in
\cite{PCS99} for hydrogen photospheres. Let
$p_{s\nu}(\Kp)\,\dd^2\Kp$ be the probability of
finding a hydrogen atom in the state $|s,\nu\rangle$ in the
element $\dd^2\Kp$ near $\bm{K}_\perp$ in the plane of
transverse pseudomomenta. Then the number of atoms in the
element $\dd^3K$ of the pseudomomentum space equals
\beq
  \dd N(\bm{K})  = N_{s\nu}\,
        \frac{\lambda_\mathrm{a} }{2\pi\hbar}\,
         \exp\left(-\frac{K_z^2}{2m_\mathrm{a}\kB T}\right)\, p_{s\nu}(\Kp) \dd^3K,
\eeq
where $m_\mathrm{a}$ is the mass of the atom, 
$\lambda_\mathrm{a} = [{2\pi\hbar^2}/({ m_\mathrm{a} \kB T})]^{1/2}$
is its thermal wavelength,
and $N_{s\nu}=\int\dd N_{s\nu}(\bm{K})$ is the total number
of atoms with given discrete quantum numbers. The distribution
 $N_{s\nu} p_{s\nu}(\Kp)$ is not known in advance, but
should be calculated in a self-consistent way by
minimization of the free energy including the nonideal terms. It
is convenient to define deviations from the Maxwell
distribution with the use of generalized occupation
probabilities 
$w_{s\nu}(\Kp)$. Then the atomic contribution
$(F_\mathrm{id}+F_\mathrm{int})$ to the free energy equals \cite{PCS99}
\beq
 \kB T \sum_{s\nu} N_{s\nu}
     \int\ln\left[n_{s\nu} \lambda_\mathrm{a}^3 
        \frac{w_{s\nu}(\Kp) }{ \exp(1) \mathcal{Z}_{s\nu}}\right]
       p_{s\nu}(\Kp)\,\dd^2\Kp,
\eeq
where
\beq
   \mathcal{Z}_{s\nu} =
\frac{ \lambda_\mathrm{a}^2 }{ (2\pi\hbar^2) } 
          \int_0^\infty w_{s\nu}(\Kp)\mathrm{e}^{E_{s\nu}(\Kp)/\kB T}
            \Kp \dd \Kp.
\label{Z-int}
\eeq
The nonideal part of the free energy that describes
atom-atom and atom-ion interactions and is responsible
for the pressure ionization has been calculated in 
\cite{PCS99} with the use of the hard-sphere model. The
plasma model included also hydrogen molecules H$_2$ and
chains H$_n$, which become stable in the strong magnetic
fields. For this purpose, approximate formulae of
Lai~\cite{Lai01} have been used, which do not take full
account of the motion effects, therefore the results of
\cite{PCS99} are reliable only when the molecular fraction
is small.

This hydrogen-plasma model underlies thermodynamic
calculations of hydrogen photospheres of neutron stars with
strong \cite{PC03} and superstrong \cite{PC04} magnetic
fields.\footnote{\raggedright Some results of these
calculations are available at
http://www.ioffe.ru/astro/NSG/Hmagnet/} Mori and Heyl 
\cite{MoriHeyl} applied the same approach with slight
modifications to strongly magnetized helium plasmas. One of
the modifications was the use of the plasma microfield
distribution from \cite{PGC02} for calculation of 
$w(\Kp)$. Mori and Heyl considered atomic and molecular
helium states of different ionization degrees. Their treated
rotovibrational molecular levels by perturbation theory and
considered the dependence of binding energies on
orientation of the molecular axis
relative to $\bm{B}$. The
$\Kp$-dependence of the energy, $E(\Kp)$, was
described by an analytical fit, based on an extrapolation of
adiabatic calculations at small $\Kp$. The motion
effects of atomic and molecular ions were not considered.

\subsection{Applicability of the LTE approximation}
\label{sect:LTE}

The models of EOS and ionization balance
usually assume that the LTE conditions are satisfied for
the atoms and ions. In particular, the Boltzmann
distribution over the Landau levels is assumed.
This assumption does not apply for free
electrons in the neutron-star atmospheres, if the
spontaneous radiative decay rate of excited Landau levels,
\beq
   \Gamr = \frac43\,\frac{e^2\omc^2}{\mel c^3}
=3.877\times10^{15}B_{12}^2\mbox{ s}^{-1},
\label{Gammar}
\eeq
exceeds the rate of their collisional de-excitation.

In a nonquantizing field, the characteristic frequency of
electron-ion Coulomb collisions equals (see, e.g., \cite{Callen})
\beq
   \Gamc = \frac{4\sqrt{2\pi}\nion Z^2 e^4
\Lambda_\mathrm{c}}{
       3\sqrt{\mel} (\kB T)^{3/2}} = 
2.2\times10^{15}\,
  \frac{Z^2}{A}\,\frac{\rho'\Lambda_\mathrm{c}}{T_6^{3/2}}\mbox{ s}^{-1},
\eeq
where $\rho'\equiv\rho/$\gcc, and $\Lambda_\mathrm{c}$ is a
Coulomb logarithm, which weakly depends on $T$ and $\rho$
and usually has an order of magnitude of 1\,--\,10. In a
quantizing field, the electrons are de-excited
from the first Landau level by electron-ion Coulomb collisions
at the rate
\bea\hspace*{-1ex}
   \Gamma_{10}
&\!\!\!\!=&\!\!\!\!
  \frac{4\sqrt{2\pi}\,\nion Z^2 e^4\tilde{\Lambda}_{10}}{
       \sqrt{\mel} \,\,(\hbar\omc)^{3/2}} =
4.2\times10^{12}\,\frac{Z^2}{A}
\,\frac{\rho'\tilde{\Lambda}_{10}}{B_{12}^{3/2}}\mbox{~s}^{-1}
\nonumber\\&&
 =
4.9\times10^{13}\,\frac{Z^2}{A}\frac{\rho'\Lambda_{10}}{B_{12}\sqrt{T_6}}
\mbox{~s}^{-1}
,
\label{Gam10}
\eea
where $\tilde{\Lambda}_{10}=\sqrt{\zete}\Lambda_{10}$ is a
new Coulomb logarithm, which has an order of unity at
$\zete\gg1$, whereas $\Lambda_{10}$ has that order at
$\zete\ll1$ \cite{PL07}. Note that the rate of the inverse
process of collisional
excitation equals
$\Gamma_{01}=\Gamma_{10}\,\mathrm{e}^{-\zete}$.
Comparing (\ref{Gammar}) and
(\ref{Gam10}), we see that in the weak-field
($B\lesssim10^{10}$~G) photospheres of isolated neutron stars,
at typical $\rho\gtrsim10^{-3}$ \gcc{} and $T_6\sim1$, the
LTE conditions are fulfilled (it may not be the case in
the magnetosphere due to the lower densities). In a strong field 
($B\gtrsim10^{11}$~G), the LTE is violated, and the
fraction of electrons on the excited Landau levels is
lower than the Boltzmann value $\mathrm{e}^{-\zete}$.
However, this does not entail any consequence for the
atmosphere models, because in the latter case
$\mathrm{e}^{-\zete}$ is vanishingly small.

For the ions, the
spontaneous decay rate of the excited Landau levels
$\Gamri$ differs from $\Gamr$
by a factor of $Z\,(Z \mel/\mion)^3 \sim 10^{-10}$. The
statistical distribution of ions over the Landau levels has
been studied in \cite{PL07}. The authors showed
that the fraction of the ions on the first excited Landau
level is accurately given by
\beq
   \frac{n_1}{n_0} = \mathrm{e}^{-\zeti} \,
   \frac{ 1 +
\epsilon\,(\Gamri/\Gamma_\mathrm{10,i})/(1-\mathrm{e}^{-\zeti}) }{
      1+ \Gamri/\Gamma_\mathrm{10,i} +
\epsilon\,(\Gamri/\Gamma_\mathrm{10,i})/(\mathrm{e}^{\zeti}-1) } ,
\label{balance10}
\eeq
where $\epsilon=J_{\omega}/\mathcal{B}_{\omega,T}$ at
$\omega=\omci$, and
$\Gamma_\mathrm{10,i}$ is the collisional frequency of the
first level, which differs from \req{Gam10} by the factor
$\sqrt{\mion/\mel}$ and the value of the Coulomb logarithm.
The parameter $\epsilon$ is small in the outer layers of the
photospheres, therefore the distribution over the levels is
determined by the ratio $\Gamri/\Gamma_\mathrm{10,i}$. If
$\Gamri/\Gamma_\mathrm{10,i}\ll1$, then the Boltzmann
distribution is recovered, that is, the LTE approximation
holds; otherwise the excited levels are underpopulated.
According to \cite{PL07},
\beq
   \frac{\Gamma_\mathrm{10,i}}{\Gamri}\sim
     \frac{ \rho'}{(B_{12}/300)^{7/2}}\,.
\eeq
In the atmospheres and at the radiating surfaces of the
ordinary neutron stars this ratio is large, because the
denominator is small, and for magnetars with
$B\lesssim10^{15}$~G the ratio is large because  $\rho'$ is
large (see~(\ref{rhoB12})). Moreover, as shown in
\cite{PL07}, even in the outer atmospheres of magnetars,
where  $\Gamri/\Gamma_\mathrm{10,i}\ll1$, deviations from
the LTE should not affect the spectral modeling. The reason
is that absorption coefficients are mainly contributed from
the second-order quantum transitions  that do not change the
Landau number $N$. Therefore the depletion of the upper
states is unimportant, so that the Kirchhoff law, which
holds at the LTE, remains approximately valid also in this
case.

\subsection{Condensed surface}
\label{sect:cond}

Ruderman \cite{Ruderman71} suggested that a strong magnetic
field can stabilize polymer chains directed along the field
lines, and that the dipole-dipole attraction of these chains
may result in a condensed phase. Later works have shown that
such chains indeed appear in the fields
$B\sim10^{12}$\,--\,$10^{13}$~G, but only for the chemical
elements lighter than oxygen, and they polymerize into a
condensed phase either in superstrong fields, or at
relatively low temperatures, the sublimation energy being
much smaller than Ruderman assumed (see \cite{MedinLai06b},
and references therein).

From the thermodynamics point of view, the magnetic
condensation is nothing but the plasma phase transition
caused by the strong electrostatic attraction between the
ionized plasma particles. This attraction gives a negative
contribution to pressure $P_\mathrm{ex}$, which is not counterbalanced at
low temperatures (at $\Gami\gtrsim1$) until the electrons
become degenerate with increasing density. In the absence of
a magnetic field, such phase transitions were studied
theoretically since 1930s (see \cite{PPT}, for a review). In
this case, the temperature of the outer layers of a neutron
star $T\gtrsim(10^5-10^6)$~K exceeds the critical
temperature $\Tc$ for the plasma phase transition. However,
we have seen in \S\,\ref{sect:EOS} that a quantizing
magnetic field lifts electron degeneracy. As a result, $\Tc$
increases with increasing $B$, which may enable such
phase transition.

Lai \cite{Lai01} estimated the condensed-surface density as
\beq
   \rho_\mathrm{s}\approx 561\,\eta\,A Z^{-3/5}
          B_{12}^{6/5}\mbox{~\gcc},
\label{rhos}
\eeq
where $\eta$ is an unknown factor of the order of unity. In
the ion-sphere model \cite{Salpeter61}, the electrons are
replaced by a uniform negative background, and the potential
energy per ion is estimated as the electrostatic
energy of the ionic interaction with the negative background
contained in the sphere of radius
$\aion=(4\pi\nion/3)^{-1/3}$. By equating
$|P_\mathrm{ex}|$ to the pressure
of degenerate electrons $P_\mathrm{e}$, one obtains \req{rhos} with
$\eta=1$. This estimate disregards the ion correlation
effects, the electron-gas polarizability, and bound
state formation. Taking account of the electron polarization
by different versions of the Thomas-Fermi method, one gets
quite different results: for example, the zero-temperature
Thomas-Fermi data for a magnetized iron at
$10^{10}\mbox{~G}\leqslant B\leqslant 10^{13}$~G
\cite{Rognvaldsson-ea} can be described by
\req{rhos} with $\eta \approx 0.2 +
0.01/B_{12}^{0.56}$, and in a finite-temperature
Thomas-Fermi model \cite{Thorolfsson-ea} there is no
phase transition at all.

At $1\lesssim B_{12}\lesssim10^3$, the EOS of
partially ionized, strongly magnetized hydrogen \cite{PCS99}
that was described in \S\,\ref{sect:Saha} predicts a phase
transition with the critical temperature
$T_\mathrm{crit}\approx3\times10^5\,B_{12}^{0.39}$~K and
critical density $\rho_\mathrm{crit} \approx
143\,B_{12}^{1.18}\mbox{~\gcc}$, which corresponds to
$\eta\approx1/4$. With decreasing temperature below
$\Tc$, the condensed-phase density increases and tends
asymptotically to \req{rhos} with $\eta\approx1/2$, while
the density of the gaseous phase quickly decreases, and the
atmosphere becomes optically thin. Lai and Salpeter
\cite{LS97} obtained qualitatively similar results from
calculations of density of saturated vapor above the
condensed surface, but with 3--4 times lower $\Tc$.
The quantitative differences may be caused by the
less accurate approximate treatment of the molecular contribution
in \cite{PCS99}, on one hand, and by the less accurate
account of the effects of atomic motion across the magnetic
field in \cite{LS97}, on the other hand.

Medin and Lai \cite{MedinLai06b} treated the condensation
energy by the density functional method. In 
\cite{MedinLai07} they calculated the equilibrium density of
a saturated vapor of the atoms and polymer chains of helium,
carbon, and iron above the respective condensed surfaces at
$1\lesssim B_{12} \leqslant 10^3$. By equating this density
to $\rhos$, they found $\Tc$ at several $B$ values. Unlike
previous authors, Medin and Lai
\cite{MedinLai06b,MedinLai07} have taken a self-consistent
account of the electron band structure in the condensed
phase. Meanwhile, they did not take account of the effects
of atomic and molecular motion across the magnetic field in
the gaseous phase and treated the
excited-states contribution rather roughly. They calculated the
condensed-surface density assuming that the linear atomic
chains, being unchanged as such, form a rectangular lattice
in the plane, perpendicular to  ${\bm B}$. As shown in
\cite{PC13}, such evaluated values of $\rho_\mathrm{s}$ can
be described by \req{rhos} with
$\eta=0.517+0.24/B_{12}^{1/5}\pm0.011$ for carbon and
$\eta=0.55\pm0.11$ for iron, and the critical temperature
can be evaluated as  $\Tc \sim
5\times10^4\,Z^{1/4}\,B_{12}^{3/4}$~K. For comparison, in
the fully-ionized plasma model
$\Tc\approx2.5\times10^5\,Z^{0.85}\,B_{12}^{0.4}$~K and
$\eta = [1+1.1\,(T/\Tc)^5]^{-1}$. Hopefully, the present
uncertainty in $\rhos$ and $\Tc$ estimates may be diminished
with an analysis of future neutron-star observations.

When magnetic field increases from $10^{12}$~G to
$10^{15}$~G, the cohesive energy, calculated in
\cite{MedinLai07} for the condensed surface, varies
monotonically from 0.07 keV to 5 keV for helium, from 0.05
keV to 20 keV for carbon, and from 0.6 keV to 70 keV for
iron. The power-law interpolation gives order-of-magnitude
estimates between these limits. The electron work function
changes in the same $B$ range from 100 eV to
$(600\pm50)$~eV. With the calculated energy values, the
authors \cite{MedinLai07} determined the conditions of
electron and ion emission in the vacuum gap above the polar
cap of a pulsar and the conditions of gap formation, and
calculated the pulsar death lines on the
$\mathcal{P}$\,--\,$\dot{\mathcal{P}}$ plane.

\section{Magnetic atmospheres}

\subsection{Radiative transfer in normal modes}
\label{sect:RTEmag}

Propagation of electromagnetic waves in magnetized plasmas
was studied in many works, the book by Ginzburg 
\cite{Ginzburg} being the most complete of them. At
radiation frequency $\omega$ much larger than the electron
plasma frequency
$\ompe=\left({4\pi e^2 n_\mathrm{e} / \mel^\ast } \right)^{1/2}$, 
where $\mel^\ast \equiv \mel \sqrt{1+\pF^2/(\mel c)^2}$ is the
effective dynamic mass of an electron at the Fermi surface,
the waves propagate in the form of two polarization modes,
extraordinary (hereafter denoted by subscript or superscript
$j=1$ or X) and ordinary ($j=2$ or O). They have different
polarization vectors $\bm{e}_j$ and different absorption and
scattering coefficients, which depend on the angle
$\theta_B$ (Fig.~\ref{fig:bending}). The modes interact
with one another through scattering. Ventura
\cite{Ventura79} performed an analysis of the
polarization modes in application to the neutron stars from
the physics point of view.
Gnedin and Pavlov \cite{GP73} formulated the radiative
transfer problem in terms of these modes. They showed that
in the strongly magnetized neutron-star atmospheres, as a
rule, except narrow frequency ranges near resonances,
a strong Faraday depolarization occurs. In this case, it is
sufficient to consider specific intensities of the two
normal modes instead of the four components of the Stokes
vector. The radiative transfer equation for these specific
intensities is a direct generalization of \req{RTEgen} \cite{KaminkerPS82}:
\bea&&\hspace*{-2em}
 \cos\theta_k \frac{\dd I_{\omega,j}(\khat)}{\dd \ycol} =
    \opac_{\omega,j}(\khat) I_{\omega,j}(\khat) 
-
 \frac12\,\opac_{\omega,j}^\mathrm{a}(\khat)
     \mathcal{B}_{\omega,T}
\nonumber\\&&-
      \sum_{j'=1}^2 \int_{(4\pi)} 
        \opac_{\omega,j'j}^\mathrm{s}(\khat',\khat)
         I_{\omega,j'}(\khat') \,\dd\khat',
\label{RTEmag}
\eea
where $\opac_{\omega,j}(\khat) \equiv
\opac_{\omega,j}^\mathrm{a}(\khat) + \sum_{j'=1}^2
\int_{(4\pi)}
\opac_{\omega,j'j}^\mathrm{s}(\khat',\khat)\,\dd\khat'$. The
dependence of the opacities $\opac$ on ray directions
$(\khat,\khat')$ is affected by the magnetic-field
direction. Therefore, the emission of a magnetized
atmosphere, unlike the nonmagnetic one, depends not only on
the angle  $\theta_k$ that determines the ray inclination to
the stellar surface, but also on the angles
$\theta_\mathrm{n}$ and $\varphi_k$ in
Fig.~\ref{fig:bending}. For hydrostatic and
energy balance, we can keep Eqs.~(\ref{barometric}),
(\ref{grad}), and (\ref{enbal}), if we put
$I_\omega=\sum_{j=1}^2 I_{\omega,j}$ by definition.

The diffusion equation for the normal modes in these
approximations was derived in
\cite{Nagel80,KaminkerPS82}. For the plane-parallel
photosphere it reads \cite{Zavlin09}
\bea&&\hspace*{-2em}
   \frac{\dd}{\dd\ycol}
D_{\omega,j}
         \frac{\dd}{\dd\ycol} J_{\omega,j} =
        \bar{\opac}_{\omega,j}^\mathrm{a}\,
       \left[ J_{\omega,j} -
           \frac{\mathcal{B}_{\omega,T}}{2} \right]
\nonumber\\&&+
 \bar{\opac}_{\omega,12}^\mathrm{s}
        \left[J_{\omega,j} - J_{\omega,3-j} \right].
\hspace*{2em}
\label{diffmag}
\eea
Here,
\bea
   J_{\omega,j} &=& \frac{1}{4\pi}
    \int_{(4\pi)}
    I_{\omega,j}(\khat)\,\dd\khat,
\nonumber
\\
   \bar{\opac}_{\omega,j}^\mathrm{a} &=& \frac{1}{4\pi}\int_{(4\pi)}
\opac_{\omega,12}^\mathrm{a}\,\dd\khat,
\nonumber\\
\bar{\opac}_{\omega,j}^\mathrm{s} &=& \frac{1}{4\pi}
\int_{(4\pi)}\dd\khat'
\int_{(4\pi)} \dd\khat\,\,
\opac_{\omega,12}^\mathrm{s}(\khat',\khat) \, ,
\nonumber
\eea
and the effective diffusion coefficient equals
\beq
   D_{\omega,j} = \frac{1}{3\opac_{\omega,j}^{\mathrm{eff}}}=
\frac{\cos^2\theta_\mathrm{n}}{3\opac_{\omega,j}^\|} +
\frac{\sin^2\theta_\mathrm{n}}{3\opac_{\omega,j}^\perp},
\eeq
where $\theta_\mathrm{n}$ is the angle between $\bm{B}$ and
intensity gradient,
\beq
\left\{
   \begin{array}{c}
 (\opac_j^\|)^{-1}
\\
   (\opac_j^{\perp})^{-1\rule{0pt}{2ex}}
   \end{array}
  \right\}
 = \frac34 \int_0^\pi
\left\{
   \begin{array}{c}
2\cos^2\theta_B \\
\sin^2\theta_B
\end{array}
  \right\}
\frac{\sin\theta_B\,\mathrm{d}\theta_B}{\opac_j(\theta_B)}\,.
\label{kappa-eff}
\eeq
The effective opacity for nonpolarized
radiation is
$
\opac^{\mathrm{eff}}={2}/(3D_{\omega,1}+3D_{\omega,2}).
$
The diffusion approximation (\ref{diffmag}) serves as a
starting point in an iterative method \cite{SZ95}, which
allows one to solve the system (\ref{RTEmag})
more accurately.

\subsection{Plasma polarizability}

In the Cartesian coordinate system with the $z$-axis along
$\bm{B}$, the plasma dielectric tensor is
\cite{Ginzburg}
\beq
  \bm{\varepsilon} = \mathbf{I} + 4\pi\chi
          = 
 \left( \begin{array}{ccc}
 \varepsilon_\perp & i \varepsilon_\wedge & 0 \\
 -i\varepsilon_\wedge & \varepsilon_\perp & 0 \\
 0 & 0 & \varepsilon_\| 
 \end{array} \right),
\label{eps-p}
\eeq
where $\mathbf{I}$ is the unit tensor, 
$\chi=\ChiH+\mathrm{i}\ChiA$ is the complex polarizability
tensor of plasma, $\ChiH$ and $\ChiA$ are its Hermitian and
anti-Hermitian parts, respectively. Under the assumption
that the electrons and ions lose their regular velocity,
acquired in an electromagnetic wave, by collisions
with an effective frequency $\nu_\mathrm{eff}$
independent of the velocities, then the cyclic
components of the polarizability tensor are  (\cite{Ginzburg},
\S\,10)
\beq
   \chi_\alpha = -\frac{1}{4\pi}\,\frac{\ompe^2}{
    (\omega + \alpha \omc)\,(\omega - \alpha \omci)
          +\mathrm{i}\omega\nu_\mathrm{eff}}\,
\label{chi-elem}
\eeq
$(\alpha=0,\pm1)$. A more rigorous kinetic theory leads to
results which cannot be described by \req{chi-elem} with the
same frequency 
$\nu_\mathrm{eff}$ for the Hermitian and anti-Hermitian
components $\chiH_\alpha$ and $\chiA_\alpha$
(\cite{Ginzburg}, \S\,6).

\begin{figure}[t]
\begin{center}
\includegraphics[width=.9\columnwidth]{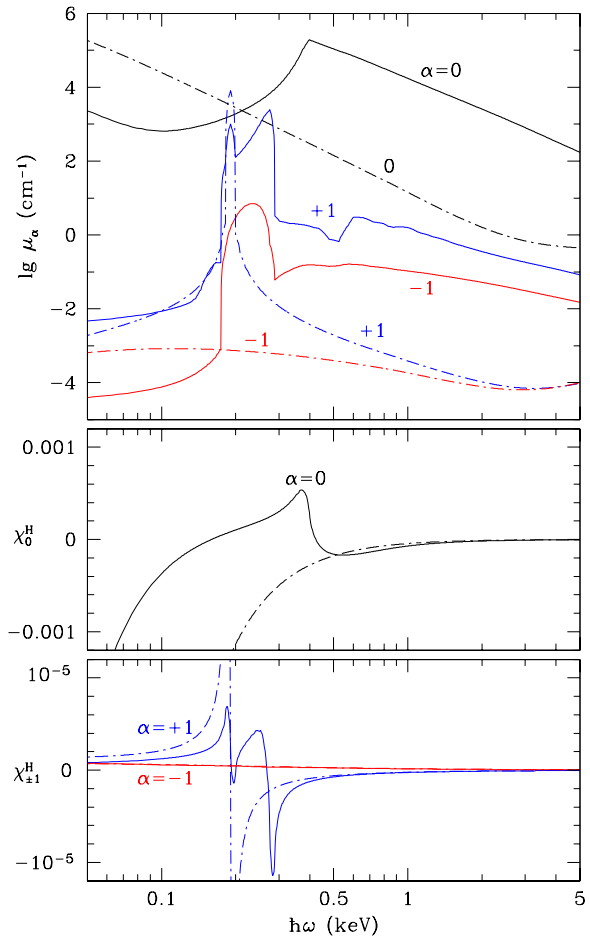}
\end{center}
\caption{
Absorption coefficients (top panel) and 
polarizability coefficients
$\chiH_{0}$ (middle panel) and $\chiH_{\pm1}$ (bottom panel)
in the partially ionized (solid curves) and fully ionized
(dot-dashed curves) plasma models at
 $B=3\times10^{13}$~G, $\rho=1$ \gcc{} and $T=3.16\times10^5$~K.
\label{fig:kk13a4}}
\end{figure}

The anti-Hermitian part of the
polarizability tensor determines the opacities:
$\opac_\alpha(\omega)= 4\pi\omega\chiA_\alpha(\omega)/(\rho
c)$. Then the Kramers-Kronig relation gives \cite{BulikPavlov,KK}
\bea
  \chiH_\alpha(\omega) &=&
   \frac{c\rho}{4\pi^2\omega}\, \bigg\{\!
    \int_0^\omega \!\big[\,\opac_\alpha (\omega+\omega')
     - \opac_\alpha (\omega-\omega')\,\big]
     \frac{\dd\omega'}{\omega'}
\nonumber\\&
   + & \int_{2\omega}^\infty \frac{\opac_\alpha(\omega')}{\omega'-\omega}
     \,\dd\omega'
   - \int_0^\infty \frac{\opac_{-\alpha}(\omega')}{\omega'+\omega}
     \,\dd\omega' \bigg\} .
\label{KK-mu}
\eea
Thus we can calculate the polarizability tensor
$\bm{\chi}$ from the opacities $\opac_\alpha(\omega)$. It
has been done in \cite{BulikPavlov} for a gas of neutral
hydrogen atoms and in \cite{KK} for partially ionized
hydrogen plasmas. 

Figure~\ref{fig:kk13a4} shows the cyclic components of absorption coefficients,
$\mu_\alpha=\rho\opac_\alpha$ in the top panel, and
corresponding polarizability components $\chiH_\alpha$ in
the middle and bottom panels, for a partially ionized
hydrogen plasma at $B=3\times10^{13}$~G, $\rho=1$ \gcc, and
$T=3.16\times10^5$~K. In this case, the neutral fraction is
89\%. For comparison we show the results of an analogous
calculation for the fully-ionized plasma model. In addition
to the proton cyclotron resonance at $\hbar\omega=0.19$~keV that
is present in both models, the absorption coefficients show
rather pronounced features due to atomic transitions in the
partially ionized plasma model. Most remarkable are the
absorption features due to bound-bound transitions at
$\hbar\omega\approx0.2$--0.3 keV for $\mu_{+1}$ and the
photoionization jump (partly smeared by the magnetic
broadening) at $\hbar\omega=0.4$ keV for $\mu_0$. These
features have clear imprints on the behavior of $\chiH_{+1}$
and $\chiH_0$.

\subsection{Vacuum polarization}
\label{sect:vacpol}

In certain ranges of density $\rho$ and frequency $\omega$,
normal-mode properties are dramatically affected by a
specific QED effect called vacuum polarization (its other
manifestation has already been considered in
\S\,\ref{sect:QED}). The influence of the vacuum
polarization on the neutron-star emission has been first
evaluated in \cite{Novick-ea77,GPS78} and studied in detail
in the review \cite{PavlovGnedin}. If the vacuum
polarization is weak, then it can be linearly added to the
plasma polarization. Then the complex dielectric tensor can
be written as
$
 \bm{\varepsilon}' = \mathbf{I} + 4\pi\chi + 4\pi\chi^\mathrm{vac},
$
where 
\beq
 \bm{\chi}^\mathrm{vac} = (4\pi )^{-1}\,
 \mathrm{diag}(\bar{a}, \bar{a}, \bar{a}+\bar{q} )
\label{chivac}
\eeq
is the vacuum polarizability tensor, and diag(\ldots)
denotes the diagonal matrix. Magnetic susceptibility of
vacuum is determined by expression 
\beq
 \bm{\mu}^{-1} = \mathbf{I} + \mathrm{diag}(\bar{a}, \bar{a},
 \bar{a}+\bar{m}).
\label{polvac}
\eeq
Adler \cite{Adler} obtained the vacuum polarizability 
coefficients  $\bar{a}$, $\bar{q}$, and $\bar{m}$ that enter
Eqs.~(\ref{chivac}) and (\ref{polvac}) in an explicit form
at $b\ll1$, Heyl and Hernquist \cite{HH97} expressed them in
terms of special functions in the limits of $b\ll 1$ and
$b\gg1$. Kohri and Yamada \cite{KohriYamada} presented their
numerical calculations. Finally, in \cite{KK} we found
simple but accurate expressions
\bea
&&
 \bar{a}= - \frac{2\alphaf}{9\pi} \ln\bigg(
 1 + \frac{b^2}{5}\,\frac{
  1+0.25487\,b^{3/4}
  }{
  1+0.75\,b^{5/4}}\bigg),
\label{fit-a}
\\&&
 \bar{q} = \frac{7\alphaf}{45\pi}\,b^2\,\frac{
 1 + 1.2\,b
 }{
 1 + 1.33\,b + 0.56\,b^2
 },
\label{fit-q}
\\&&
 \bar{m} = - \frac{\alphaf}{3\pi} \, \frac{ b^2 }{
 3.75 + 2.7\,b^{5/4} + b^2 }.
\label{fit-m}
\eea
The coefficients (\ref{fit-a})\,--\,(\ref{fit-m}) are not small
at $B\gtrsim 10^{16}$~G, therefore the vacuum refraction
coefficients substantially differ from unity. In such strong
fields, the vacuum that surrounds a neutron star acts as a
lens, distorting its radiation
\cite{ShavivHL99,HeylShaviv02,vanAdelsbergPerna09}. At
smaller $B$, the vacuum polarization results in
a resonance, which manifests in the coincidence of the
normal-mode polarization vectors at a certain frequency,
depending on plasma density. In the photospheres with $B
\gtrsim 10^{13}$~G, this resonance falls in the range
$\sim0.1$\,--\,1~keV and affects the thermal spectrum.

\subsection{Polarization vectors of the normal modes}

Shafranov \cite{Shafranov} obtained the polarization vectors
$\bm{e}_j$ for fully ionized plasmas. Ho and Lai
\cite{HoLai03} presented their convenient expressions in
terms of the coefficients $\varepsilon_\perp$,
$\varepsilon_\|$, $\varepsilon_\wedge$, $\bar{a}$,
$\bar{q}$, and $\bar{m}$, including the contributions of
electrons, ions, and vacuum polarization.
In the Cartesian coordinate system ($xyz$) with the $z$-axis
along the wave vector $\bm{k}$ and with $\bm{B}$ in the plane 
$x$--$z$, one has
\beq
\bm{e}_j=
\left(\begin{array}{c}
  e^j_x \\ e^j_y \\ e^j_z
  \end{array}\right)=\frac{1}{\sqrt{1+K_j^2+K_{z,j}^2}} \,
\left(\begin{array}{c}
 \mathrm{i} K_j \\ 1 \\ \mathrm{i} K_{z,j}
  \end{array}\right) ,
\label{eq:e}
\eeq
where
\bea
&&\hspace*{-2em}
  K_j = \beta \left\{
   1 + (-1)^j \left[ 1 + \frac{1}{\beta^2} 
   + \frac{\bar{m}}{1+\bar{a}} \frac{\sin^2\theta_B}{\beta^2}\right]^{1/2}
   \right\},
\label{eq:K}
\hspace*{2em}\\&&
  K_{z,j} = - \frac{ 
   (\varepsilon_\perp' - \varepsilon_\|') K_j \cos\theta_B + \varepsilon_\wedge
   }{
   \varepsilon_\perp' \sin^2\theta_B + \varepsilon_\|' \cos^2\theta_B } 
   \, \sin\theta_B,
\label{eq:Kz}
\hspace*{2em}\\&&\hspace*{-2em}
 \beta = \frac{\varepsilon_\|' - \varepsilon_\perp' + \varepsilon_\wedge^2/\varepsilon_\perp' + \varepsilon_\|'
 \,\bar{m}/(1+\bar{a})
   }{
   2 \, \varepsilon_\wedge }
   \,\, \frac{ \varepsilon_\perp'}{\varepsilon_\|'}
   \,\,\frac{\sin^2\theta_B}{\cos\theta_B},
\hspace*{2em}
\eea
$\varepsilon_\perp' = \varepsilon_\perp + \bar{a}$, 
and
$\varepsilon_\|' = \varepsilon_\| + \bar{a} +
\bar{q}$.
If the plasma and vacuum
polarizabilities are small
($|\chiH_\alpha| \ll (4\pi)^{-1}$ and
$|\bar{a}|,\bar{q},|\bar{m}| \ll 1$), as usual,
\beq
  \beta \approx \frac{2\chiH_0 - \chiH_{+1} - \chiH_{-1} +
  (\bar{q}+\bar{m})/(2\pi)
  }{
  2\,(\chiH_{+1} - \chiH_{-1}) }\,\frac{\sin^2\theta_B}{\cos\theta_B} .
\label{beta-approx}
\eeq

\subsection{Opacities}
\label{sect:opac}

In the approximation of isotropic scattering, at a given
frequency $\omega$, the opacities can be presented in the form
\bea
&&\hspace*{-.7em}
   \opac_j^\mathrm{a} = \sum_{\alpha=-1}^1
     |e_{j,\alpha}(\theta_B)|^2 \,
        \frac{\sigma_\alpha^\mathrm{a}}{\mion},
\\&&\hspace*{-4em}
 \opac_{jj'}^\mathrm{s} \!=\!\!
     {\frac34}
\!\!
  \sum_{\alpha=-1}^1 \!\!
     |e_{j,\alpha}(\theta_B)|^2 \,
     \frac{\sigma_\alpha^\mathrm{s}}{\mion}\int_0^\pi \!\!\!
       |e_{j',\alpha}(\theta_B')|^2\sin\theta_B'\,\mathrm{d}\theta_B',
\eea
where $\sigma_\alpha$ are the cross sections for the three
basic polarizations according to \req{sumalpha}. The partial
cross sections  $\sigma_\alpha^\mathrm{a,s}$ include
contributions of photon interaction with free
electrons or ions (free-free transitions) as well as  with
bound states of atoms and ions (bound-bound and bound-free
transitions). The latter implies, in
particular, averaging of the cross sections of photon and
atom absorption over all values of $\Kp$. Since the
distribution over $\Kp$ is continuous for the atoms and 
discrete for the ions, such averaging for atoms
reduces to an integration over $\Kp$, analogous to
\req{Z-int}, whereas for ions it implies summation with
an appropriate statistical weight. To date, such calculation
has been realized for atoms of hydrogen \cite{PC03,PC04} and
helium \cite{MoriHeyl}. 

\begin{figure}[t]
\begin{center}
\includegraphics[width=.85\linewidth]{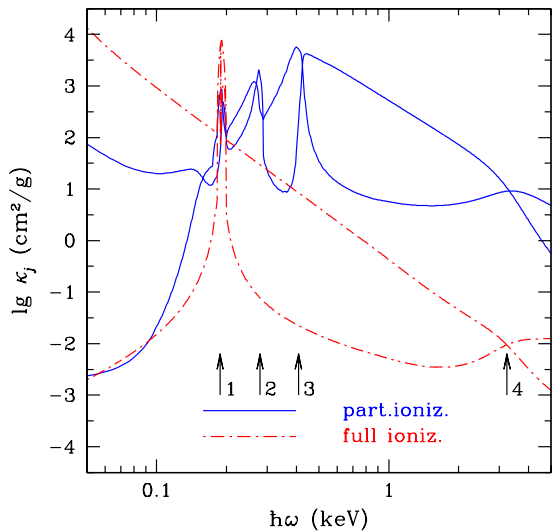}
\caption{
Logarithm of spectral opacities ($\log\opac_j$) for two normal modes,
propagating at the angle
$\theta_B=10^\circ$ to the magnetic field lines in a
hydrogen plasma at
$B=3\times10^{13}$~G, $T=3.16\times10^5$~K, $\rho=1$
\gcc. Solid curves: partially ionized plasma model; 
dot-dashed curves: fully-ionized plasma model. The lower
(upper)
curve of each type corresponds to the extraordinary
(ordinary) wave. The arrows indicate the features at 
resonant frequencies: 1-- the ion cyclotron resonance
$\omega=\omci$; 2 -- energy threshold for a transition
between the lowest two levels
$\hbar\omega=|E_{0,0}^{(0)}-E_{1,0}^{(0)}|$;  3 -- the
ground-state binding energy $\hbar\omega=|E_{0,0}^{(0)}|$; 4
-- the vacuum resonance.
\label{fig:ang13a4}}
\end{center}
\end{figure}

Figure~\ref{fig:ang13a4} presents opacities for the two
normal modes propagating at the angle $\theta_B=10^\circ$ to
the magnetic field under the same physical conditions as in
Fig.~\ref{fig:kk13a4}. One can clearly distinguish the
features reflecting the peaks at the ion cyclotron frequency
and the resonant atomic frequencies, and the line crossings
related to the behavior of the plasma polarizability as
function of frequency. For comparison, we show also
opacities for the fully ionized plasma model under the same
conditions. They miss the features related to the atomic
resonances, and their values is underestimated by orders of
magnitude in a wide frequency range.

\subsection{Spectra of magnetic photospheres}
\label{sect:atmodels}

\begin{figure}[t]
\begin{center}
\includegraphics[width=.83\linewidth]{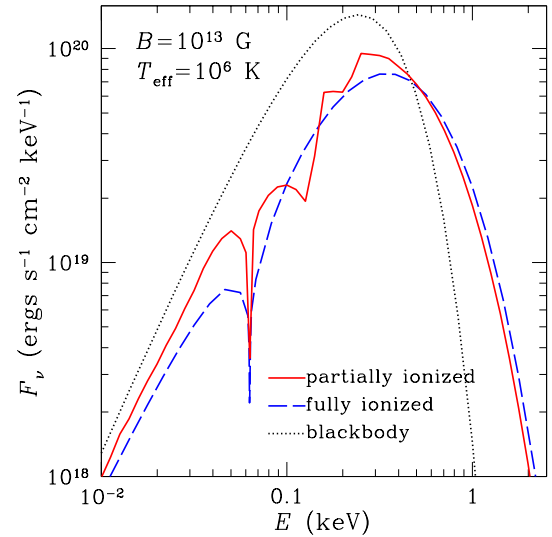} 
\caption{
Local spectrum of hydrogen photosphere with $B=10^{13}$~G
(the field is normal to the surface) and $\Teff=10^6$~K. The
solid line presents a self-consistent model of a partially
ionized photosphere, the dashed line presents the
fully-ionized atmosphere model, and the dots show the
blackbody spectrum. (The figure is provided by W.~C.~G.~Ho.)
\label{fig:f131}}
\end{center}
\end{figure}

Shibanov and coworkers \cite{Shibanov-ea92} were the first
to perform detailed calculations of the spectra of radiation
formed in the strongly magnetized neutron-star photospheres,
using the fully ionized plasma model, and created a database
of magnetic hydrogen spectra
\cite{Pavlov-ea95}.\footnote{Model \textsc{NSA} in the
\textit{XSPEC} database \cite{XSPEC}.} They have shown that
the spectra of magnetic hydrogen and helium atmospheres are
softer than the respective nonmagnetic spectra, but harder
than the blackbody spectrum with the same temperature. In addition to the spectral
energy distribution, these authors have also studied the
polar diagram and polarization of the outgoing emission,
which proved to be quite nontrivial because of
redistribution of energy between the normal modes.  The
thermal radiation of a magnetized photosphere is strongly
polarized, and the polarization sharply changes at the
cyclotron resonance with increasing frequency. At contrast
to the isotropic blackbody radiation, radiation of a
magnetic photosphere consists of a narrow ($<5^\circ$) pencil beam
along the magnetic field and a broad fan beam
with typical angles  $\sim20^\circ-60^\circ$
\cite{Zavlin-ea95} (see also \cite{vanAdelsbergLai}). These
calculations have thus fully confirmed the early analysis by
Gnedin and Sunyaev \cite{GnedinSunyaev74}.

Later, analogous calculations were performed by other
research groups \cite{Zane-ea01,HoLai03,vanAdelsbergLai}.
They paid special attention to manifestations of the ion
cyclotron resonance in observed spectra in the presence of
superstrong magnetic fields, which was prompted by tentative
magnetar discoveries. It was shown in
\cite{LaiHo02} that the vacuum polarization
leads in the superstrong fields to a conversion of the
normal modes, when a photon related to one mode transforms,
with certain probability, into a photon of the other mode
while crossing a surface with a certain critical density.
The latter density is related to the photon energy as
\beq
  \rho = 0.00964\,(A/Z)\,
     (\hbar\omega/\mbox{keV})^2\, B_{12}^2/f_B^2 ~\gcc,
\label{conversion}
\eeq
where
$f_B^2= \alphaf b^2 / [15\pi(\bar{q}+\bar{m})]$, while $\bar{q}$
and $\bar{m}$ are given by Eqs.~(\ref{fit-q}), (\ref{fit-m});
$f_B$ weakly depends on $B$, and $f_B\approx1$ at $B\lesssim10^{14}$~G.
The energy $\hbar\omega$ in \req{conversion} corresponds to
the line crossing in Fig.~\ref{fig:ang13a4}, indicated by
arrow 4. It follows from \req{conversion} that in the field
of $B\sim10^{14}$~G this energy coincides with the ion
cyclotron energy at the density where the atmosphere is
optically thin for the extraordinary mode, but optically
thick for the ordinary mode. Under such conditions, the
mode conversion strongly suppresses the ion cyclotron
feature in the emission spectrum.

In the first computations of partially ionized photospheres
of neutron stars with magnetic fields 
$B\sim10^{12}$\,--\,$10^{13}$~G that were presented in
\cite{Miller92} and \cite{RRM97}, the properties of the
atoms in magnetic fields were calculated by the adiabatic
Hartree-Fock method (\S\,\ref{sect:atoms}). The atomic motion
was either ignored \cite{Miller92}, or treated approximately
by the perturbation theory \cite{RRM97}.

In \cite{KK}, a hydrogen photosphere model has been
constructed beyond the framework of the adiabatic
approximation, taking the full account of the partial
ionization as well as the atomic motion effects in the
strong magnetic fields. Figure~\ref{fig:f131} gives an
example of radiation spectrum going out of such photosphere
with $B=10^{13}$~G. We see a narrow absorption line at the
proton cyclotron energy $E=0.063$~keV and the features at
higher energies, related to atomic transitions. For
comparison, a spectrum calculated in the fully-ionized
plasma model and the Planck spectrum are shown. The
comparison shows that the two photospheric models have
similar spectral shapes, but the model that allows for the
partial ionization has additional features. The spectral
maximum of both models is shifted to higher energies
relative to the Planck maximum. This demonstrates that an
attempt of interpretation of the hydrogen spectra with the
blackbody model would strongly overestimate the effective
temperature, while the fully-ionized photosphere model
yields a more realistic temperature, but does not reproduce
the spectral features caused by atomic transitions.

\begin{figure}[t]
\begin{center}
\includegraphics[width=.85\linewidth]{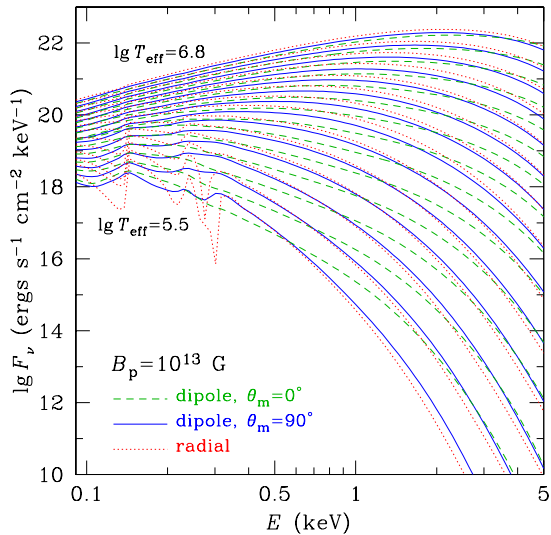}
\caption{
Integral spectra of a hydrogen atmosphere of a neutron star
with $M=1.4\,M_\odot$, $R=12$ km, and
with different effective temperatures $\Teff$ ($\log\Teff$ (K)
from 5.5
to 6.8 with step 0.1). The dashed and solid lines represent
the model with a dipole field of strength
$B_\mathrm{p}=10^{13}$~G at the pole and oriented along and
across the line of sight, respectively. For comparison, the
dotted curve shows the model with a constant field
$B=10^{13}$~G, normal to the surface.
\label{fig:spectr13}}
\end{center}
\end{figure}

Magnetic fields and temperatures of neutron stars vary from
one surface point to another. In order to reproduce the
radiation spectrum that comes to an observer, one can use
\req{Fintegral}. The problem is complicated, because the
surface distributions of the magnetic field and the
temperature are not known in advance. As a fiducial model
one conventionally employs the relativistic dipole model
(\ref{eq:dipole}), (\ref{eq:dipole2}), while the temperature
distribution, consistent with the magnetic-field
distribution, is found from calculations of heat transport
in neutron-star envelopes (e.g., \cite{PYCG03}). Results of
such calculations, performed in \cite{HoPC}, are shown in
Fig.~\ref{fig:spectr13}. We see that the spectral features
are strongly smeared by the averaging over the surface, and
the spectrum depends on the magnetic axis orientation 
$\theta_\mathrm{m}$. When the star rotates, the latter
dependence leads to pulsations of the measured spectrum.

\begin{figure}[t]
\begin{center}
\includegraphics[width=.85\linewidth]{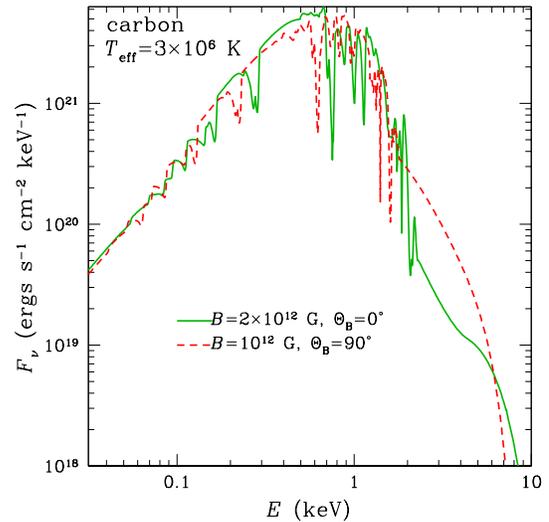}
\caption{
Local spectra at the magnetic pole (solid curve) and
equator (dashed curve) for a neutron star with carbon
atmosphere, the dipole field with polar strength of
$B_\mathrm{p}=2\times10^{12}$~G (neglecting the relativistic
corrections) and uniform effective temperature
$3\times10^6$~K. (Fig.~20 from \cite{MoriHo},
reproduced with permission of the authors and \copyright\,Oxford
University Press.)
\label{fig:MoriHo}}
\end{center}
\end{figure}

Mori et al.~\cite{MoriHailey06,MoriHo} calculated model
spectra of neutron-star photospheres composed of the atoms
and ions of elements with $\Znuc\lesssim10$. They calculated
the quantum-mechanical properties of the atoms and ions by
the method of Mori and Hailey \cite{MoriHailey02} and
treated the atomic motion effects by the perturbation theory
(\S\,\ref{sect:motion}). The equation of state and
ionization equilibrium were determined by the methods
described in \S\,\ref{sect:Saha}, the plasma polarizability
was calculated by \req{KK-mu}, and the opacities were
treated according to \S\,\ref{sect:opac}. As an example,
Fig.~\ref{fig:MoriHo} demonstrates local spectra of the
carbon photosphere with magnetic field $B=2\times10^{12}$~G,
normal to the surface, and the field $B=10^{12}$~G parallel
to the surface, which approximately (with account of neither
relativistic corrections nor temperature nonuniformity)
corresponds to the local spectra at the magnetic pole and
equator of a star with a dipole magnetic field. By analogy
to the case of hydrogen photosphere, the integration over the
surface between the pole and equator should smear the
spectral features between the two limiting curves shown in
the figure. 

The results described in this section have been used to
produce databases of spectra of partially ionized,
strongly magnetized neutron-star photospheres composed of
hydrogen \cite{HoPC} and heavier elements up to neon
\cite{MoriHo}.\footnote{Models \textsc{NSMAX} and
\textsc{NSMAXG} \cite{Ho14}
in the database \textit{XSPEC} \cite{XSPEC}.}

\section{Spectra of neutron stars with condensed surfaces}
\label{sect:NScond}

\subsection{Radiation of a naked neutron star}
\label{sect:surfem}

As we have seen in \S\,\ref{sect:cond}, the stars with a
very low effective temperature and a superstrong
magnetic field can have a liquid or solid condensed surface.
In this case, thermal emission can escape directly from the
metallic surface without transformation in a gaseous
atmosphere, and then the spectrum is determined by the
emission properties of this surface. Formation of thermal
spectra at a condensed surface of a strongly magnetized
neutron star depends on its reflection properties, which
were considered in 
\cite{Itoh75,LenzenTruemper,Brinkmann,TurollaZD04,surfem,PerezAMP05,reflefit}.
The first works \cite{Itoh75,LenzenTruemper} gave
order-of-magnitude estimates. A method of detailed
calculation of the reflectivity was proposed in
\cite{Brinkmann} and then was used with some modifications
in \cite{Brinkmann,TurollaZD04,surfem,PerezAMP05,reflefit}.
It is as follows. First, the normal-mode
polarization vectors $\bm{e}_{1,2}^\mathrm{(t)}$ in the
medium under the surface,
Eqs.~(\ref{eq:e})\,--\,(\ref{eq:Kz}), and the complex
refraction coefficients are expressed as functions of the
angles $\theta_k$ and $\varphi_k$ that determine the direction
of a reflected ray (Fig.~\ref{fig:bending}), using the
standard dispersion equation for the transmitted wave and
the Snell's law. Second, the complex electric amplitudes of the
incident, reflected, and transmitted waves are expanded over
the respective basic polarization vectors
$\bm{e}_{1,2}^\mathrm{(i,r,t)}$. Then the Maxwell boundary
conditions yield a system of equations, which determine the
coefficients of these expansions. These
reflected-wave expansion coefficients form the reflection
matrix $\{r_{jj'}\}$ and determine the surface reflectivity
for each incident-wave polarization,
$r_{\omega,j}=\sum_{j'}r_{\omega,jj'}$.
Then the total emissivity
$\emissivity=1-\frac12(r_{\omega,1}+r_{\omega,2})$.

The early works assumed that the ions are firmly fixed at
the crystalline lattice sites in the metal. In
\cite{surfem,PerezAMP05,reflefit} the authors have
considered not only this model, but also the opposite
limit of free ions. It is assumed \cite{surfem} that
the real reflectivity of the surface lies between the
limits given by these two models, although this problem has
not yet been definitely solved.

\begin{figure}[t]
\begin{center}
\includegraphics[width=\linewidth]{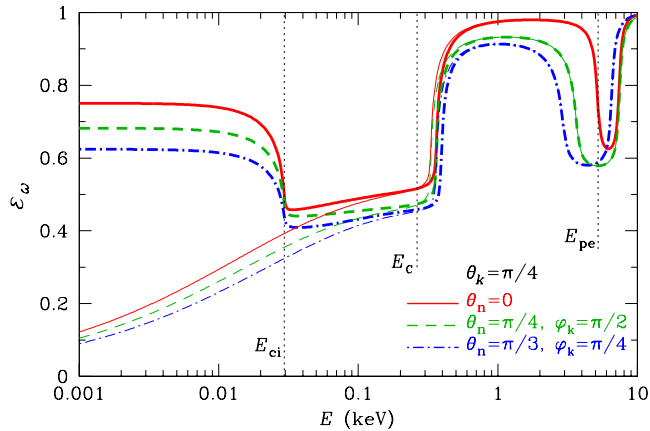}
\caption{
Emissivity of a condensed iron surface at $B=10^{13}$~G and
$T=10^6$~K, averaged over polarizations, is shown as a
function of energy of a photon emitted at the angle
$\theta_k=45^\circ$, for different magnetic-field
inclination angles $\theta_\mathrm{n}$ and azimuthal angles
$\varphi_k$. The thick and thin curves are obtained,
respectively, in the models of free and fixed ions. Vertical
dotted lines mark positions of the characteristic energies:
the ion cyclotron energy $E_\mathrm{ci}=\hbar\omci$, the
electron plasma energy $E_\mathrm{pe}=\hbar\ompe$, and the
hybrid energy $E_\mathrm{C}$.
\label{fig:rb13}}
\end{center}
\end{figure}

Figure \ref{fig:rb13} shows examples of the emissivity
$\emissivity$, normalized to the blackbody emissivity, as a
function of photon energy $E=\hbar\omega$, according to the
free- and fixed-ions models, for different values of the
angles $\theta_\mathrm{n}$, $\theta_k$, and
$\varphi_k$ that are defined in Fig.~\ref{fig:bending}.
The characteristic energies
$E_\mathrm{ci}=\hbar\omci$, $E_\mathrm{pe}=\hbar\ompe$, and
$E_\mathrm{C}=E_\mathrm{ci}+E_\mathrm{pe}^2/\hbar\omc$ are
marked. The spectral features near these energies are
explained in \cite{surfem}. For instance, the emissivity
suppression at $E_\mathrm{ci}\lesssim E\lesssim
E_\mathrm{C}$ is due to the strong damping of one of the two
normal modes in the plasma in this energy range. In the
fixed-ions mode, $\omci\to0$, therefore there is no kink of
the spectrum at $E\approx E_\mathrm{ci}$ in this model. The
results almost coincide in the two alternative
models at $E\gg E_\mathrm{ci}$, but strongly differ at
$E\lesssim E_\mathrm{ci}$, which may be important for
magnetar spectra. Near the electron plasma energy
$E_\mathrm{pe}=\hbar\ompe$, there is a resonant absorption, depending on
the directions of the incident wave and the magnetic field.

The local flux density of radiation from a condensed surface
is equal to the Planck function $\mathcal{B}_{\omega,T}$
(\ref{Bomega}), multiplied by the normalized emissivity
$\emissivity$. Since $\emissivity$ depends on the frequency $\omega$ and on
the angles $\theta_\mathrm{n}$, $\theta_k$, and $\varphi_k$
(Fig.~\ref{fig:bending}), thermal radiation depends on the
frequency and angles in a nontrivial way. In
Fig.~\ref{fig:rb13}, the emissivity is averaged over
polarizations. But $r_{\omega,1}\neq r_{\omega,2}$,
hence the thermal emission of a condensed surface is
polarized, the polarization depending in an
equally nontrivial way on the frequency and angles. For
example, the degree of linear polarization can reach tens
percent near the frequencies $\omci$ and $\ompe$, which
makes promising the polarization diagnostics of neutron
stars with condensed surfaces. Both the intensity and the
polarization degree can be evaluated using analytical
expressions, which have been constructed in \cite{reflefit}
for the reflectivity matrix of a condensed iron surface for
$B=10^{12}$\,--\,$10^{14}$~G.

\subsection{Thin and layered atmospheres}
\label{sect:thin}

Motch, Zavlin, and Haberl \cite{MotchZH03} suggested that
some neutron stars can possess a hydrogen atmosphere of a
finite thickness above the solid iron surface. If the
optical depth of such atmosphere is small for some
wavelengths and large for other ones, this should lead to a
peculiar spectrum, different from the spectra of thick
atmospheres. Such spectra were calculated in
\cite{Ho-ea07,SuleimanovPW09,Suleimanov-ea10} using
simplified boundary conditions for the radiative transfer
equation at the inner boundary of the atmosphere. More
accurate boundary conditions have been suggested in
\cite{reflefit}, where the authors have taken into account
that an extraordinary or ordinary wave, falling from outside
on the interface, gives rise to reflected waves of both
polarizations, whose intensities add to the respective
intensities of the waves emitted by the condensed surface:
\bea
   I_{\omega,j}(\theta_k,\varphi) &=& 
   \sum_{j'=1,2} r_{\omega,jj'}(\theta_k,\varphi)\,
        I_{\omega,j'}(\pi-\theta_k,\varphi)
\nonumber\\&&
        +  \textstyle\frac12
        \big[1-r_{\omega,j}(\theta_k,\varphi)\big]
          \,\mathcal{B}_{\omega,T}.
\label{XX}
\eea
In Ref.~\cite{reflefit}, the reflectivity matrix was
calculated and fitted for linear polarizations, and then
converted into the reflectivity matrix  $\{r_{\omega,jj'}\}$
for normal modes pertinent to \req{XX}, using an
approximate relation valid for a sufficiently rarefied
photosphere.

\begin{figure}[t]
\begin{center}
\vspace*{1ex}
\includegraphics[width=\linewidth]{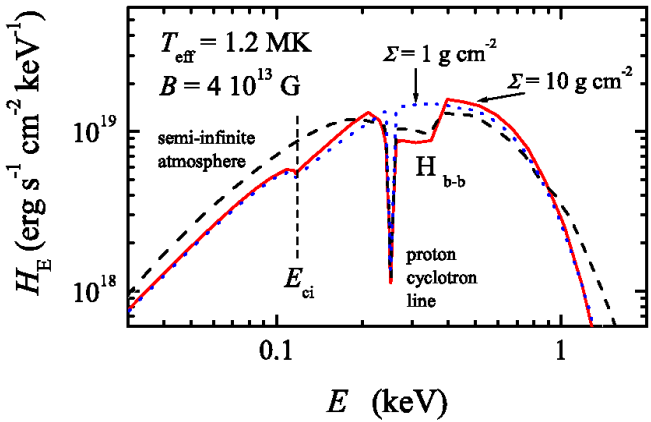}
\caption{
Comparison of the radiation spectrum of a neutron star with
a partially ionized thick hydrogen photosphere (dashed line)
with the spectra that are formed at hydrogen column
densities of 1 g cm$^{-2}$ (dots) and 10 g cm$^{-2}$ (solid line) over
the iron surface of the star (Fig.~12 from \cite{reflefit},
provided by V.~F. Suleimanov, reproduced with permission of
the author and
\copyright ESO.)
\label{fig:sp3a}}
\end{center}
\end{figure}

In Fig.~\ref{fig:sp3a} we show local spectra of radiation
emitted by hydrogen atmospheres of different thicknesses over
the iron neutron-star surface with the magnetic field
$B=4\times10^{13}$~G, normal to the surface with effective
temperature $\Ts=1.2\times10^6$~K. The narrow absorption
line corresponds to the proton cyclotron resonance in the
atmosphere. The feature to the right of it is related to
atomic transitions (H$_\mathrm{b-b}$). It has a large width
because of the motion effects (\S\,\ref{sect:motion}). This
feature is formed mainly at depths $\sim2$ g~cm$^{-2}$, that
is why it is almost invisible in the spectrum of the
thinnest atmosphere that has the column density of 1
g~cm$^{-2}$. The kink at $E_\mathrm{ci}=0.12$ keV
corresponds to the ion cyclotron energy of iron, therefore
it is absent for the pure hydrogen atmosphere. The
spectrum of the moderately deep atmosphere (10 g~cm$^{-2}$)
reveals all the three features. At high energies
($E\gtrsim1$ keV), the spectrum is determined by the
condensed-surface emission, because both finite atmospheres
are almost transparent at such energies.
The spectrum of the pure hydrogen atmosphere is harder in
this spectral range (cf.~\S\,\ref{sect:atmodels}).

The origin of the thin atmospheres remains hazy. Ho
\etal~\cite{Ho-ea07} discussed three possible scenarios.
First, it is the accretion from the interstellar medium. But
its rate should be very low, in order to accumulate the 
hydrogen mass $4\pi R^2\ycol \sim 10^{-20} M_\odot$ in
$\sim10^6$ years. Another scenario assumes diffusive nuclear
burning of a hydrogen layer, fell back soon after the
formation of the neutron star \cite{ChangBildsten03}. But
this process is too fast at the early cooling epoch, when
the star is relatively hot, and would have rapidly consumed
all the hydrogen on the surface \cite{ChangBildsten04}. The
third possibility is a self-regulating mechanism that is
driven by nuclear spallation in collisions with
ultrarelativistic particles at the regions of open field
lines, which leads to creation of protons and
alpha-particles. The estimate (\ref{magnetocoulomb}) for the
penetration depth of the magnetospheric accelerated
particles indicates that this process could create
a hydrogen layer of the necessary thickness $\ycol\sim1$ g
cm$^{-2}$.

It is natural to consider also an atmosphere having a helium
layer beneath the hydrogen layer. Indeed, all three
scenarios assume that a hydrogen-helium mixture appears
originally at the surface, and the strong gravity quickly
separates these two elements. Such ``sandwich atmosphere''
was considered in \cite{SuleimanovPW09}, where the authors
showed that its spectrum can have two or three absorption
lines in the range $E\sim(0.2$\,--\,1) keV at $B\sim10^{14}$~G.

\section{Theoretical inter\-pret\-ation of ob\-serv\-ed spectra}
\label{sect:obs}

As we have seen in \S\,\ref{sect:NMA}, theoretical models of
nonmagnetic atmospheres are successfully applied to analyses
of spectra of many neutron stars with relatively weak
magnetic fields $B\lesssim10^9$~G. There are only a few such
examples for the stars with strong magnetic fields. They
will be discussed in this section. At the end of the section
we will give a general compilation of modern estimates of
masses and radii of neutron stars with weak and strong
magnetic fields, based on the photosphere models.

\subsection{RX~J1856.5--3754}
\label{sect:1856}

As we discussed in \S\,\ref{sect:INS}, there is no
satisfactory description of the spectrum of the ``Walter
star'' RX~J1856.5--3754 based on nonmagnetic atmosphere
models. Simple models of magnetic atmospheres 
also failed to solve this problem. It was
necessary to explain simultaneously the form of the spectrum
in the X-ray and optical ranges that reveal substantially
different color temperatures $\Tbb^\infty$, along with the
complete absence of absorption lines or other spectral
features that was confirmed at a high significance level. To
solve this problem, Ho~\cite{Ho-ea07,Ho07} involved the
model of a partially ionized hydrogen atmosphere of finite
thickness above a condensed iron surface with a strong
magnetic field. He managed to reproduce the
measured spectrum of RX~J1856.5--3754 in the entire range
from X-rays to optical within observational errorbars.
The best agreement between the theoretical and observed
spectra has been achieved at the atmosphere column density
$\ycol=1.2$ g cm$^{-2}$,
$B\sim(3$\,--\,$4)\times10^{12}$~G,
$\Teff^\infty=(4.34\pm0.03)\times10^5$~K, $z_g=0.25\pm0.05$,
and $R_\infty=17.2^{+0.5}_{-0.1}\,D_{140}$  km. Here, the
errors are given at the $1\sigma$ significance level, and
$D_{140}\equiv D/(140$ pc). Note that a fit of the observed
X-ray spectrum with the Planck function yields a 70\% higher
temperature and a 3.5 times smaller radius of the emitting
surface. Such huge difference exposes the importance of a
correct physical interpretation of an observed spectrum for
evaluation of neutron-star parameters.

With the aid of expressions
(\ref{Tinfty})\,--\,(\ref{Rinfty})
and \req{r_g}, we obtain from these estimates
$\Teff=(5.4\pm1.1)\times10^5$~K,
$R=13.8^{+0.9}_{-0.6}\,D_{140}$ km, and
$M=1.68^{+0.22}_{-0.15}\,D_{140}\,M_\odot$. Forgetting for a
moment the factor $D_{140}$, one might conclude that this
radius is too large for such mass. However, the distance to
the star is not very accurately known. The value $D=140$ pc
was adopted in \cite{Ho-ea07} from \cite{Kaplan-ea02} and
lies between alternative estimates
$D\approx117$ pc \cite{WL02} and $D\approx(160$\,--\,170) pc
\cite{vKK07,Kaplan-vKA07}. More recently, a more accurate
estimate of the distance was obtained, $D=123^{+11}_{-15}$
pc \cite{Walter-ea10}. With the latter estimate, we obtain
$R=12.1^{+1.3}_{-1.6}$ km and
$M=1.48^{+0.16}_{-0.19}\,M_\odot$, which removes
all the contradictions. Nevertheless, the given
interpretation of the spectrum is not indisputable, since it
does not agree with the magnetic-field estimate
$B\approx1.5\times10^{13}$~G that has been obtained for
this star from \req{PPdot} in \cite{vKK08}. 

Using the same thin-atmosphere model, Ho \cite{Ho07}
analyzed the light curve of RX~J1856.5--3754 and obtained
constraints on the angles $\alpha$ and $\zeta$
(Fig.~\ref{fig:bending}). It turned out that the light curve
can be explained if one of these angles is small
($<6^\circ$), while the other angle lies between  $20^\circ$
and $45^\circ$. In this case, the radio emission around the
magnetic poles does not cross the line of sight. As
noted in \cite{Ho07}, this may explain the
non-detection of this star as a radio pulsar
\cite{Kondratiev-ea09}.

\subsection{RBS 1223}
\label{sect:RBS1223}

Hambaryan \etal~\cite{Hambaryan-ea11} analyzed the spectrum
of the X-ray source RBS 1223, by a method analogous to the
case of
RX~J1856.5--3754 described in \S\,\ref{sect:1856}. RBS 1223
reveals a complex structure of the 
X-ray spectrum, which can be described by a wide absorption
line centered around $\hbar\omega=0.3$~keV, superposed on
the Planck spectrum, with the line parameters depending on
the stellar rotation phase. Using all 2003\,--\,2007
\textit{XMM-Newton} observations of this star, the
authors~\cite{Hambaryan-ea11} obtained a set of X-ray
spectra for different rotation phases. They tried to
interpret these spectra with different models, assuming
magnetic fields $B\sim10^{13}$\,--\,$10^{14}$~G, different
atmosphere compositions, possible presence of a condensed
surface and a finite atmosphere. Different surface
temperature distributions were  described by a
self-consistent parametric model of
Ref.~\cite{PerezAzopinMP06}.

As a result, the authors \cite{Hambaryan-ea11} managed to describe
the observed spectrum and its rotational phase dependence
with the use of the model of the iron surface covered by
partially ionized hydrogen atmosphere with
$\ycol\sim1$\,--\,10 g~cm$^{-2}$, with mutually consistent
asymmetric bipolar distributions of the magnetic field and the
temperature, with the polar values
$B_\textrm{p1}=B_\textrm{p2}=(0.86\pm0.02)\times10^{14}$~G,
$T_\mathrm{p1}=1.22^{+0.02}_{-0.05}$ MK, and
$T_\mathrm{p2}=1.15\pm0.04$ MK.
The magnetic field and
temperature proved to be rather smoothly distributed over
the surface. When compared to the theoretical model
\cite{PerezAzopinMP06}, it implies the absence of a
superstrong toroidal component of the crustal magnetic
field. The integral effective temperature is 
$\Teff\approx0.7$ MK. The gravitational redshift is
estimated to be $z_g=0.16^{+0.03}_{-0.01}$, which
converts into $(M/M_\odot)/R_6=0.87^{+0.13}_{0.05}$ and
suggests a stiff EOS of the neutron-star matter.

We must note that the paper \cite{Hambaryan-ea11} preceded
the work \cite{reflefit}, which was discussed in
\S\,\ref{sect:NScond}. For this reason, the authors of
\cite{Hambaryan-ea11} used rough approximations for the
iron-surface emissivity, published before, and simplified
boundary conditions for the radiative transfer equations. An
analysis of the same spectra with the use of the improved
results for the emissivity and more accurate boundary
conditions, described in \S\,\ref{sect:NScond}, remains to
be done in the future.

\begin{table*}[t]
\begin{center}
\vspace*{-1ex}
\caption{
Estimates of neutron-star masses and radii based on
atmosphere models.}
\label{tab:obs}
\begin{tabular}{lccccc}
\hline
\hline\rule[-1.2ex]{0pt}{3.5ex}
~~~~Object & $R$ (km)
             & $M$ ($M_\odot$) 
             & $D$ (kpc)
             & Ref.
             & Notes
 \rule[-1.2ex]{0pt}{3.5ex}\\
\hline
qLMXB X7 in 47 Tuc  & $14.5^{+1.8}_{-1.6}$
             & $1.4$ [a]
             & 4.85 [a]
             &  \cite{Heinke-ea06} & [b]
\rule{0pt}{3ex}\\
qLMXB XTE 1701--462& $10.5\pm2.5$ [c]
             & 1.4 [a]
             & 8.8 [a]
             & \cite{Fridriksson-ea11} & [b]
  \rule{0pt}{3ex}\\
qLMXB EXO 0748--676 & $13.7^{+1.0}_{-2.7}$
             & $1.8^{+0.4}_{-0.6}$
             & 7.1 [a]
             & \cite{DiazTrigo-ea11} & [b]
\rule{0pt}{3ex}  \\
same object & $11.8^{+0.7}_{-2.2}-15.2^{+1.5}_{-3.0}$
             & $1.5^{+0.5}_{-0.5}-2.1^{+0.4}_{-0.8}$
             & $5.9-8.3$ 
             & \cite{DiazTrigo-ea11} & [d], [e]
 \rule{0pt}{0pt}\\
qLMXB in M28  & $10.5^{+2.0}_{-2.9}$
             & $1.25^{+0.54}_{-0.63}$
             & $5.5$ [a]
             & \cite{Guillot-ea13} & [b]
  \rule{0pt}{3ex}\\
same object& $9^{+3}_{-3*}$ 
             & $1.4^{+0.4}_{-0.9*}$
             & $5.5$ [a]
             & \cite{Servillat-ea12} & [b] \rule{0pt}{0pt}\\
same object& $14^{+3}_{-8*}$ 
             & $2.0^{+0.5}_{-1.5*}$
             & $5.5$ [a]
             & \cite{Servillat-ea12} & [e] \rule{0pt}{0pt}\\
qLMXB in NGC 6397     & $6.6^{+1.2}_{-1.1}$
             & $0.84^{+0.30}_{-0.28}$
             & $2.02$ [a]
             & \cite{Guillot-ea13} & [b]
 \rule{0pt}{3ex}\\
qLMXB in M13  & $10.1^{+3.7}_{-2.8}$
             & $1.27^{+0.71}_{-0.63}$
             & $6.5$ [a]
             & \cite{Guillot-ea13} & [b]
 \rule{0pt}{3ex}\\
same object& $10.6^{+2.1}_{-2.2}$
             & 1.4 [a]
             & 7.7 [a]
             & \cite{Catuneanu-ea13} & [b] \rule{0pt}{0pt}\\
same object& $14.6^{+3.5}_{-3.1}$
             & 1.4 [a]
             & 7.7 [a]
             & \cite{Catuneanu-ea13} & [e] \rule{0pt}{0pt}\\
qLMXB in $\omega$ Cen & $20.1^{+7.4}_{-7.2}$
             & $1.8^{+1.0}_{-1.1}$
             & $4.8$ [a]
             & \cite{Guillot-ea13} & [b]
 \rule{0pt}{3ex}\\
qLMXB in NGC 6304 & $9.6^{+4.9}_{-3.4}$
             & $1.16^{+0.90}_{-0.56*}$
             & $6.22$ [a]
             & \cite{Guillot-ea13} & [b]
  \rule{0pt}{3ex}\\
CCO in Cas A  & $15.6^{+1.3}_{-2.7}$
             & $1.4$ [a] 
             & 3.4 [a]
             & \cite{HoHeinke09} & [f]
  \rule{0pt}{3ex}\\
same object & $8-17$ & $1.5-2.4$ & $3.3-3.7$
             & \cite{HoHeinke09} & [d], [f] \rule{0pt}{0pt}\\
CCO in HESS J1731  &
$12.6^{+2.1}_{-5.3}-15.6^{+3.6}_{-5.3}$
             & $1.5^{+0.4}_{-0.6}-
2.2^{+0.3}_{-0.9}$
             & $3.2-4.5$
             & \cite{Klochkov-ea13} & [d], [f]
\rule{0pt}{3ex}\\
Burster 4U 1724--307 & $14.7\pm0.8$
             & $1.9\pm0.4$
             & 5.3--7.7
             & \cite{Suleimanov-ea11} &[b]
\rule{0pt}{3ex}\\
same object & $18\pm3.5$
             & $1.05^{+0.55}_{-0.4}$
             & 5.3--7.7
             & \cite{Suleimanov-ea11} &[e]
\rule{0pt}{0pt}\\
Burster GS 1826--24 & $<8.2$
             & $<1.3$
             & $<4.3$ [g]
             & \cite{ZamfirCG12} &[b]
\rule{0pt}{3ex}\\
same object & $<19.8$
             & $<2.8$
             & $<9.7$ [g]
             & \cite{ZamfirCG12} &[e]
\rule{0pt}{3ex}\\
PSR J0437--4715 & $ > 11.1$ ($3\sigma$) & 1.76 [a]
                & 0.1563 [a]
                & \cite{Bogdanov13}
                    & [b]; see \S\,\ref{sect:msPSR}
\rule[-1.4ex]{0pt}{4ex}\\
\hline
XDINS RX~J1856 & $12.1^{+1.3}_{-1.6}$
             & $1.48^{+0.16}_{-0.19}$
             & $0.123^{+0.011}_{-0.015}$ [h]
             & \cite{Ho-ea07} & [c], [i], [j]; see \S\,\ref{sect:1856}
  \rule{0pt}{3ex}\\
XDINS RBS 1223 & $16^{+1}_{-2}$
             & 1.4 [a]
             & $0.380^{+0.015}_{-0.030}$
             & \cite{Hambaryan-ea11}
                         & [c], [i], [k]; see \S\,\ref{sect:RBS1223}
 \rule[-1.2ex]{0pt}{4.2ex}\\
\hline
\hline
\end{tabular}
\end{center}
\small{Notes: Errors are listed at significance level
90\%, unless otherwise stated. The asterisk at a value of
an error signifies that a hard limit of a model was reached.
[a] The parameter is fixed.
[b] Nonmagnetic H atmosphere.
[c] Errors at the significance level $1\sigma$ (68\%).
[d] Results for selected limiting $D$ from a range of
possible values.
[e] Nonmagnetic He atmosphere.
[f] Nonmagnetic  C atmosphere.
[g] Constraint on $D\xi_\mathrm{b}^{1/2}$ is given, where $\xi_\mathrm{b}$ is the
anisotropy factor.
[h] $D$ is adopted from \cite{Walter-ea10}.
[i] Partially ionized thin H atmosphere 
over iron surface.
[j] $\ycol=1.2$ g cm$^{-2}$, $B\sim(3$\,--\,$4)\times10^{12}$~G.
[k] $\ycol\sim1$ g cm$^{-2}$, mutually consistent
distributions of magnetic field
$B\sim8\times10^{13}$~G and temperature $\Ts\sim0.7$ MK.
}
\end{table*}

\subsection{1E~1207.4--5209}

The discovery of absorption lines in the spectrum of CCO 1E
1207.4--5209 at energies $E\sim0.7\,N$~keV ($N=1,2,\ldots$)
immediately entrained the natural assumption that they are
caused by cyclotron harmonics \cite{Bignami-ea03}. As we
have seen in \S\,\ref{sect:crosssect}, such harmonics can be
only electronic, as the ion harmonics are unobservable.
Therefore, this interpretation implies
$B\approx7\times10^{10}$~G. Mori \etal~\cite{MoriCH} showed
that only the first and second lines in the spectrum of 1E
1207.4--5209 are statistically significant, but some authors
take also the third and fourth lines into account. This
hypothesis was developed in \cite{SuleimanovPW12}, where the
authors include in the treatment both types of the electron
cyclotron harmonics that were discussed in
\S\,\ref{sect:crosssect}: the quantum oscillations of the
Gaunt factor and the relativistic thermal harmonics. It is
possible that the analogous explanation of the shape of the
spectrum may be applied also to CCO PSR J0821--4300
\cite{GotthelfHA13}.

Mori \etal~\cite{MoriHailey06,MoriHo} have critically
analyzed the earlier hypotheses about the origin of the
absorption lines in the spectrum of 1E 1207.4--5209 and
suggested their own explanation. They analyzed and rejected
such interpretations as the lines of molecular hydrogen
ions, helium ions, and also as the cyclotron lines and their
harmonics. One of the arguments against the latter
interpretation is that the fundamental cyclotron line should
have much larger depth in the atmosphere spectrum than
actually observed. Another argument is that the cyclotron
lines and harmonics have small widths at a fixed $B$,
therefore their observed width in the integral spectrum is
determined by the $B$ distribution. Thus their width should
be the same, in contradiction to observations
\cite{MoriHailey06}. These arguments were neglected in
\cite{SuleimanovPW12}. It has to be noted that in
\cite{MoriHailey06}, as well as in \cite{SuleimanovPW12},
the authors studied the cyclotron harmonics in spectra of
fully ionized plasmas. The effect of the partial ionization
on the model spectrum remains unexplored.

As an alternative, Mori \etal{} \cite{MoriHailey06,MoriHo}
suggested models of atmospheres composed of mid-$Z$
elements. An example of such spectrum is shown in
Fig.~\ref{fig:MoriHo}. Its convolution with the telescope
point-spread function smears the line groups, producing wide
and shallow suppressions of the spectral flux, similar to
the observed ones. Integration of the local spectrum over
the stellar surface, whose necessity we mentioned in
\S\,\ref{sect:atmodels}, should lead to an additional
smearing of the spectral features. The authors \cite{MoriHo}
found that an oxygen atmosphere with magnetic field
$B=10^{12}$~G provides a spectrum similar to the observed
one. However, the constraint $B<3.3\times10^{11}$~G that was
obtained in \cite{HalpernGotthelf} disagrees with this
model, but rather favors the cyclotron interpretation of the
lines.

Unlike the cases of RX~J1856.5--3754 and RBS 1223 that were
considered above, there is no published results of a
detailed fitting of the observed spectrum of 1E~1207.4--5209
with a theoretical model. Thus the applicability of
any of them remains hypothetical.

\subsection{PSR J1119--6127}
\label{sect:1119}

Recently, the partially ionized, strongly magnetized
hydrogen photosphere model \cite{HoPC} has been successfully
applied to interpret the observations of pulsar J1119--6127
\cite{Ng-ea12}, for which the estimate (\ref{PPdot}) gives
an atypically high field $B=4\times10^{13}$~G. In the X-ray
range, it emits pulsed radiation, which has
apparently mostly thermal nature. At fixed $D=8.4$ kpc
and $R=13$, the bolometric flux gives an estimate of the
mean effective temperature $\Teff\approx1.1$~MK. It
was difficult to explain, however, the large pulsed
fraction ($48\pm12$\%) by the thermal emission.  The
authors  \cite{Ng-ea12} managed to reproduce the X-ray light
curve  of this pulsar assuming that one of its magnetic
poles is surrounded by a heated area, which occupies 1/3 of
the surface, is covered by hydrogen and heated to
$1.5$~MK, while the temperature of the opposite
polar cap is below $0.9$~MK.
\vspace*{-1em}

\subsection{Masses and radii: the results}
\label{sect:MR}

Table~\ref{tab:obs} presents modern estimates of
neutron-star masses and radii, obtained from analyses of
their thermal spectra with the atmosphere models. The estimates
that fixed surface gravity in advance are not listed here,
because they are strongly biased, as shown, e.g., in
\cite{Heinke-ea06}.

In most cases determination of the neutron-star radii
remains unreliable. The estimates are done, as a rule, at a
fixed distance $D$. An evaluation of $R$ is also often
performed for a fixed mass $M$. In most cases it is
stipulated by the fact that a joint evaluation of $R$ and
$M$ (let alone $R$, $M$, and $D$) from the currently
available thermal spectra leaves too large uncertainties and
almost does not constrain $M$ and $D$. A comparison of the
results obtained for the same objects with different
assumptions on $D$ values readily shows that a choice of $D$
can drastically affect the $R$ estimate. In addition,
the estimate of $R$ is strongly affected by assumptions on
the photosphere composition, as one can see, for example,
from a comparison of the results obtained by assuming
hydrogen and helium photospheres for the qLMXBs in globular
clusters M28 \cite{Servillat-ea12} and M13
\cite{Catuneanu-ea13}.

\section{Conclusions}
\label{sect:concl}

We have considered the main features of neutron-star
atmospheres and radiating surfaces and outlined the current
state of the theory of the formation of their spectra. The
observations of bursters and neutron stars in low-mass X-ray
binaries are well described by the nonmagnetic atmosphere
models and yield ever improving  information on the key
parameters such as the neutron-star masses, radii, and
temperatures.  The interpretation of observations enters a
qualitatively new phase, unbound from the blackbody spectrum
or the ``canonical model'' of neutron stars. Absorption
lines have been discovered in thermal spectra of strongly
magnetized neutron stars. On the agenda is their detailed
theoretical description, which provides information on the
surface composition, temperature and magnetic field
distributions. Indirectly it yields information on heat
transport and electrical conductivity in the crust, neutrino
emission, nucleon superfluidity, and proton
superconductivity in the core. In order to clear up this
information, it still remains to solve a number of problems
related to the theory of the magnetic atmospheres and
radiating surfaces. Let us mention just a few of them.

First, the calculations of the quantum-mechanical properties
of atoms and molecules in strong magnetic fields beyond the
adiabatic approximation have been so far performed only for
atoms with $\Znuc\lesssim10$ and for one- and two-electron
molecules and molecular ions. The thermal motion effect on
these properties has been rigorously treated only for the
hydrogen atom and helium ion, and approximately for the
heavier atoms. It is urgent to treat the finite nuclear mass
effects for heavier atoms, molecules, and their ions,
including not only binding energies and characteristic
sizes, but also cross sections of interaction with
radiation. This should underlie computations of photospheric
ionization equilibrium and opacities, following the
technique that is already established for the hydrogen
photospheres. In the magnetar photospheres, one can
anticipate the presence of a substantial fraction of exotic
molecules, including polymer chains. The properties of such
molecules and their ions are poorly known. In particular,
nearly unknown are their radiative cross sections that are
needed for the photosphere modeling.

Second, the emissivities of condensed
magnetized surfaces have been calculated in frames of
the two extreme models of free and fixed
ions. It will be useful to do similar calculations
using a more realistic description of ionic bonding
in a magnetized condensed matter. This should be particularly
important in the frequency range $\omega\lesssim\omci$,
which is observable for the thermal spectrum
in the superstrong magnetic fields.

Third, the radiative transfer theory, currently used for
neutron-star photospheres, implies the electron plasma
frequency to be much smaller than photon frequencies. In
superstrong magnetic fields, this condition is violated
in a substantial frequency range. Thus the theory of
magnetar spectra requires a more general treatment of
radiative transfer in a magnetic field.

In conclusion, I would like to thank my colleagues, with
whom I had a pleasure to work on some of the problems
described in this review: V.G.~Bezchastnov, G.~Chabrier,
W.C.G.~Ho, D.~Lai, Z.~Medin, G.G.~Pavlov, Yu.A.~Shibanov,
V.F.~Suleimanov, M.~van Adelsberg, J.~Ventura, K.~Werner. My
special thanks are to Vasily Beskin, Wynn Ho, Alexander
Kaminker, Igor Malov, Dmitry Nagirner, Yuri Shibanov, and Valery Suleimanov for useful
remarks on preliminary versions of this article. This work
is partially supported by the Russian Ministry of Education
and Science (Agreement 8409, 2012), Russian Foundation for
Basic Research (Grant 11-02-00253), Programme for Support
of the Leading Scientific Schools of the Russian Federation
(Grant NSh--294.2014.2), and PNPS (CNRS/INSU, France).

\newcommand{\artref}[4]{\textit{#1} \textbf{#2} #3 (#4)}
\newcommand{\AandA}[3]{\artref{Astron.\ Astrophys.}{#1}{#2}{#3}}
\newcommand{\AnnPhysNY}[3]{\artref{Ann.\ Phys. (N.Y.)}{#1}{#2}{#3}}
\newcommand{\ApJ}[3]{\artref{Astrophys.\ J.}{#1}{#2}{#3}}
\newcommand{\ApJS}[3]{\artref{Astrophys.\ J.\ Suppl.\ Ser.}{#1}{#2}{#3}}
\newcommand{\ApSS}[3]{\artref{Astrophys.\ Space Sci.}{#1}{#2}{#3}}
\newcommand{\ARAA}[3]{\artref{Annu.\ Rev.\ Astron.\ Astrophys.}{#1}{#2}{#3}}
\newcommand{\JPB}[3]{\artref{J.\ Phys.\ B: At.\ Mol.\ Opt.\ Phys.}{#1}{#2}{#3}}
\newcommand{\jpb}[3]{\artref{J.\ Phys.\ B: At.\ Mol.\ Phys.}{#1}{#2}{#3}}
\newcommand{\AL}[3]{\artref{Astron.\ Lett.}{#1}{#2}{#3}}
\newcommand{\SvAL}[3]{\artref{Sov.\ Astron.\ Lett.}{#1}{#2}{#3}}
\newcommand{\SvA}[3]{\artref{Sov.\ Astron.}{#1}{#2}{#3}}
\newcommand{\MNRAS}[3]{\artref{Mon.\ Not.\ R.\ astron.\ Soc.}{#1}{#2}{#3}}
\newcommand{\PL}[4]{\artref{Phys.\ Lett. #1}{#2}{#3}{#4}}
\newcommand{\PR}[4]{\artref{Phys.\ Rev. #1}{#2}{#3}{#4}}
\newcommand{\PRL}[3]{\artref{Phys.\ Rev.\ Lett.}{#1}{#2}{#3}}
\newcommand{\RMP}[3]{\artref{Rev.\ Mod.\ Phys.}{#1}{#2}{#3}}
\newcommand{\SSRv}[3]{\artref{Space Sci.\ Rev.}{#1}{#2}{#3}}

\renewcommand{\refname}{References}
\makeatletter
\renewcommand\@biblabel[1]{#1.}
\makeatother

\label{sect:bib}

\end{document}